\documentclass[11pt,a4paper,oneside]{article}
\pdfoutput=1

\usepackage[affil-it]{authblk}

\usepackage[a4paper,textwidth=16.5cm,textheight=24cm,centering]{geometry}

\usepackage[]{amsmath}
	\numberwithin{equation}{section}
\usepackage[]{amssymb}
\usepackage{amsfonts}
\usepackage{mathrsfs}
\usepackage{amsthm}

\usepackage{esint}

\usepackage[matrix,arrow,color]{xy}

\usepackage{bbm}

\theoremstyle{definition}

	\newtheorem{lem}[equation]{Lemma}
	\newtheorem{prop}[equation]{Proposition}
	\newtheorem{cor}[equation]{Corollary}
	\newtheorem{defin}[equation]{Definition}

\DeclareMathOperator{\R}{\mathbb{R}}
\DeclareMathOperator{\C}{\mathbb{C}}
\DeclareMathOperator{\N}{\mathbb{N}}
\DeclareMathOperator{\Z}{\mathbb{Z}}

\DeclareMathOperator{\ct}{\mathbb{T}}

\newcommand{\mz}{\mathcal{Z}}

\newcommand{\sA}{\mathsf{A}}
\newcommand{\sB}{\mathsf{B}}
\newcommand{\sC}{\mathsf{C}}
\newcommand{\sO}{\mathsf{O}}
\newcommand{\sK}{\mathsf{K}}
\newcommand{\sZ}{\mathsf{Z}}

\newcommand{\sP}{\mathsf{P}}

\newcommand{\dd}{\mathrm{d}}
\newcommand{\ii}{\mathrm{i}}
\newcommand{\e}{\mathrm{e}}
\newcommand{\id}{\mathbbm{1}}
\newcommand{\tr}{\mathrm{Tr}}
\newcommand{\hd}{\ast\! }

\newcommand{\UN}{\mathrm{U}( N )}
\newcommand{\SUN}{\mathrm{SU}( N)}
\newcommand{\PSUN}{\mathrm{PSU}( N )}
\newcommand{\wSUN}{\widetilde{\mathrm{SU}}( N )}
\newcommand{\Uu}{\mathrm{U}( 1 )}
\newcommand{\SpiN}{\mathrm{Spin}( N )}
\newcommand{\SON}{\mathrm{SO}( N )}

\newcommand{\ON}{\mathrm{O}( N )}
\newcommand{\PiN}{\mathrm{Pin}^{+}( N )}

\makeatletter
\newcommand*{\doublerightarrow}[2]{\mathrel{
  \settowidth{\@tempdima}{$\scriptstyle#1$}
  \settowidth{\@tempdimb}{$\scriptstyle#2$}
  \ifdim\@tempdimb>\@tempdima \@tempdima=\@tempdimb\fi
  \mathop{\vcenter{
    \offinterlineskip\ialign{\hbox to\dimexpr\@tempdima+1em{##}\cr
    \rightarrowfill\cr\noalign{\kern.1ex}
    \rightarrowfill\cr}}}\limits^{\!#1}_{\!#2}}}
\newcommand*{\triplerightarrow}[1]{\mathrel{
  \settowidth{\@tempdima}{$\scriptstyle#1$}
  \mathop{\vcenter{
    \offinterlineskip\ialign{\hbox to\dimexpr\@tempdima+1em{##}\cr
    \rightarrowfill\cr\noalign{\kern.5ex}
    \rightarrowfill\cr\noalign{\kern.5ex}
    \rightarrowfill\cr}}}\limits^{\!#1}}}
\makeatother

\usepackage[utf8]{inputenc}
\usepackage[english]{babel}
\usepackage[pdftex]{graphicx}
\usepackage[dvipsnames]{xcolor}
\usepackage[toc,page]{appendix}
\usepackage[nottoc]{tocbibind}
\usepackage{caption}
\usepackage{enumerate}
\usepackage{ytableau}

\usepackage{tikz}
		\usetikzlibrary{er,positioning,arrows.meta}
\usepackage{tqft}

\definecolor{colU}{rgb}{0.76, 0.23, 0.13}
\definecolor{colC}{rgb}{0.0, 0.5, 0.5}

\usepackage[linktocpage=true,colorlinks=true,citecolor=Blue,linkcolor=blue,urlcolor=Blue]{hyperref}

\usepackage[numbers,compress,square,comma]{natbib}
\newcommand\aafootnote[1]{%
  \begingroup
  \renewcommand\thefootnote{}\footnote{#1}%
  \addtocounter{footnote}{-1}%
  \endgroup
}

\begin{document}
\bibliographystyle{myJHEP}
\captionsetup[figure]{labelfont={bf,small},labelformat={default},labelsep=period,font=small}
\captionsetup[table]{labelfont={bf,small},labelformat={default},labelsep=period,font=small}

{\pagenumbering{roman}

	\title{\huge\textbf{Higher form symmetries and orbifolds\protect\\[1mm] of two-dimensional Yang--Mills theory}}

\author[$\diamond$]{Leonardo Santilli}
\affil[$\diamond$]{\small Yau Mathematical Sciences Center, Tsinghua University, Beijing 100084, China}

\author[$\ast$,$\dagger$]{Richard J. Szabo}
\affil[$\ast$]{\small School of Mathematical and Computer Sciences, Heriot--Watt University, Edinburgh EH14 4AS, U.K.}
\affil[$\dagger$]{\small Maxwell Institute for Mathematical Sciences, Edinburgh, U.K.}
	
	\date{ \hspace{8pt} }
	
	\maketitle
	\thispagestyle{empty}
	
\aafootnote{$^{\diamond}$santilli@tsinghua.edu.cn}	
\aafootnote{$^{\ast}$R.J.Szabo@hw.ac.uk}

	\begin{abstract}
	\noindent
We undertake a detailed study of the gaugings of two-dimensional Yang--Mills theory by its intrinsic charge conjugation 0-form and centre 1-form global symmetries, elucidating their higher algebraic and geometric structures, as well as the meaning of dual lower form symmetries. Our derivations of orbifold gauge theories make use of a combination of standard continuum path integral methods, networks of topological defects, and techniques from higher gauge theory. We provide a unified description of higher and lower form gauge fields for a $p$-form symmetry in the geometric setting of $p$-gerbes, and derive reverse orbifolds by the dual $(-1)$-form symmetries. We identify those orbifolds in which charge conjugation symmetry is spontaneously broken, and relate the breaking to mixed anomalies involving $(-1)$-form symmetries. We extend these considerations to gaugings by the non-invertible 1-form symmetries of two-dimensional Yang--Mills theory by introducing a notion of generalized $\theta$-angle.
	\end{abstract}

	\clearpage
	{
	\baselineskip=14pt
	\tableofcontents
}}	
		
	\clearpage
	\pagenumbering{arabic}
	\setcounter{page}{1}
		\renewcommand*{\thefootnote}{\arabic{footnote}}
		\setcounter{footnote}{0}
\setlength{\parskip}{\medskipamount}

\section{Introduction}

The realization that symmetries in quantum field theory are completely described using topological defect operators has led to a shift of paradigm, paving the way to wide generalizations of the standard notion of symmetry \cite{Gaiotto:2014kfa}. The two avenues for defining generalized symmetries consist in abandoning the restriction that the topological defects have codimension one, leading to higher form symmetries~\cite{Gaiotto:2014kfa}, as well as abandoning the requirement that the fusion of these operators should satisfy a corresponding group law, leading to non-invertible symmetries \cite{Bhardwaj:2017xup,Chang:2018iay}; see~\cite{Cordova:2022ruw,Schafer-Nameki:2023jdn,Luo:2023ive} for reviews. These are organised algebraically by higher groups and more general structures from category theory.
Generalized symmetries have proven to yield very concrete applications in gauge theory and particle physics phenomenology \cite{Gaiotto:2017yup,Hidaka:2020iaz,Choi:2022jqy,Cordova:2022ieu,Choi:2022rfe,Cordova:2022fhg,Karasik:2022kkq,Cordova:2022qtz,Brennan:2023kpw,Putrov:2023jqi,Damia:2023ses,Copetti:2023mcq,Damia:2023gtc,Choi:2023pdp,Cordova:2023her,Cordova:2024ypu}; we refer to the reviews~\cite{Brennan:2023mmt,Shao:2023gho} for an introduction to applications of higher form and non-invertible symmetries to the standard model and beyond.

It is a classic result that gauging a finite symmetry in two spacetime dimensions produces a dual, or `quantum', symmetry \cite{Vafa:1989ih}; the reverse orbifold by this dual symmetry undoes the gauging and brings one back to the original field theory. This statement admits a generalization: gauging a $p$-form symmetry in $d$ spacetime dimensions, implemented by topological defects of codimension~$p+1$ (which can wrap $p$-dimensional charged operators), produces a dual $(d-p-2)$-form symmetry~\cite{Tachikawa:2017gyf}, not necessarily invertible. In this language the result of \cite{Vafa:1989ih} refers to an `invertible 0-form symmetry'.

Let us explain this fact in the example of pure Yang--Mills theory with simply-connected gauge group $G$ in $d$ dimensions. The centre $\sZ (G)$ of $G$ is a global symmetry of the theory. It leaves the local gauge-invariant operators unaffected, while the Wilson lines of Yang--Mills theory transform with a non-trivial charge under the centre symmetry. Therefore $\sZ (G)$ is an electric 1-form symmetry, and gauging it leads to $d$-dimensional Yang--Mills theory with centreless gauge group $\mathrm{P}G = G/\sZ(G)$ and a magnetic $(d-3)$-form symmetry.

\subsubsection*{Gauging the $\boldsymbol{(d-1)}$-form symmetry}

The highest possible symmetry in $d$ spacetime dimensions is the $(d-1)$-form symmetry, whose generators are topological pointlike operators. One of the motivations for the present paper is to understand the algebraic and geometric structures that emerge upon gauging $(d-1)$-form symmetries in $d$ dimensions, as well as their physical implications. In this case the dual symmetry is a $(-1)$-form symmetry, and the question arises as to the precise meaning and nature of this lower form symmetry.

The pragmatic definition of a $(-1)$-form symmetry relies on considering the $d$-dimensional topological operators in the theory; as these fill the spacetime they are sometimes referred to as `background operators'. There are, however, two roadblocks along this path. First and foremost, as we shall show, background $(-1)$-form gauge fields modify the parameters of the theory. Thus this is not a symmetry in any canonical sense. The study of orbifolds by the $(d-1)$-form symmetry prompts us to rethink the definition of symmetry that accommodates the lower form symmetry. There are two possibilities:

\begin{itemize}
\item Consider the full family of quantum field theories fibred over the parameter space, and enlarge the standard notion of symmetry to encompass automorphisms of this fibration. In this picture, $p$-form symmetries act fibrewise if $p \ge 0$ and on the base if $p<0$. They may generically combine into higher groups.

\item Abandon the idea that the orbifold by a $(d-1)$-form symmetry produces a dual symmetry. This is however in conflict with the widespread definition of symmetries as topological defect operators because, upon gauging the $(d-1)$-form symmetry, the orbifold theory \emph{does} possess $d$-dimensional topological operators.\footnote{Topological invariance for defect operators of the same dimension as the spacetime is made precise in Subsection~\ref{sec:dualgauging}.}
\end{itemize}

In this paper we will adopt the first approach, as was also done recently in \cite{Heckman:2024oot} for applications to holography. Our results remain valid in the second approach, in which case the $(-1)$-form symmetry should be reinterpreted accordingly.

Resolving the tension with lower form symmetries becomes especially relevant in light of the `decomposition conjecture' of~\cite{Hellerman:2006zs,Sharpe:2014tca}. This is the expectation that a $d$-dimensional quantum field theory with a $(d-1)$-form symmetry is equivalent to the disjoint union of theories in which the symmetry is gauged using local topological operators. See \cite{Sharpe:2019ddn,Sharpe:2021srf,Robbins:2022wlr} for a sample of recent developments related to the present work.

We encounter a second puzzle in the treatment of $(-1)$-form symmetries. It is expected that the higher structure capturing the full set of symmetries of a $d$-dimensional quantum field theory is a $(d-1)$-category \cite{Johnson-Freyd:2020usu,Bhardwaj:2022lsg}. The proposal is backed by the physical reasoning that the $p$-morphisms of the higher category are realized in the theory by the $(d-p-1)$-dimensional topological symmetry operators, for all $p=0,1, \dots, d-1$. This claim covers all higher form symmetries and has passed several consistency checks; however, it cannot accommodate $d$-dimensional topological operators. Thus the current proposals utilizing higher category theory do not include the lower form symmetries and, as a consequence, are lacking an intrinsic explanation of the operation of gauging the $(d-1)$-form symmetry. Reconciling the higher categorical picture with the  formulation of lower form symmetries presented in this paper is a challenging task left for future work.

\subsubsection*{The lower form symmetry and its applications}

The first systematic study of $(-1)$-form symmetries can be traced back to \cite{Cordova:2019jnf}. Their emergence from gauging a 2-group global symmetry was partly discussed in \cite{Yu:2020twi}, akin to the treatment of~\cite{Tachikawa:2017gyf} but including the lower form symmetries of two-dimensional quantum field theories. Consider a 1-form symmetry in three dimensions based on a finite group $\sB$, and a finite 0-form symmetry group $\sA$, which together mix to form a 2-group $\underline{\mathsf{\Gamma}}$ through an extension of 2-groups
\begin{equation*}
1\longrightarrow \mathrm{B}\sB\longrightarrow \underline{\mathsf{\Gamma}} \longrightarrow \sA \longrightarrow 1 \ .
\end{equation*}
According to~\cite{Tachikawa:2017gyf}, after gauging $\sB$ one obtains a 0-form symmetry \smash{$\sA \times \widehat{\sB}$} with a mixed anomaly. The analysis of \cite{Yu:2020twi} goes along the same lines, except that it is carried out in $d=2$: the dual symmetry \smash{$\widehat{\sB}$} is a $(-1)$-form symmetry, and \cite{Yu:2020twi} obtains a non-trivial anomaly also in such case. Our findings in Section~\ref{sec:anomalies} fit within this framework.

The $(-1)$-form symmetry in $d=2$ dimensions was also considered in \cite{Vandermeulen:2022edk}, with a concentrated focus on decomposition. Dynamical implications of $(-1)$-form $\Uu$ symmetries were discussed in \cite{Brennan:2020ehu,Damia:2022seq,Aloni:2024jpb}. The action of $(-1)$-form gauge transformations on the space of couplings provides additional support to the conjecture that quantum gravity does not admit free parameters \cite{McNamara:2020uza}.

One of the purposes of the present paper is to fully elucidate the lower form symmetry in the simple example of pure Yang--Mills theory in two dimensions, where the $(-1)$-form symmetry is a magnetic symmetry. The simplest example is pure $\Uu$ Yang--Mills theory, in which case the 1-form and $(-1)$-form $\Uu$ symmetries are the immediate analogues of the electric and magnetic $\Uu$ symmetries of Maxwell theory in four dimensions. This motivating example has also been considered recently in~\cite{Aloni:2024jpb}, which more generally agrees with our treatment of the periodicity of the $\theta$-angle as invariance under $(-1)$-form background gauge transformations; in particular, our unified geometric definition of higher and lower form $\Uu$ gauge symmetries encompasses the operational definition of $(-1)$-form symmetry from~\cite{Aloni:2024jpb}.

In this paper our main focus is the extension of these ideas to non-abelian Yang--Mills theory. Our constructions highlight the necessity, originating in considerations from higher gauge theory, of including the lower form symmetries. We shall furthermore study the gauging of non-invertible 1-form symmetries, thus extending these considerations beyond the cases that satisfy a group law.

\subsubsection*{Summary of results}

In this paper we shall undertake a detailed study of the algebraic and geometric structures that emerge when gauging a 1-form symmetry in two dimensions, as well as its physical implications. To this aim, we focus on the tractable example of pure Yang--Mills theory with gauge group $G$, which enables us to substantiate our results with explicit computations.

The orbifolds of Yang--Mills theory by an intrinsic finite abelian 1-form symmetry group are derived, utilizing both the standard continuum path integral formalism and techniques from higher gauge theory. One of the upshots of our analysis is that lower form symmetries naturally fit into the hierarchy of higher form symmetries, and can thereby be treated on the same footing. The topological structure of $p$-gerbes automatically includes $p$-form symmetries for any two-dimensional quantum field theory, with $p\in\{1,0,-1\}$, thereby laying the geometric foundation which underlies our novel definition of gauge fields for the $(-1)$-form symmetry. We  derive orbifolds by the dual $(-1)$-form symmetry, utilizing both path integral techniques and by summing over topological surface operators,  demonstrating that they recover the anticipated partially gauged theories.

Charge conjugation is an intrinsic finite 0-form symmetry of pure Yang--Mills theory, which participates with the intrinsic 1-form symmetries in the natural global automorphism 2-group of the gauge theory.
We show by an explicit computation that the theory after gauging a 1-form symmetry, with gauge group $G=\SUN/\Z_k$ where $N$ is even and $k$ is any power of 2 that divides $N$, has two degenerate vacua at $\theta=\pi$, and that charge conjugation symmetry is spontaneously broken. Additionally, we rule out the possibility of such spontaneous symmetry breaking for several other gauge groups. We complement the analysis with a study of the orbifolds by charge conjugation symmetry.

We relate this spontaneous symmetry breaking to the presence of an anomaly involving the $(-1)$-form symmetry, which parallels and extends the anomaly in the space of couplings found by~\cite{Cordova:2019jnf}. This means that it is not possible to preserve background gauge invariance when backgrounds for both the 1-form and $(-1)$-form symmetries are turned on. Therefore not only does the language of gerbes includes the treatment of $(-1)$-form symmetries, but it also allows for the extension of the notion of (mixed) 't Hooft anomalies to such cases.

The abelian 1-form symmetry group arising from the centre of the gauge group $G$ is the invertible part of the full 1-form symmetry of two-dimensional Yang--Mills theory. We also consider orbifolds by the non-invertible higher form symmetry. We introduce a generalization of the $\theta$-parameter to this non-invertible case, and show that it is given by an isomorphism class of irreducible representations of $G$. Equipped with this notion, we extend our previous results to the orbifolds by non-invertible symmetries. We present an in-depth study of the partition functions, discussing various gaugings with different generalized $\theta$-parameters. In particular, we identify those orbifolds that lead to spontaneous  breaking of charge conjugation symmetry in this generalized setting.

These results initiate the study of orbifolds by non-invertible higher form symmetries, and yield the first explicit realization of spontaneous symmetry breaking in such orbifold theories. 

\subsubsection*{Outline}

Throughout we frame the main results as Proposition and Corollary for emphasis.
The remainder of this paper is organized as follows.

Section \ref{sec:GTpre} serves to set the stage. We discuss geometric features of gauge fields and aspects of representation theory needed for our treatment of two-dimensional Yang--Mills theory. 
Aspects of higher gauge theory are elucidated in Subsection \ref{sec:prelim}. Using the language of $p$-gerbes  with $(p+1)$-connection, phrased in the framework of differential cohomology and bundle gerbes, we provide a uniform definition of higher and lower form gauge fields which play a prominent role throughout. We further explain how the charge conjugation 0-form symmetry and the centre 1-form symmetry participate in the natural 2-group global symmetry of the gauge theory. 

In Section \ref{sec:orbi} we study orbifolds of two-dimensional Yang--Mills theory. The derivation for the centre 1-form symmetries in the standard continuum path integral formalism is given in Subsection \ref{subsec:pathintB}. The equivalent derivation in higher gauge theory is given in Subsection \ref{sec:deriv}, based on a 2-group central extension which serves as the gauge 2-group of a principal 2-bundle. In Subsection~\ref{sec:WittenLoc} we compare our novel derivation based on the path integral over 2-connections on a non-abelian gerbe with Witten's more traditional approach based on the combinatorial quantization of two-dimensional Yang--Mills theory~\cite{Witten:1992xu}, through cut-and-paste techniques involving surfaces with boundary. The effect of gauging the charge conjugation 0-form symmetry is discussed in Subsection~\ref{sec:gaugeC}. 

Section \ref{sec:chargeconj} contains a detailed study of spontaneous breaking of the intrinsic finite 0-form symmetry in the orbifold theories.

Section \ref{sec:NIhigher} is devoted to the non-invertible higher form symmetry. The orbifolds by this larger symmetry and its subsets are thoroughly analyzed in Subsection \ref{sec:orbiNI}, whilst Subsection \ref{sec:SSBNI} extends the results on spontaneous symmetry breaking to these generalized orbifolds. 

Mixed anomalies in the orbifold gauge theories, involving $(-1)$-form symmetries, are discussed in Section \ref{sec:anomalies}. In particular, we exemplify the point of view that mixed 't~Hooft anomalies involving the $2\pi$ periodicity of the $\theta$-angle have their origin in an anomalous $(-1)$-form symmetry.

Finally, we showcase some applications of our results to a diverse range of problems in Section~\ref{sec:applications}.
  
Two appendices at the end of the paper complement the analysis of the main text. For the reader's convenience, we summarise our notation and conventions in Appendix~\ref{app:notation}. Appendix~\ref{app:Ortho} provides a detailed analysis of discrete $\theta$-angles, mixed anomalies, orbifolds and spontaneous breaking of charge conjugation symmetry for Yang-Mills theories based on the gauge algebra $\mathfrak{so}(N)$.

\subsubsection*{Acknowledgements}

LS thanks Jeremias Aguilera Damia, Chi-Ming Chang and Jing-Yuan Chen for helpful discussions, and the School of Mathematical and Computer Sciences at Heriot--Watt University for hospitality at the early stages of this project. RJS is grateful to Lukas M\"uller for helpful discussions and for sharing his notes on 1-form symmetries. We also thank Eric Sharpe for comments and correspondence. This article is based upon work from COST Actions CaLISTA CA21109 and THEORY-CHALLENGES CA22113 supported by COST (European Cooperation in Science and Technology). The work of LS is supported by the Shuimu Scholars program of Tsinghua University  and by the Beijing Natural Science Foundation project IS23008 ``Exact Results in Algebraic Geometry from Supersymmetric Field Theory''. The work of RJS was supported in part by Perimeter Institute for Theoretical Physics. Research at Perimeter Institute is supported by the Government of Canada through the Department of Innovation, Science, and Economic Development, and by the Province of Ontario through the Ministry of Colleges and Universities.

\section{Gauge theory preliminaries}
\label{sec:GTpre}

In this section we provide a brief review of various gauge theoretic notions to set the stage. Our notation is summarized in Appendix \ref{app:notation}.

\subsection{Geometry of gauge fields in two dimensions}
\label{sec:prelim}

We start by reviewing basic geometric definitions and how they extend to higher form symmetries. We then provide a definition of lower form gauge transformations, on the same footing of the more widely studied higher form gauge transformations.

\subsubsection*{Geometric setup}

Let $\Sigma$ be a closed, connected and oriented Riemann surface with Euler characteristic $\chi$ and K\"ahler 2-form $\omega$, which we assume to be normalized to unit area
\begin{equation*}
	\int_{\Sigma}\, \omega = 1 \ .
\end{equation*}

Let $G$ be a semi-simple, compact and connected Lie group, and let
\begin{equation}
\label{PoverSigma}
	P \longrightarrow \Sigma
\end{equation}
be a principal $G$-bundle with connection $A$. We denote the right action $P\times G\longrightarrow P$ of $G$ on $P$ as $(p,g)\longmapsto p\cdot g$. As usual, we write $\mathrm{ad}\, P := P\times_G\mathrm{ad}\,\mathfrak{g}$ for the vector bundle associated to \eqref{PoverSigma} by the adjoint representation of $G$ on its Lie algebra $\mathfrak{g}$. The curvature of $A$ is denoted $F$ and the first Chern class of $\mathrm{ad}\, P$ is represented by the Chern--Weil form
\begin{equation*}
	c_1 (\mathrm{ad}\, P)= \tr \, \frac{F}{2\pi}  \ ,
\end{equation*}
where $\tr$ is a suitable trace on $\mathfrak{g}$.

We denote by $\mathscr{A}$ the affine space of $G$-connections $A$ on $\Sigma$, leaving the dependence on $G$ and $\Sigma$ implicit to reduce clutter. If $G$ is not simply-connected, then $\mathscr{A}$ is a disjoint union of connected components, one for each isomorphism class of $G$-bundles \eqref{PoverSigma} labelled by elements of the fundamental group $\pi_1(G)$ of $G$.

\subsubsection*{The $\boldsymbol\theta$-term}

Gauge theories in two dimensions admit a $\theta$-term when the gauge group $G$ is not simply-connected. It is an element of (see e.g.~\cite{Hori:2013gga,Sharpe:2014tca})
\begin{equation*}
	\mathrm{Hom} \big( \pi_1 (G), \Uu \big)^{\pi_0 (G)}  \ ,
\end{equation*}
where the restriction to the $\pi_0 (G)$-invariant part stems from gauge invariance. The fundamental group of any compact Lie group $G$ is abelian and finitely generated, thus we can restrict to two classes:
\begin{enumerate}[(a)]
\item $\pi_1 (G) \cong \Z$, which is the case when the gauge group $G$ is $\Uu$ or $\UN$. Group homomorphisms $\Z \longrightarrow \Uu$ are classified by phases $\e^{\,\ii\, \theta} \in \Uu$, where the parameter $\theta \in [0, 2 \pi)$ is called a $\theta$-angle or $\theta$-parameter.
\item $\pi_1 (G) \cong \Z_k$ is purely torsional, for some $k \in \N$. Group homomorphisms $\Z_k \longrightarrow \Uu$ are classified by $k^{\text{th}}$ roots of unity $\e^{\, 2 \pi\,\ii\, \kappa/k} \in \Z_k \subset \Uu$. The parameter $\theta_{\kappa}:= \frac{2 \pi\, \kappa}{k}  \in [0, 2 \pi) \cap \frac{2\pi}{k} \Z$ is called a discrete $\theta$-angle, sometimes referred to as discrete torsion.
\end{enumerate}
The most general possibility $\pi_1 (G) \cong \Z^{r} \oplus \Z_{k_1} \oplus \cdots \oplus \Z_{k_s} $ is a synthesis of these two cases.

\subsubsection*{The 2-group global symmetry}

Recall that a groupoid  is a small category in which every homomorphism is invertible. A (Lie) 2-group is a (Lie) groupoid equipped with a (smooth) monoidal structure that obeys the usual group axioms. In many instances of interest this is understood in a weak sense, i.e. up to natural isomorphism. 2-groups are equivalently presented as crossed modules of groups, which are more commonly used in applications to physics; we direct the reader to~\cite{Baez:2009as} for an extensive introduction to 2-groups. Lie 2-groups arise in e.g.~field theories with first order reducible gauge symmetries. Together with their homomorphisms and natural transformations, Lie 2-groups form a bicategory; see e.g.~\cite{JohnsonYau} for an introduction to bicategories and 2-categories.

There is a natural 2-group of automorphisms of the stack of $G$-bundles with connection, which we may present as the action groupoid
\begin{align*}
\mathsf{Aut}(G)\,\big/\!\!\big/\,G = \big(G \times\mathsf{Aut}(G) \doublerightarrow{\ \ }{ \ } \mathsf{Aut}(G)\big) \ , 
\\[-9mm]
\end{align*}
where $g\in G$ acts on a group isomorphism $\varphi:G\longrightarrow G$ by conjugation $\varphi\longmapsto\mathrm{Ad}(g)\circ\varphi$. The set of isomorphism classes of objects is the group
\begin{equation*}
	\pi_0\big(\mathsf{Aut}(G)\,\big/\!\!\big/\,G\big) = \mathsf{Out}(G) =: \sO \ , 
\end{equation*}
where $\mathsf{Out}(G)\subset\mathsf{Aut}(G)$ is the subgroup of outer automorphisms of $G$. The set of automorphisms of any object projecting to $\id\in\sO$ is the group
\begin{equation*}
	\pi_1\big(\mathsf{Aut}(G)\,\big/\!\!\big/\,G\big) = \sZ(G) =: \sZ = \pi_1(\mathrm{B}\sZ) \ , 
\end{equation*}
where $\sZ(G)$ is the centre of $G$, and \smash{$\mathrm{B}\sZ=1\,/\!\!/\,\sZ=(\sZ \doublerightarrow{\ \ }{ \ } 1)$} is the delooping groupoid of $\sZ$.

The automorphism 2-group is an extension
\begin{align*}
1\longrightarrow \mathrm{B}\sZ\longrightarrow \mathsf{Aut}(G)\,\big/\!\!\big/\,G \longrightarrow \sO \longrightarrow 1
\end{align*}
in the bicategory of 2-groups. This is alternatively encoded in Hoang data through the four-term exact sequence
\begin{align*}
1\longrightarrow\sZ\longrightarrow G\xrightarrow{ \ \mathrm{Ad} \ }\mathsf{Aut}(G)\longrightarrow\sO\longrightarrow 1
\end{align*}
in the category of groups, where $\big(\mathsf{Aut}(G),G,\mathrm{Ad},\id_{\mathsf{Aut}(G)}\big)$ is the crossed module of groups defining the automorphism 2-group.

This implies that the stack of $G$-bundles with connection has an $\sO$ group symmetry, i.e. a 0-form symmetry $\sO^{\scriptscriptstyle (0)}$ acting by automorphisms of a principal $G$-bundle \eqref{PoverSigma} with connection, and a $\mathrm{B}\sZ$ 2-group symmetry, i.e. a 1-form symmetry~$\sZ^{\scriptscriptstyle (1)}$ acting by natural automorphisms of the identity and by the centre symmetry $\sZ(G)$ on $P\longrightarrow\Sigma$. Note that the 0-form symmetry acts on the 1-form symmetry.

\subsubsection*{Charge conjugation}

For $G$ a simple compact Lie group with Lie algebra $\mathfrak{g}$ and non-trivial outer automorphism group, the elements of the charge conjugation group $\sC^{\scriptscriptstyle (0)} = \Z_2 ^{\scriptscriptstyle (0)}\subseteq\sO^{\scriptscriptstyle (0)}$ are involutive outer automorphisms $\mathfrak{g} \longrightarrow \mathfrak{g}$. We denote by $\mathcal{C}$ the non-trivial element of $\sC^{\scriptscriptstyle (0)}$, which acts on $\mathfrak{g}$-valued fields by flipping their sign.

By the CP theorem in two dimensions, charge conjugation $\sC$ is identified with parity $\sP$. In two Euclidean dimensions, the non-trivial element in $\sP$ is an involution that leaves scalar fields unchanged, and reflects the K\"ahler form of $\Sigma$ as $\omega \longmapsto - \omega$.

\subsubsection*{Higher and lower form symmetries}

A $p$-form symmetry acts on extended operators of dimension $\ge p$. This means that a $p$-form symmetry is generated by topological defect operators of codimension $p+1$, which can link non-trivially with the charged operators. On $\Sigma$, the set of allowed codimensions is $\left\{ 2,1,0 \right\}$, thus we will deal with $1$-form, 0-form and $(-1)$-form symmetries.

A 0-form (i.e. ordinary) $\Uu^{\scriptscriptstyle (0)}$ gauge transformation shifts a scalar field by a 0-form gauge parameter $\alpha^{\scriptscriptstyle (0)} \in \Omega^0 (\Sigma, \mathfrak{u}(1))$, and locally acts on the connection $A$ of the principal bundle \eqref{PoverSigma} as
\begin{equation*}
	A \longmapsto A + \dd \alpha^{\scriptscriptstyle (0)} \  .
\end{equation*}
The 0-form gauge parameter $\alpha^{\scriptscriptstyle (0)} $ is subject to the semi-classical quantization condition
\begin{equation}\label{eq:semi0}
	\frac{1}{2\pi}\,\oint_{\gamma}\, \dd \alpha^{\scriptscriptstyle (0)} \  \in \ \ii \Z 
\end{equation}
for every homology 1-cycle $[\gamma] \in H_1 (\Sigma, \Z) \cong \Z^{2-\chi}$. This condition is non-trivial because $\dd \alpha^{\scriptscriptstyle (0)} $ is only defined locally. By splitting the integral into a sum of integrals over open subsets $U_i$ covering $\gamma$, the transition functions on the various overlaps $U_i \cap U_j$ produce a factor in~$2 \pi\, \ii \Z$. Globally, $\exp\alpha^{\scriptscriptstyle (0)} \in\Omega^0(\Sigma,\Uu)$ is a $\Uu$-valued function on $\Sigma$ and the integers in \eqref{eq:semi0} are its winding numbers.
 
Higher form gauge transformations are the immediate generalization of this idea. A 1-form $\Uu^{\scriptscriptstyle (1)}$ gauge transformation shifts the connection 
\begin{align*}
A \longmapsto A + \alpha^{\scriptscriptstyle (1)}
\end{align*}
by a 1-form gauge parameter $\alpha^{\scriptscriptstyle (1)} \in \Omega^1 (\Sigma, \mathfrak{u}(1))$ subject to 
\begin{equation}\label{eq:semi1}
	\frac{1}{2\pi}\,\int_{\Sigma}\, \dd \alpha^{\scriptscriptstyle (1)} \ \in \ \ii \Z  \ ,
\end{equation}
where again the expression $\dd \alpha^{\scriptscriptstyle (1)} $ is meant locally; globally it is replaced with the curvature of a connection $\alpha^{\scriptscriptstyle (1)} $ on a $\Uu$-bundle over $\Sigma$ of first Chern class \eqref{eq:semi1}. For both 0-form and 1-form symmetries, one can consider the action of a cyclic subgroup $\Z_k \subset \Uu$, in which the gauge parameter $\alpha^{\scriptscriptstyle (p)}$ is not arbitrarily valued in $\R/\Z$ but is in the additive group of $\Z_k$, namely it exponentiates to a $k^{\text{th}}$ root of unity.

It remains to address the nature of a $(-1)$-form symmetry, as there is no such thing as a $(-1)$-form parameter. However, the discussion thus far suggests that the answer is phrased in terms of the topological surface operators mediating the symmetry transformations, with an integrality condition on the curvature of the gauge parameter. The curvature of the would-be $(-1)$-form is a 0-form, that is, a scalar $\varphi$. For it to be closed means $\dd \varphi =0$, hence it is constant on every connected component of the spacetime. The semi-classical quantization condition on the $\Uu^{\scriptscriptstyle (-1)}$ gauge transformation becomes formally 
\begin{equation*}
	 \frac{1}{2\pi} \,  \varphi (\wp)  \ \in \ \ii \Z 
\end{equation*}
for every homology 0-cycle $[\wp] \in H_0 (\Sigma, \Z)$. As we are interested in closed and connected Riemann surfaces $\Sigma$, $ H_0 (\Sigma, \Z) \cong \Z$ and the quantization condition simply means that $\varphi \in 2 \pi\, \ii \Z$ is a constant integer multiple of~$2\pi\, \ii$.

\subsubsection*{Abelian gauge fields}

Given a $p$-form symmetry, a $(p+1)$-form gauge field is a local $(p+1)$-form defined on a good open cover of $\Sigma$, whose transition functions involve $p$-form gauge transformations. In other words, it is a $(p+2)$-cocycle for the  differential cohomology of $\Sigma$, whose gauge transformations are generated by $(p+1)$-cocycles; see e.g.~\cite{Szabo:2012hc} for an elaborate review. The curvature $F\in\Omega^{p+2}_{\Z}(\Sigma)$ of a differential $(p+2)$-cocycle is the closed $(p+2)$-form with integer periods which represents its characteristic class in the charge group $H^{p+2}(\Sigma,\Z)$ for the $p$-form symmetry. Since $p$-form gauge parameters generally come with gauge transformations themselves, gauge fields for a $p$-form symmetry naturally assemble into the structure of a $(p+1)$-groupoid, whose set of $(p+1)$-isomorphism classes has the structure of a $(p+1)$-group. Geometrically, a $p$-form symmetry gives rise to a $p$-gerbe on $\Sigma$ with $(p+1)$-connection.
In two dimensions there are four possibilities for differential cocycles: 
\begin{enumerate}[(i)]
\item For $p=-2$, a differential 0-cocycle is identified with its characteristic class (the component class) in the group $H^0(\Sigma,\Z)$ of connected components of $\Sigma$; its curvature $F$ assigns an integer to each component. Since we assume $\Sigma $ is closed and connected, $H^0(\Sigma,\Z)\cong\Z$. 
\item For $p=-1$, a differential 1-cocycle is a $\Uu$-valued function $f\in\Omega^0(\Sigma,\Uu)$, with gauge transformations generated by $H^0(\Sigma,\Z)\cong\Z$, acting as shifts of $\frac1{2\pi\,\ii}\log f$. Its characteristic class (the degree class) lies in the $(-1)$-form charge group $H^1(\Sigma,\Z)\cong\Z^{2-\chi}$, given by the winding numbers of $f$ and represented by the curvature $F=\dd\log f$. \label{item:1cocycle}
\item For $p=0$, a differential 2-cocycle is a principal $\Uu$-bundle on $\Sigma$ with connection $A$, whose gauge transformations are generated by $\Omega^0(\Sigma,\Uu)$ and which assemble into a group symmetry. Its characteristic class (the first Chern class) lies in the $0$-form charge group $H^2(\Sigma,\Z)\cong\Z$ and is represented (locally) by its curvature $F=\dd A$. \label{item:2cocycle}
\item For $p=1$, a differential 3-cocycle is a $\Uu$-gerbe on $\Sigma$ with 2-connection $(A,B)$, whose gauge transformations are generated by $\Uu$-bundles with connection, which themselves come with $\Omega^0(\Sigma,\Uu)$ gauge transformations and assemble into a 2-group symmetry. Its characteristic class (the Dixmier--Douady class) is trivial because $H^3(\Sigma,\Z)=0$ for any Riemann surface $\Sigma$. Hence any differential 3-cocycle on $\Sigma$ is trivial. In particular, we may set the connection 1-form to $A=0$ as the curving $B\in\Omega^2(\Sigma,\mathfrak{u}(1))$ is a globally defined 2-form on $\Sigma$.
\end{enumerate}

In the following we will insist on the nomenclature `$p$-form gauge field' to emphasize the difference between, for example, a 0-form gauge field and an ordinary scalar. In distinguishing between gauge and global symmetries, we will refer to the corresponding gauge fields as dynamical and background gauge fields respectively.

\subsubsection*{Reduction to cyclic symmetries}

The hierarchy implied by differential cohomology culminates in $\Uu$-gerbes with 2-connection. These are equivalent to bundle gerbes with connection; see \cite{Murray:2007ps} for a review. For an abelian group $\sB$, let $\underline{\sB}$ denote the sheaf of smooth functions on $\Sigma$ valued in $\sB$. Bundle gerbes are characterized by a class in the degree~2 sheaf cohomology \smash{$H^2 (\Sigma, \underline{\Uu})$} \cite{Carey:1997xm,Murray:2007ps}. 

Consider the short exact sequence of sheaves on $\Sigma$ given by
\begin{equation}
\label{sessheaves}
	0 \longrightarrow \Z  \longrightarrow \underline{\R}  \longrightarrow  {\underline{\Uu}}  \longrightarrow 0 \ ,
\end{equation}
where we have used the fact that smooth functions into a discrete set are constant.
The associated long exact sequence in cohomology reads
\begin{equation*}
	\cdots \longrightarrow H^{p+1}(\Sigma,\underline{\R}) \longrightarrow H^{p+1}(\Sigma,\underline{\Uu}) \xrightarrow{ \ \delta \ } H^{p+2}(\Sigma,\Z) \longrightarrow \cdots \ .
\end{equation*}
Since $\underline{\R}$ is a fine sheaf, $H^{p+1}(\Sigma,\underline{\R})=0$ and so the coboundary map $\delta $ is an isomorphism. Therefore, at $p=1$, the characteristic class of the bundle gerbe is identified with the Dixmier--Douady class defined previously. In the case at hand, this is trivial because $\Sigma$ is two-dimensional.

We now reduce the structure group $\Uu$ of the bundle gerbe to a cyclic subgroup $\Z_k \subset \Uu$, adapting from \cite[Section~6.1]{Carey:1997xm}. We extend \eqref{sessheaves} into the commutative diagram of sheaves on $\Sigma$ given by
\begin{equation}
\label{diagsheaves}
\begin{split}
\xymatrix{
0 \ar[r] &\Z  \ar[r] &\underline{\R}  \ar[r]  &{\underline{\Uu}}  \ar[r] & 0 \\
0 \ar[r] &\Z \ar[r] \ar[u]^{\id}  &\frac{1}{k} \Z  \ar[r] \ar[u] &\Z_k \ar[r] \ar[u] & 0
}
\end{split}
\end{equation}
where the first vertical arrow is the identity map, while the second and third vertical arrows are immersions. 

Taking the induced long exact sequence in cohomology associated to the bottom row of \eqref{diagsheaves}, the coboundary map 
\begin{equation*}
		\mathtt{Bock}  :  H^{p+1}(\Sigma,\Z_k) \, \longrightarrow \, H^{p+2}(\Sigma,\Z)
\end{equation*}
is the Bockstein homomorphism (or more precisely its pullback to $\Sigma$), whose image consists of torsion classes. Therefore the $\Z_k$-bundle gerbes and their lower degree predecessors are treated by working in $H^{p+2} (\Sigma, \Z)$, exactly as for $\Uu$-bundle gerbes. Eventually we restrict to the image of $	\mathtt{Bock} $, that is, to torsion classes in integer cohomology.

\subsubsection*{Definition of a 2-form $\mathbb{Z}_k$ gauge field}

In this paper, by a $2$-form gauge field $B$ for the finite abelian group $\mathbb{Z}_k$ we mean that $\frac{k\, B}{2\pi}$ is a de~Rham representative of an integer cohomology class in $H^2(\Sigma,\Z)\cong\Z$. 
In practice, this guarantees that 
\begin{equation}
\label{Bperiod}
	\exp \Big(  \int_{\Sigma}\, k\, B \Big) =1  \quad \Longleftrightarrow \quad \int_{\Sigma}\, B \ \in \ \tfrac{2 \pi\, \ii }{k}\, \mathbb{Z} \ .
\end{equation}

To describe such a $2$-form gauge field, we let $B$ be a $2$-form $\Uu$  gauge field on $\Sigma$, and enforce the quantization condition on its cohomology class via a Lagrange multiplier, which is a dynamical periodic scalar field
\begin{equation*}
	\lambda \ \in \ \Omega ^0 (\Sigma, \R/\Z )  \ .
\end{equation*}
Explicitly
\begin{equation}
\label{Zkmultiplier}
	\sum_{\ell \in \mathbb{Z}}\, \delta \Big( \int_{\Sigma}\, \frac{k\, B}{2 \pi}  - \ii\, \ell \Big)  = \int_{ \Omega ^0 (\Sigma, \R/\Z ) }\, \mathscr{D} \lambda \ \exp \Big(  2 \pi\, k\, \int_{\Sigma}\, \lambda\, \frac{B}{2 \pi} \Big) \ .
\end{equation}
Note that in our conventions $B$ is valued in $\mathfrak{u}(1) \cong \ii \R$ while $\lambda$ is real, thus the exponent in the right-hand side of \eqref{Zkmultiplier} takes values in $2 \pi\, \ii \R$. One can easily check that \eqref{Zkmultiplier} is invariant under the 1-form gauge symmetry transformations
\begin{align} \label{eq:1formtransf}
B\longmapsto B+\dd\alpha^{\scriptscriptstyle (1)} \ ,
\end{align}
where $\alpha^{\scriptscriptstyle (1)}$ is a 1-form $\Uu$ gauge field on $\Sigma$.

In the bundle gerbe description, consider the infinitesimal analogue of \eqref{diagsheaves} which is given by
\begin{equation}
\label{diagsheaf2}
\begin{split}
\xymatrix{
0 \ar[r] &\Z  \ar[r] &\underline{\R}  \ar[r]  &{\underline{\R/\Z}}  \ar[r] & 0 \\
0 \ar[r] &\Z \ar[r] \ar[u]^{\id} &\frac{1}{k} \Z  \ar[r] \ar[u]^{\iota} &\Z_k \ar[r] \ar[u] & 0
}
\end{split}
\end{equation}
It is more convenient for calculations to work with a $\Uu$-gerbe, equipped with a 2-connection which provides a 2-form gauge field $B$ associated to the upper short exact sequence. The 2-form $\Z_k$ gauge field is the pullback $\iota^{\ast}B$ of the differential form $B$. This last step is enforced in practice by the Lagrange multiplier $\lambda$.

\subsubsection*{Definition of a 0-form $\Z_k$ gauge field}

Unravelling the definition \eqref{item:1cocycle} of differential 1-cocycle from above, a 0-form $\Uu$ gauge field $a$ is a scalar subject to 
\begin{equation*}
	\oint_{\gamma}\, \frac{\dd a }{2 \pi} \ \in \ \ii \Z 
\end{equation*}
for all $[\gamma] \in H_1 (\Sigma, \Z)\cong\Z^{2-\chi}$. This is analogous to the usual definition~\eqref{item:2cocycle} of a 1-form $\Uu$ gauge field~$A$ as a differential 2-cocycle.

To reduce the structure group from $\Uu$ to $\Z_k$, we proceed as before. From \eqref{diagsheaves} and the subsequent discussion, it follows that the curvature of a $0$-form $\Z_k$ gauge field is the image of the Bockstein homomorphism in $H^1 (\Sigma,\Z)$. Using  \eqref{diagsheaf2} again, throughout we work with the 0-form $\Uu$ gauge field $a$ defined above. The 0-form $\Z_k$ gauge field is the pullback $\iota^{\ast}a = a\circ\iota$.

\subsection{Representation theory}
\label{sec:repreview}

The properties of irreducible representations of $G$ are central to the study of Yang--Mills theory on $\Sigma$, thus we briefly review them here. 

\subsubsection*{$\boldsymbol{\SUN}$ representations}

For $G=\SUN$, irreducible representations up to isomorphism are in one-to-one correspondence with partitions $R =(R_1, R_2, \dots, R_{N-1})$ subject to
\begin{equation*}
	R_1 \ge R_2 \ge \cdots \ge R_{N-1} \ge 0 \  .
\end{equation*} 
Each partition $R$ is identified with a Young diagram whose $j^{\text{th}}$ row consists of $R_j$ boxes, for each $j=1, \dots, N-1$. For example, the fundamental representation is 
\begin{equation*}
	\text{fund}=(1,0, \dots ,0) = \Box  \ . 
\end{equation*}
It is customary to denote by $|R|$ the total number of boxes in the Young diagram corresponding to the irreducible $\SUN$ representation $R$:
\begin{equation*}
	\lvert R \rvert := \sum_{j=1}^{N-1}\, R_j \ .
\end{equation*}

Isomorphism classes of irreducible representations are equivalently specified by their Dynkin labels $[w_1,w_2, \dots, w_{N-1}]$. The two bases are related through
\begin{equation*}
	w_j = R_{j}-R_{j+1}  \ ,
\end{equation*}
for all $j=1, \dots, N-1$, with $R_N:=0$.

\subsubsection*{Charge conjugation of representations}

Charge conjugation $\mathcal{C}$ maps a representation $R$, corresponding bijectively to a Young diagram $R=(R_1, R_2 , \dots, R_{N-1},0)$, to its conjugate representation $\overline{R}$. The latter is identified with the Young diagram obtained by removing the Young diagram $R$ from the rectangular lattice of size $R_1 \times N$, and then rotating by $180^{\circ}$. For example
\begin{equation*}
\begin{aligned}
	& R = (5,4,2) \\ &
\begin{color}{colC}\overline{R} = (5,3,1)\end{color} 
\end{aligned} \qquad \qquad \ytableausetup{centertableaux,smalltableaux}
\begin{aligned} & \begin{ytableau} \ & \ & \ & \ & \ \\ \ & \ & \ & \ & *(colC) \ \\ \ & \ & *(colC) \ & *(colC) \ & *(colC) \ \\ *(colC) \ & *(colC) \ & *(colC) \ & *(colC) \ 
& *(colC) \ \end{ytableau} \end{aligned}
\end{equation*}
Equivalently, in terms of Dynkin labels the map is given by
\begin{equation*}
	\mathcal{C} \ : \ [w_1, w_2 , \dots, w_{N-2},w_{N-1}] \ \longmapsto \ [w_{N-1}, w_{N-2}, \dots, w_2,w_1] \ .
\end{equation*}

The dimension of an $\SUN$ representation $R$ and its conjugate $\overline{R}$ are equal:
$$\dim R = \dim \overline{R} \ . $$
For example, consider the rank~$r$ symmetric representation and its complex conjugate (shown below for $N=4$ and $r=5$):
\begin{equation*}
\ytableausetup{centertableaux,smalltableaux}
\begin{ytableau} \ & \ & \ & \ & \ \\  *(colC) \ & *(colC) \ & *(colC) \ & *(colC) \ & *(colC) \  \\  *(colC) \ & *(colC) \ & *(colC) \ & *(colC) \ & *(colC) \  \\ *(colC) \ & *(colC) \ & *(colC) \ & *(colC) \ & *(colC) \ \end{ytableau} 
\end{equation*}
Their dimensions are given respectively by
\begin{align*}
\ytableausetup{centertableaux,smalltableaux}
	\dim \,  \begin{ytableau} \ & \ & \ & \ & \ \end{ytableau}  & \, = \, \prod_{j=2}^{N}\,  \frac{r+j-1}{j-1}  \  , \\[4pt]
	\dim \, \begin{ytableau} *(colC) \ & *(colC) \ & *(colC) \ & *(colC) \ & *(colC) \  \\  *(colC) \ & *(colC) \ & *(colC) \ & *(colC) \ & *(colC) \  \\ *(colC) \ & *(colC) \ & *(colC) \ & *(colC) \ & *(colC) \ \end{ytableau}  & \, = \,  \prod_{i=1}^{N-1}\,  \frac{r+N-i}{N-i}  \ ,
\end{align*}
which are equal upon relabelling $j=N-i +1$.

\subsubsection*{Representation category}

The finite-dimensional representations of $G$ and $G$-equivariant maps between them form a $\C$-linear symmetric monoidal category. In the following we will work with the Grothendieck ring $(\mathfrak{R}(G), \otimes, \oplus)$ of this category: the Grothendieck completion of the coarse moduli space of isomorphism classes
\begin{equation*}
	\mathfrak{R}(G) \ = \ \Z[\,G\text{-representations}\,] \ \big/\, \cong 
\end{equation*}
is equipped with the tensor product $\otimes$ and the direct sum $\oplus$ of representations, while the unit is the one-dimensional trivial representation $\emptyset=(0,\dots,0)$.

\subsubsection*{Characters}

Let $\chi_R:G\longrightarrow\C$ denote the character of a representation $R$ of $G$, with $\dim R=\chi_R(\id)$ the dimension of $R$. They can be regarded as the image of a homomorphism $R\longmapsto\chi_R$ from the Grothendieck ring $\mathfrak{R}(G)$ to the ring $\mathfrak{C}(G)$ of class functions on $G$, which preserves the conjugation involutions:
\begin{align*}
\overline{\chi_R} = \chi_{\overline R} \ ,
\end{align*}
with $\overline{\chi_R(g)} = \chi_R(g^{-1})$ for $g\in G$ (because of unitarity).
That is, they are additive:
\begin{align*}
\chi_{R} + \chi_{R'} = \chi_{R\oplus R'} \ ,
\end{align*}
and multiplicative:
\begin{align*}
\chi_{R} \, \chi_{R'} = \chi_{R\otimes R'} \ .
\end{align*} 

In particular, the characters of irreducible representations generate the ring $\mathfrak{C}(G)$. They are orthonormal with respect to integration over the bi-invariant normalized Haar measure $\dd g$ on~$G$:
\begin{align*}
\int_{G} \, \dd g \ \overline{\chi_{R_1}(g)} \, \chi_{R_2}(g) = \delta_{R_1,R_2} \ .
\end{align*}
Let \smash{$ \mathtt{N}^{R_3}_{R_1,R_2}\geq0$} denote the fusion numbers appearing as multiplicities in the decomposition of the tensor product $R_1\otimes R_2 = \bigoplus_{R_3}\, \mathtt{N}^{R_3}_{R_1,R_2} \, R_3$ into irreducible representations. Then by multiplicativity
\begin{align*}
\chi_{R_1} \, \chi_{R_2} = \sum_{R_3} \, \mathtt{N}^{R_3}_{R_1,R_2} \ \chi_{R_3} \ .
\end{align*}
Using orthonormality, the fusion numbers can be expressed as
\begin{align*}
 \mathtt{N}^{R_3}_{R_1,R_2} = \int_{G} \, \dd g \ \chi_{R_1}(g) \, \chi_{R_2}(g) \, \overline{\chi_{R_3}(U)} \ .
\end{align*}

\subsubsection*{Frobenius--Schur indicator}

For any representation $R$ of $G=\SUN$, let $\mathrm{FS}_R$ be its Frobenius--Schur indicator given by
\begin{equation*}
	\mathrm{FS}_R := \int_{\SUN}\,\dd U \ \chi_R (U^2) \ .
\end{equation*}
If $R$ is irreducible, then
\begin{equation}
\label{eq:FSRirrep}
	\mathrm{FS}_R = \begin{cases} \ 1 & \text{if } R \text{ is real}  \ , \\ \ -1 & \text{if } R \text{ is pseudo-real} \ , \\ \ 0 & \text{if } R \text{ is complex} \ . \end{cases}
\end{equation}
In particular, $\SUN$ does not admit pseudo-real irreducible representations if $N-2 \notin 4\Z$. 

For later convenience, we introduce the indicator 
\begin{equation}
\label{eq:deltafr}
	\delta^{\mathrm{FS}}_R := (\mathrm{FS}_R)^2 = \begin{cases} \ 1 & \text{if } R \cong \overline{R} \ , \\ \ 0 & \text{otherwise} \ . \end{cases}
\end{equation}
Examples of irreducible $\SUN$-representations $R$ with $\delta^{\mathrm{FS}}_R =1$ are listed in Table \ref{tab:Creps}.

\begin{table}[htb]
\centering
\ytableausetup{centertableaux,smalltableaux}
\begin{tabular}{c|c|c}
		Partition & Dynkin labels & Young diagram \\
		\hline
		$(2,\underbrace{1, \dots, 1}_{N-2})$ & $[1, 0, \dots , 0,1]$ & \begin{ytableau} \ & \ \\ \ \\ \ \\ \ \\ \ \end{ytableau}   \\
		\hline 
		$(\underbrace{1, \dots, 1}_{\frac{N}{2}},\underbrace{0, \dots, 0}_{\frac{N}{2}-1})$ & $[0, \dots , 0,1,0, \dots ,0]$ &  \begin{ytableau} \ \\ \  \\ \  \end{ytableau}  \\
		\hline 
		$(3,\underbrace{2, \dots, 2}_{\frac{N}{2}-1},\underbrace{1, \dots, 1}_{\frac{N}{2}-1})$ & $[1,0, \dots , 0,1,0, \dots ,0,1]$ &  \begin{ytableau} \ & \ & \ \\ \ & \  \\ \ & \ \\ \ \\ \  \end{ytableau}  \\
		\hline 
		$(\underbrace{p, \dots, p}_{\frac{N}{2}},\underbrace{0, \dots, 0}_{\frac{N}{2}-1})$ & $[0, \dots , 0,p,0, \dots ,0]$ &  $\overbrace{\begin{ytableau} \ & \ & \  \\ \ & \ & \  \\ \ & \ & \  \end{ytableau}}^{p} $ \\
		\hline 
		$(p+2q,\underbrace{p+q, \dots, p+q}_{\frac{N}{2}-1},\underbrace{q, \dots, q}_{\frac{N}{2}-1})$ & $[q, 0,\dots , 0,p,0, \dots ,0,q]$ &  $\overbrace{\begin{ytableau} \ & \ & \  & \ & \ & \ & \ \\ \ & \ & \  & \ & \ \\ \ & \ & \  & \ & \ \\ \ & \ \\ \ & \  \end{ytableau}}^{p+2q} $ \\
		\hline 
		$(p+2q,p+2q,\underbrace{p+q, \dots, p+q}_{\frac{N}{2}-2},\underbrace{q, \dots, q}_{\frac{N}{2}-2},0)$ & $[0,q,\dots , 0,p,0, \dots ,q,0]$ &  $\overbrace{\begin{ytableau} \ & \ & \  & \ & \ & \ & \ \\ \ & \ & \  & \ & \ & \ & \ \\ \ & \ & \  & \ & \ \\ \ & \  \end{ytableau}}^{p+2q} $ \\
		\hline
	\end{tabular}
\caption{Some irreducible $\SUN$-representations that satisfy $R \cong \overline{R}$. The Young diagrams are displayed for $N=6$. The Young diagrams in the last three rows are displayed for $p=3$ and $q=2$.}
\label{tab:Creps}
\end{table}

\subsubsection*{U\texorpdfstring{${\boldsymbol{(N)}}$}{N} representations from SU\texorpdfstring{${\boldsymbol{(N)}}$}{N} representations}

The group central extension
\begin{equation*}
	1 \longrightarrow \Z_N \longrightarrow \SUN  \times \Uu \longrightarrow \UN \longrightarrow 1 
\end{equation*}
relates $\UN$ and $\SUN \times \Uu$ representations. An irreducible $\UN$-representation $R^{\,\textrm{e}}$ decomposes into
\begin{equation} \label{eq:R'RQ}
	R^{\,\textrm{e}}= (R, Q) \qquad \text{with} \quad  Q = q\, N + \lvert R \rvert \ , \ q \in \Z \ ,
\end{equation}
where $R$ and $q$ are, respectively, irreducible representations of $\SUN$ and $\Uu$. The Young diagram corresponding to $R^{\,\textrm{e}}$ is obtained by adding $q$ boxes to each of the $N$ rows of the Young diagram for $R$. The quadratic Casimir invariants $C_2^{\,\textrm{e}}$ of $\UN$ and $C_2$ of $\SUN$ are related through
\begin{equation}
\label{C2hatC2}
	C_2^{\,\textrm{e}} (R^{\,\textrm{e}}) =  {C}_2 (R ) + \tfrac{Q^2}{N} \ ,
\end{equation}
whereas 
\begin{align*}
\dim R^{\,\textrm{e}} = \dim R \ .
\end{align*}

\subsection{Yang--Mills theory in two dimensions}

\subsubsection*{The gauge theory}

We begin by considering pure Yang--Mills theory, with simply-connected gauge group $G$ admitting outer automorphisms, on the Riemann surface $\Sigma$. The intrinsic discrete symmetries of this theory participate in the automorphism 2-group $\mathsf{Aut}(G)\,/\!\!/\,G$ discussed in Subsection~\ref{sec:prelim}: charge conjugation $\sC^{\scriptscriptstyle (0)} = \Z_2 ^{\scriptscriptstyle (0)}\subseteq \sO^{\scriptscriptstyle (0)}:=\mathsf{Out}(G)$ and the centre 1-form symmetry $\sZ ^{\scriptscriptstyle (1)} := \sZ (G)$.

The partition function of Yang--Mills theory with gauge group $G$ is defined by the path integral
\begin{equation}
\label{pathintSUN}
\mz_{G} = \mz_{G}[\Sigma] = \int_{\mathscr{A}} \, \mathscr{D} A \  \exp \Big(\frac{1}{2g^2 }\, \int_{\Sigma}\, \tr\, F \ast F  \Big)
\end{equation}
with the suitably normalized gauge-invariant symplectic measure $\mathscr{D} A$ over the affine space $\mathscr{A}$ of $G$-connections on $\Sigma$. In the action functional, $g$ is the gauge coupling and $\hd$ is the Hodge star operator which is compatible with the K\"ahler form $\omega$ on $\Sigma$.
 
It is often convenient to introduce an additional scalar field $\phi \in \Omega^0 (\Sigma , \mathfrak{g})$ in the adjoint representation of $G$, and write \eqref{pathintSUN} in a first order formulation as 
\begin{equation}
\label{YMphi}
\mz_{G} = \int_{\mathscr{A}} \, \mathscr{D} A \ \int_{\Omega^0 (\Sigma, \mathfrak{g})} \, \mathscr{D} \phi \  \exp \, \int_{\Sigma}\, \tr\, \Big(  \frac{g^2}{2 }\,\phi \ast \phi + \ii \, \phi \, F   \Big)   \ .
\end{equation}
The Gaussian integral over $\phi$ gives back \eqref{pathintSUN}. 

On the other hand, the $BF$-type path integral \eqref{YMphi} can be evaluated exactly, resulting in the Migdal--Rusakov formula~\cite{Migdal:1975zg,Rusakov:1990rs,Witten:1991we,Blau:1993hj}
\begin{equation}
\label{eq:MRformula}
	\mz_{G} = \sum_{R}\, ( \dim R )^{\chi } \ \e^{- \frac{g^2}{2}\, C_2  (R) } \ ,  
\end{equation}
where $C_2$ is the quadratic Casimir invariant of $G$, and the sum runs over isomorphism classes $R$ of irreducible representations of $G$. More precisely, the sum in \eqref{eq:MRformula} is understood to run over the generators of the Grothendieck ring $\mathfrak{R}(G)$, which describes the Wilson lines of the gauge theory and thus distinguishes between Yang--Mills theories with different gauge groups that integrate the same Lie algebra $\mathfrak{g}$~\cite{Aharony:2013hda}.

\subsubsection*{The 1-form symmetry}

In any dimension, pure Yang--Mills theory has an intrinsic 1-form symmetry $\sZ ^{\scriptscriptstyle (1)} := \sZ (G)$ based on the centre of $G$. We focus on $G= \SUN$ for concreteness. Locally, the action of \smash{$\sZ^{\scriptscriptstyle (1)} = \Z_N ^{\scriptscriptstyle (1)}$} shifts \smash{$A \longmapsto A+ \alpha^{\scriptscriptstyle (1)}$} for a 1-form $\alpha^{\scriptscriptstyle (1)}$ subject to 
\begin{equation*}
	\oint_{\gamma}\, \frac{N\,\alpha^{\scriptscriptstyle (1)}}{2\pi} \ \in \, \ii \Z 
\end{equation*}
for all homology 1-cycles $[\gamma] \in H_1 (\Sigma, \Z)$. 

The Wilson loop in the $G$-representation $R$ supported on a closed oriented curve $L\subset\Sigma$ is given by
\begin{equation*}
	W_R (L) := \tr_R\ \mathtt{P}\, \exp\,  \oint_L\, A  \  .
\end{equation*}
The operator $W_R(L)$ is charged under $\sZ ^{\scriptscriptstyle (1)} $, because it acquires a non-trivial phase under a shift transformation of the gauge connection $A$:
\begin{equation*}
	A \ \longmapsto \ A+ \alpha^{\scriptscriptstyle (1)} \quad \Longrightarrow \quad W_R (L) \ \longmapsto \ \e^{ \,\lvert R \rvert \, \oint_L \alpha^{(1)} } \ W_R (L) \ .
\end{equation*}

\section{Orbifolds of two-dimensional Yang--Mills theory}
\label{sec:orbi}

In this section we undertake a detailed analysis of the orbifolds of pure Yang--Mills theory on $\Sigma$ by its intrinsic discrete global symmetries.

\subsection{The orbifold gauge theories}

Let us focus on $G= \SUN$, and start with the 1-form part $\sZ^{\scriptscriptstyle (1)} = \Z_N ^{\scriptscriptstyle (1)}$ of the automorphism 2-group $\mathsf{Aut}(\SUN)\,/\!\!/\,\SUN$. Let $k \in \mathbb{N}$ be a divisor of $N$, including the limiting cases $k \in \left\{1,N \right\}$. In later sections, we will focus on the case where $k$ is a non-negative integer power of $2$ strictly smaller than $N$, but for now we leave $k$ arbitrary. We define 
\begin{equation*}
	\sB^{\scriptscriptstyle (1)} := \Z^{\scriptscriptstyle (1)} _k \ \subseteq \ \Z^{\scriptscriptstyle (1)} _N \ .
\end{equation*}

Gauging the subgroup \smash{$\sB^{\scriptscriptstyle (1)}$} of the $1$-form symmetry \smash{$\sZ^{\scriptscriptstyle (1)}$} produces the orbifold theory with gauge group $G=\SUN / \Z_k$. As opposed to the original $\SUN$ theory, the orbifold gauge theory in two dimensions has non-trivial fundamental group $\pi_1 (G)\cong \Z_k$, and therefore it involves non-trivial gauge bundles as well as topological terms which are parametrized by discrete $\theta$-angles
\begin{equation*}
\theta_{\kappa} = \frac{2 \pi\, \kappa }{k} \quad , \quad \kappa \in \{ 0, 1 , \dots, k-1 \} \  .
\end{equation*}

\begin{prop}\label{prop:Zorbi}
The partition function of the orbifold $\SUN/\Z_k$ pure Yang--Mills theory on $\Sigma$ is given by
\begin{align}
	\mz^\kappa_{\SUN / \Z_k} & = \frac{1}{k}\, \sum_{\beta=0} ^{k-1}\, \e^{\,\ii\, \theta_{\kappa} \, \beta }  \ \sum_{R  }\, ( \dim R )^{\chi } \ \e^{- \frac{g^2}{2}\, C_2 (R) - 2 \pi\,\ii\, \beta\, \lvert R \rvert / k } \label{ZorbiSum} \\[4pt]
		 & = \sum_{R  } \, ( \dim R )^{\chi } \ \e^{- \frac{g^2}{2}\, C_2  (R) } \ \delta ( \lvert R \rvert - \kappa  \ \mathrm{mod} \ k ) \ .  \label{ZorbiDelta}
\end{align}
\end{prop}

The sum over $\beta$ in Proposition~\ref{prop:Zorbi} is understood as a sum over $\mathbb{Z}_k$-valued isomorphism classes of gauge bundles on $\Sigma$, and \eqref{ZorbiDelta} follows immediately from \eqref{ZorbiSum} after performing the sum over $\beta$. That is, the effect of the gauging is to truncate the series \eqref{eq:MRformula} to those irreducible representations $R$ of $\SUN$ whose Young diagrams contain $\lvert R \rvert = \kappa  \ \mathrm{mod} \ k$ boxes.

The orbifold theory has a dual $(-1)$-form symmetry \smash{$\sK^{\scriptscriptstyle (-1)} := \Z_k ^{\scriptscriptstyle (-1)}$}, obtained from the Pontryagin dual group of the gauged subgroup $\sB^{\scriptscriptstyle (1)}$:
\begin{equation*} 
	\sK = \widehat{\sB} = \widehat{\Z_k} =  \mathrm{Hom} \big(\Z_k, \Uu \big) \cong \Z_k \ .
\end{equation*}
At the level of abelian gauge fields, this is a consequence of Pontryagin--Poincar\'e duality of  differential cohomology~\cite{Szabo:2012hc}.
By gauging this dual symmetry, we expect to get back the $\SUN$ theory we started with. We will more generally consider gauging subgroups \smash{$\Z^{\scriptscriptstyle (-1)}_m \subseteq \sK^{\scriptscriptstyle (-1)} $}, for any $m$ that divides $k$.

Proposition \ref{prop:Zorbi} was first proven by Witten \cite{Witten:1992xu} using the standard combinatorial approach to quantization of two-dimensional Yang--Mills theory, and rederived more recently in \cite{Aminov:2019hwg,Sharpe:2019ddn} from the point of view of gauging the 1-form symmetry. The main novelty that we introduce in the following is a first principles path integral derivation of \eqref{ZorbiSum}, as well as of the twice-orbifolded partition function, through coupling to a gauge field for the 1-form symmetry. This approach has the advantage of being tailored to the study of anomalies, which we shall undertake in Section~\ref{sec:anomalies}. 

We conclude this section by offering a simplified treatment and new perspective on the orbifold by the 0-form part $\sC^{\scriptscriptstyle (0)} := \sO^{\scriptscriptstyle (0)} = \Z_2^{\scriptscriptstyle (0)}$ of the automorphism 2-group $\mathsf{Aut}(\SUN)\,/\!\!/\,\SUN$, considered in~\cite{Muller:2019tnp}. The gauging is performed by summing over insertions of networks of topological line defects for the charge conjugation symmetry $\sC$, which produces the orbifold Yang--Mills theory on $\Sigma$ with gauge group $\SUN\rtimes\sC$. We do not discuss the gauging of the dual 0-form symmetry $\widehat{\sC}^{\scriptscriptstyle (0)}\cong\Z_2^{\scriptscriptstyle (0)}$ here; the reverse generalized orbifold construction is presented in~\cite{Muller:2019tnp} by summing over topological networks of Wilson line operators.

\subsection{The path integral coupled to a background field}
\label{subsec:pathintB}

In the following we prove Proposition \ref{prop:Zorbi} as well as the gauging of the dual symmetry. We begin here with a more direct approach, and then reformulate it in Subsection \ref{sec:deriv} via a lift of the $G$-bundles to gerbes when a gauge field for the 1-form symmetry is turned on.

We first couple the theory to a background 2-form gauge field $B$ for the 1-form symmetry $\Z^{\scriptscriptstyle (1)} _k$, and then promote it to a dynamical field. The path integral over $B$, which we compute in Subsection~\ref{subsec:gauging1form}, will enforce the gauging.

\subsubsection*{Coupling to a background field}

We couple pure Yang--Mills theory to a background 2-form $\Z_k$ gauge field $B$ for the 1-form symmetry $\sB ^{\scriptscriptstyle (1)} := \Z^{\scriptscriptstyle (1)} _k$. The gauge field $B$ has been properly defined at the end of Subsection \ref{sec:prelim}. We use the standard minimal coupling to the electric 1-form gauge symmetry \eqref{eq:1formtransf}, by replacing \cite{Gaiotto:2017yup,Komargodski:2017dmc}
\begin{equation*}
	F \longmapsto F-B\otimes \id_N
\end{equation*}
in the Lagrangian of \eqref{pathintSUN}, where $\id_N $ is the $N {\times} N$ identity matrix. This gives
\begin{align*}
 \mz_{\SUN} [B] &= \int_{\mathscr{A}}\, \mathscr{D} A \ \int_{ \Omega ^0 (\Sigma, \R/\Z ) }\,\mathscr{D} \lambda \ \exp \bigg( \frac{1}{2g^2}\, \int_{\Sigma}\, \tr\big[ (F - B \otimes \id_N) \ast  (F - B \otimes \id_N)\big] \\
 & \hspace{8cm}+ \int_{\Sigma}\, \lambda\, k\,  B \bigg)\ .
\end{align*}

We may also introduce a counterterm 
\begin{equation}
\label{kappaB}
	\kappa\, \int_{\Sigma}\, B 
\end{equation}
for the background field $B$, which will add discrete torsion to the orbifold. It follows from \eqref{Bperiod} that $\kappa$ is subject to the periodic identification $\kappa \sim \kappa + k$. Moreover \eqref{kappaB} is properly quantized and invariant under background $\sB$-gauge transformations only if $\kappa \in \Z$. That is, we are left with the coefficient $\kappa \in  \left\{ 0,1, \dots, k-1 \right\}$.

Under charge conjugation, $B$ flips sign and therefore \eqref{kappaB} breaks the $\sC^{\scriptscriptstyle (0)} = \Z_2 ^{\scriptscriptstyle (0)}$ symmetry explicitly, unless $k$ is even and \smash{$\kappa=\frac{k}{2}$}, in which case the change of sign can be compensated by a shift $\kappa \longmapsto \kappa - k$.

We would now like to pass to the first order formalism as in \eqref{YMphi}. Note, however, that we should not use an $\mathfrak{su}(N)$-valued scalar $\phi$ as before, because now $F$ has been shifted by a diagonal element. We thus need $\phi$ to take values in $\mathfrak{su}(N) \oplus \mathfrak{u}(1)$. With this in mind, we get 
\begin{align*}
\begin{split}
 \mz_{\SUN} ^{\kappa} [B] &= \int_{\mathscr{A}} \, \mathscr{D} A \ \int_{\Omega^0(\Sigma,\mathfrak{su}(N)\, \oplus\, \mathfrak{u}(1))} \, \mathscr{D} \phi \  \int_{ \Omega ^0 (\Sigma, \R/\Z ) }\,  \mathscr{D} \lambda \ \exp \bigg( \int_{\Sigma}\, \lambda\, k \, B  \bigg) \\
 & \hspace{4cm} \times \exp \, \int_{\Sigma}\,  \ii \, \tr \big(  \phi\, ( F - B \otimes \id_N )  \big) + \frac{g^2}{2} \tr \left( \phi \ast \phi\right) + \kappa\, B  \ .
\end{split}
\end{align*}
In the following, we will assume that $k$ is even, and therefore that $N$ is even.

\subsubsection*{Abelianization}

At this point, the dependence on the $\SUN$ curvature $F$ is decoupled from the background field $B$ and we can run the abelianization procedure of \cite{Blau:1993hj}. We skip the details because, for the $\mathfrak{su}(N)$ part of $\phi$ that couples only to $F$, they are exactly as in \cite{Blau:1993hj}. We denote by
\begin{equation*}
	\phi_i \ , \quad  i=1, \dots, N-1
\end{equation*}
the components of $\phi$ in a Cartan subalgebra of $\mathfrak{su}(N)$, and by $\phi_{\mathfrak{u}(1)}$ the residual $\mathfrak{u}(1)$ part which couples to $B$ but not to $F$. As we have begun with an expression which is invariant under $\sB$-gauge transformations, imposing this invariance to the resulting expression leads to the identification 
\begin{equation*}
	N\, \phi_{\mathfrak{u}(1)} = \sum_{i=1} ^{N-1}\, \phi_i + \ii\, k\, q \ ,
\end{equation*}
for $q \in \Z$. 

One first integrates out the off-diagonal modes of the gauge field $A$ to produce a Jacobian factor~\cite{Blau:1993hj}. Then one integrates out the residual modes of $A$ that take values in a Cartan subalgebra of $\mathfrak{su}(N)$, which are connections on torus bundles over the Riemann surface $\Sigma$. The term
\begin{equation*}
	 \ii\, \int_{\Sigma}\, \tr\, \phi\, F 
\end{equation*}
in the action functional, with the trace over $\mathfrak{su}(N)$, imposes the constraint that $\phi_i \in \ii \Z $ are purely imaginary integers which moreover can be written in terms of the new variables \cite{Blau:1993hj}
\begin{equation*}
	\ii\, \phi_i = R_i - i + \tfrac{N}{2} \ .
\end{equation*}

We now use the identities  
\begin{align*}
	\ii\, \sum_{i=1}^{N-1}\, \phi_i & = \sum_{i=1}^{N-1}\, R_i = \lvert R \rvert \ , \\[4pt]
	- \sum_{i=1}^{N-1}\, \phi_i ^2 & = \sum_{i=1}^{N-1}\, R_i\, (R_i -2i + N) + \sum_{i=1}^{N-1}\, \left( i- \frac{N}{2} \right)^2 = C_2 (R) + \frac{N\,(N^2-3N+2)}{12}
\end{align*}
to write the result as a sum over representations:
\begin{align*}
 \mz_{\SUN} ^{\kappa} [B] =  \sum_{R}\, ( \dim R )^{\chi } \ \e^{- \frac{g^2}{2}\, \left[ C_2  (R)- \frac{N}{12}\, (N^2-3N+2)\right] } \ \int_{ \Omega ^0 (\Sigma, \R/\Z ) }\, \mathscr{D} \lambda \ \exp \bigg( \int_{\Sigma}\, B\,  ( \lambda\, k + \kappa - \lvert R \rvert )    \bigg) \ .
\end{align*}
Here we are omitting the sum over $q \in \Z$ as it is reabsorbed in the intergal over the periodic scalar $\lambda$. We may get rid of the overall $N$-dependence with the gravitational counterterm 
\begin{equation*}
	 \exp \left[ - \frac{g^2}{24}\, N\,(N^2-3N+2)\, \int_{\Sigma}\, \omega \right] \ .
\end{equation*}

In conclusion, we obtain
\begin{align}
\label{eq:ZBbg}
 \mz_{\SUN} ^{\kappa} [B] =  \sum_{R}\, ( \dim R )^{\chi } \ \e^{- \frac{g^2}{2}\, C_2  (R) } \ \exp \bigg( \int_{\Sigma}\, B\,  (  \kappa - \lvert R \rvert ) \bigg) \ ,
\end{align}
with the understanding that $B$ is subject to \eqref{Bperiod}.

\subsection{Derivation in higher gauge theory}
\label{sec:deriv}

We shall now provide an alternative derivation of \eqref{eq:ZBbg} which emphasizes the role of higher structures at play. 

\subsubsection*{The gauge 2-group}

In implementing \eqref{Zkmultiplier}, the background 2-form gauge field $B$ for the 1-form symmetry $\sB^{\scriptscriptstyle (1)}$ is promoted to a 2-form $\Uu$ gauge field. Coupling $\SUN$ Yang--Mills theory to it corresponds to considering a central extension of $\SUN$ by $\Uu$ in the category of Lie groups:
\begin{equation} \label{eq:UNextension}
	1 \longrightarrow \Uu \longrightarrow \UN \longrightarrow \SUN \longrightarrow 1 \ .
\end{equation}
The main novel point here is that the extension mixes 0-form and 1-form symmetries, which algebraically requires promoting \eqref{eq:UNextension} to the higher structure of a 2-group~\cite{Sharpe:2015mja,Cordova:2018cvg,Benini:2018reh,Kang:2023uvm}. 

Following~\cite{Schommer-Pries:2011vyj}, the relevant non-abelian 2-group $\mathbbm{U}(N)$ arises as a central extension of $\SUN$ in the bicategory of Lie 2-groups:
\begin{align*}
1\longrightarrow\mathrm{B}\Uu\longrightarrow\mathbbm{U}(N)\longrightarrow \SUN\longrightarrow 1 \ ,
\end{align*}
where the strict abelian 2-group \smash{$\mathrm{B}\Uu = (\Uu \doublerightarrow{\ \ }{ \ } 1)$} is presented as the delooping groupoid of $\Uu$. In particular, $\mathbbm{U}(N)$ is a principal $\mathrm{B}\Uu$-bundle, i.e. a $\Uu$-gerbe, on the Lie group $\SUN$, with suitable multiplicative structure. The extension is classified by the cohomology $H^4(\mathrm{B}\SUN,\Z)\cong H^3(\SUN,\Z)\cong\Z$ of the classifying space of $\SUN$, which labels the Dixmier--Douady class of the $\Uu$-gerbe. The crossed module of Lie groups defining the Lie 2-group is given from \eqref{eq:UNextension} by the trivial group homomorphisms $\Uu\longrightarrow\SUN$ and $\SUN\longrightarrow\mathsf{Aut}(\Uu)\cong\Z_2$, whose Hoang data encodes $\mathbbm{U}(N)$ in the four-term exact sequence
\begin{align*}
1\longrightarrow\Uu\xrightarrow{ \ \id \ }\Uu\longrightarrow \SUN \xrightarrow{\ \id \ }\SUN \longrightarrow 1
\end{align*}
in the category of Lie groups.

The group of isomorphism classes of objects of the Lie 2-group $\mathbbm{U}(N)$ is $$\pi_0\big(\mathbbm{U}(N)\big) = \SUN \ . $$ The automorphism group of any object projecting to $\id_N\in\SUN$ is $$\pi_1\big(\mathbbm{U}(N)\big) = \Uu = \pi_1\big(\mathrm{B}\Uu\big) \ . $$ 
Hence the extension involves a non-abelian $\SUN^{\scriptscriptstyle (0)}$ 0-form symmetry acting trivially on an abelian $\Uu^{\scriptscriptstyle (1)}$ 1-form symmetry.

\subsubsection*{The principal 2-bundle}

Since $H^4(\Sigma,\Z)=0$, we can lift the principal $\SUN$-bundle \eqref{PoverSigma} to a principal 2-bundle on $\Sigma$ with structure $2$-group $\mathbbm{U}(N)$, i.e. a non-abelian gerbe~\cite{Nikolaus:2011ag} over $\Sigma$, by endowing its total space with a $\Uu$-gerbe
\begin{align} \label{eq:P2bundle}
\begin{split}
\xymatrix{
 \mathbbm{P} \ar[d] & \\ P\ar[r] & \Sigma
}
\end{split}
\end{align}
together with an isomorphism of $\Uu$-gerbes $ \mathbbm{P}_p\,\otimes\, \mathbbm{U}(N)_g\longrightarrow \mathbbm{P}_{p\cdot g}$, implemented by tensoring with a line bundle $L\longrightarrow P\times\SUN$, as well as a 2-isomorphism implemented by a homomorphism of line bundles over $P\times\SUN\times\SUN$, see~\cite[Section~4.1]{Tellez-Dominguez:2023wwr} for details. Upon choosing a trivialisation $P\cong \Sigma\times\SUN$, $ \mathbbm{P}$ descends to a  $\Uu$-gerbe on $\Sigma$, which we denote by the same symbol. Since $H^3(\Sigma,\Z)=0$, the Dixmier--Douady class of $ \mathbbm{P}$ vanishes and hence $ \mathbbm{P}$ is trivial.

We view $B$ as a curving on the $\Uu$-gerbe $ \mathbbm{P}\longrightarrow\Sigma$ satisfying $B_p\otimes \varTheta_g=B_{p\cdot g}$ on $P\times\SUN$, where 
\begin{align*}
\varTheta_g = n\,\tr\Big(g^{-1}\,\dd g \, \wedge \, \frac{\id_N+\mathrm{Ad}(g)}{\id_N-\mathrm{Ad}(g)} \, g^{-1}\,\dd g\Big) 
\end{align*}
is the Maurer--Cartan curving $\varTheta\in\Omega^2(\SUN,\mathfrak{u}(1))$ on the $\Uu$-gerbe $\mathbbm{U}(N)\longrightarrow\SUN$, whose bi-invariant curvature 3-form $H_g = \dd\varTheta_g = n\,\tr\big((g^{-1}\,\dd g)^3\big)$ represents the Dixmier--Douady class $n\in\Z$. Pick a flat connection on the line bundle $L\longrightarrow P\times\SUN$ inducing a flat 2-isomorphism of line bundles over $P\times\SUN\times\SUN$. Then the pair $(A,B)$ from Subsection~\ref{subsec:pathintB} can be regarded as the data of a $2$-connection on the trivial principal $\mathbbm{U}(N)$-bundle \eqref{eq:P2bundle}, see~\cite[Section~4.2]{Tellez-Dominguez:2023wwr}. The total curvature of the 2-connection $(A,B)$ is simply the curvature $F$ of the $\SUN$ gauge field $A$. 

The groupoid of 2-connections is an affine $\Omega^2(\Sigma,\mathfrak{u}(1))$-bundle over the space $\mathscr{A}$ of $\SUN$-connections on $P\longrightarrow\Sigma$. 
Infinitesimal 0-form and 1-form gauge transformations, parametrized by \smash{$(\alpha^{\scriptscriptstyle (0)},\alpha^{\scriptscriptstyle (1)})\in\Omega^0(\Sigma,\mathfrak{su}(N))\oplus\Omega^1(\Sigma,\mathfrak{u}(1))$}, act on a 2-connection in a manner reminiscent of the Green--Schwarz mechanism as~\cite{Tellez-Dominguez:2023wwr}
\begin{align} \label{eq:2connvar}
(A,B) \longmapsto \big(A+\dd_A \alpha^{\scriptscriptstyle (0)}\ ,\ B+\dd\alpha^{\scriptscriptstyle (1)}+\tr(A\wedge\dd\alpha^{\scriptscriptstyle (0)}) - 2\,n\,\tr(\dd\alpha^{\scriptscriptstyle (0)}\wedge\mathrm{ad}(\alpha^{\scriptscriptstyle (0)})^{-1}\,\dd\alpha^{\scriptscriptstyle (0)}) \big) \ ,
\end{align}
where $\dd_A\alpha^{\scriptscriptstyle (0)} = \dd\alpha^{\scriptscriptstyle (0)}+[A,\alpha^{\scriptscriptstyle (0)}]$.

Let us now introduce the extended curvature
\begin{equation}
\label{Fprimedef}
	F^{\,\textrm{e}}  = F + B \otimes \id_N \ ,
\end{equation}
which combines the $\SUN$ gauge curvature $F$ and the background 2-form $\Uu$ gauge field $B$ into a $\UN$ curvature.
Taking the trace on both sides of \eqref{Fprimedef} gives
\begin{equation}
\label{traceFprime}
	\tr\, F^{\,\textrm{e}}  = N\, B \ ,
\end{equation}
because $F$ is an $\SUN$ curvature. One way to enforce this condition in the path integral is via integration over a Lagrange multiplier $\xi \in \Omega^0 (\Sigma,\R)$, and by adding the term 
\begin{equation}
\label{tracemultiplier}
	\int_{\Sigma}\, \xi\, \tr \big( F^{\,\textrm{e}}  - B \otimes \id_N  \big) 
\end{equation}
to the action functional. Note that in our conventions, $\xi$ is real while $F^{\,\textrm{e}} $ and $B$ are anti-Hermitian, whence \eqref{tracemultiplier} is purely imaginary.

From \eqref{traceFprime} together with the fact that $N$ is an integer multiple of $k$ and $\frac{k\,B}{2\pi}$ represents an integer cohomology class by definition, it follows that 
\begin{equation}
\label{FprimeCW}
	\tr\bigg( \frac{k}{N}\, \frac{F^{\,\textrm{e}} }{2 \pi} \bigg)
\end{equation}
represents an integer cohomology class in $H^2(\Sigma,\Z)\cong\Z$. Through Chern--Weil theory it can thus be identified as the first Chern class $c_1(\text{ad}\,P^{\,\textrm{e}} )$ of the adjoint bundle associated to a principal $\UN$-bundle 
\begin{equation}
\label{Pprimebundle}
	P^{\,\textrm{e}}  \longrightarrow \Sigma \ .
\end{equation}
This can be chosen to cover the $\SUN$-bundle \eqref{PoverSigma}, because the lifting $\Uu$-gerbe on $\Sigma$ associated to the group central extension \eqref{eq:UNextension} is necessarily trivial (again since $H^3(\Sigma,\Z)=0$). 

The $\UN$-bundle \eqref{Pprimebundle} has the following meaning in terms of the description of the bicategory of $\Uu$-gerbes from~\cite{Fuchs:2009dms,Bunk:2016rta}. Due to \eqref{traceFprime}, the vector bundle of rank $N$ with connection $A^{\,\textrm{e}} $ associated to \eqref{Pprimebundle} by the fundamental representation of $\UN$ generates a homomorphism from the curving $\frac kN\,B\in\Omega^2(\Sigma,\mathfrak{u}(1))$ to the zero curving on the trivial $\Uu$-gerbe  $ \mathbbm{P}\longrightarrow\Sigma$. Note that this is not a gauge transformation of $B$, because a vector bundle of rank $N>1$ with connection is not invertible and so is not an isomorphism of $\Uu$-gerbes with connection.

\subsubsection*{Change of path integration variable}

Any $\SUN$-bundle over a Riemann surface can be trivialized. The path integral measure is $\mathscr{D} A$, where $A\in\Omega^1(\Sigma,\mathfrak{su}(N))$ is related to the $\SUN$ curvature $F$ via $F= \dd A + A \wedge A$. However, this is not necessarily true for the extended curvature $F^{\,\textrm{e}} $, since the $2$-form $k\,B$ is closed but generally only locally exact: On a good open cover $\left\{ U_{i} \right\}$ of $\Sigma$, we may write $k\,B= \dd \alpha_i$, where $\alpha_i \in \Omega^1 (U_i,\mathfrak{u}(1))$ satisfies $\alpha_i-\alpha_j = \dd\log g_{ij}$ on overlaps $U_i\,\cap\,U_j$ such that $\{U_i,g_{ij}\}$ is a cocycle for the \v{C}ech cohomology $\check{H}^1(\Sigma,\Uu)\cong\Z$.

We can then \emph{locally} make the change of variable 
\begin{equation}
\label{changevarA}
	A^{\,\textrm{e}}  = A + \tfrac{1}{k}\, \alpha \otimes \id_N \ ,
\end{equation}
understood in the sense that the restriction of $A^{\,\textrm{e}} $ to $U_i$ satisfies \eqref{changevarA} for some $1$-form $\alpha_i$ locally defined on $U_i$ which defines a \v{C}ech 1-cocycle. The extended curvature \eqref{Fprimedef} is then given (locally) by $F^{\,\textrm{e}} =\dd A^{\,\textrm{e}}  + A^{\,\textrm{e}} \wedge A^{\,\textrm{e}} $. The affine space of connections on the $\UN$-bundle \eqref{Pprimebundle} is denoted~$\mathscr{A}^{\,\textrm{e}}$.

\subsubsection*{The kinetic term}

In terms of the extended curvature $F^{\,\textrm{e}} $, the Yang--Mills action functional of \eqref{pathintSUN} coupled to a background 2-form gauge field $B$ is
\begin{align}
\begin{split}
&	\frac{1}{2 g^2 }\, \int_{\Sigma}\, \tr \big[( F^{\,\textrm{e}}  -  B \otimes \id_N ) \hd\, ( F^{\,\textrm{e}}   - B \otimes \id_N )\big] \\[4pt]
& \hspace{3cm} =   \frac{1}{2 g^2 }\, \int_{\Sigma}\, \left[\tr ( F^{\,\textrm{e}}  \hd\, F^{\,\textrm{e}}  ) - 2 \hd B\, \tr ( F^{\,\textrm{e}}  ) + N\, B \hd\, B \right] \\[4pt]
& \hspace{6cm}=   \frac{1}{2 g^2 }\, \int_{\Sigma}\, \left[ \tr ( F^{\,\textrm{e}}  \hd\, F^{\,\textrm{e}}  ) - N\, B \hd\, B \right] \ . \label{actionchangevar}
	 	\end{split}
\end{align}
In the last step we have used \eqref{traceFprime}. We are using here the kinetic term for $F$ alone, which was already present before turning on $B$, rather than for $F^{\,\textrm{e}} $; this ensures invariance under the higher gauge transformations  \eqref{eq:2connvar}. The difference between the two choices is a counterterm quadratic in $B$, which will eventually cancel out below. 

\subsubsection*{The $\boldsymbol\theta$-term}

Dealing with the $\UN$ gauge curvature now, we add a $\theta$-term 
\begin{equation}
\label{thetaterm}
	\theta\, \int_{\Sigma}\, \tr\left( \frac{k}{N}\, \frac{F^{\,\textrm{e}} }{2 \pi} \right) = \theta\, \int_{\Sigma}\, \frac{k\, B}{2 \pi} \ .
\end{equation}
The coefficient $\frac{k}{N}$ on the left-hand side follows from the discussion around \eqref{FprimeCW}, as does the usual $2 \pi$-periodicity of $\theta$:
\begin{equation}
\label{thetaperiod}
	\frac{\theta\, k}{2 \pi} \sim \frac{\theta\, k}{2 \pi} + k \ .
\end{equation}
The identification \eqref{thetaperiod} can also be seen from the compactness of $\lambda$. In fact, \eqref{thetaterm} and the $\lambda$-dependence combine into 
\begin{equation*}
	 \int_{\Sigma}\, \left( \frac{\theta}{2 \pi} + \lambda\,  \right)\, k\, B \ ,
\end{equation*}
in which a shift of $\theta$ by $2\pi$ is reabsorbed by the identification $\lambda \sim \lambda + 1$.

At this stage we impose 1-form gauge invariance by shifting the 2-form field $B$ under a $\Uu^{\scriptscriptstyle (1)}$ background gauge transformation \eqref{eq:1formtransf}. Gauge invariance of \eqref{thetaterm} under these $\Uu^{\scriptscriptstyle (1)}$ transformations requires 
\begin{equation*}
\frac{\theta\, k}{2 \pi} \ \in \ \mathbb{Z} \ .
\end{equation*}
We thus combine this restriction with \eqref{thetaperiod} and write 
\begin{equation*}
\frac{\theta\, k}{2 \pi} =: \kappa \ \in \ \left\{ 0, 1 , \dots, k-1 \right\} \ .
\end{equation*}

\subsubsection*{The path integral}

We now plug the terms \eqref{actionchangevar} and \eqref{thetaterm} in the appropriate path integral and use the Lagrange multipliers \eqref{Zkmultiplier} and \eqref{tracemultiplier} to enforce the constraints on $k\,B$ and $\tr\, F^{\,\textrm{e}} $:
\begin{align}
\label{pathintBG}
\begin{split}
	\mz^\kappa_{\SUN} [B] &=\int_{\mathscr{A}^{\,\textrm{e}}}\, \mathscr{D} A^{\,\textrm{e}}  \ \int_{ \Omega ^0 (\Sigma, \R/\Z ) }\, \mathscr{D} \lambda  \ \int_{ \Omega ^0 (\Sigma,\R ) }\, \mathscr{D} \xi \\
	& \hspace{3cm} \times  \exp \bigg[ \frac{1}{2 g^2 }\, \int_{\Sigma}\, \bigg( \tr ( F^{\,\textrm{e}}  \hd\, F^{\,\textrm{e}}  ) - N\, B \hd\, B \bigg) \\
	&\hspace{6cm} +  \int_{\Sigma}\, \bigg( \xi\, \tr  ( F^{\,\textrm{e}}  ) - \xi\, N\, B + \kappa\, B + \lambda\, k\, B\bigg) \bigg]\ .
	\end{split}
\end{align}

We rewrite the $F^{\,\textrm{e}} $-dependence in the first order formalism by introducing an auxiliary scalar field $\phi^{\,\textrm{e}}$ in the adjoint representation of $\mathfrak{u}(N)$~\cite{Blau:1993hj}:
\begin{align}
\label{pathint1stO}
\begin{split}
\hspace{-0.15cm} \mz^\kappa_{\SUN} [B] &= \int_{\mathscr{A}^{\,\textrm{e}}} \, \mathscr{D} A^{\,\textrm{e}}  \ \int_{\Omega^0(\Sigma,\mathfrak{u}(N))} \, \mathscr{D} \phi^{\,\textrm{e}} \  \int_{ \Omega ^0 (\Sigma, \R/\Z ) }\, \mathscr{D} \lambda \ \int_{ \Omega ^0 (\Sigma,\R) }\, \mathscr{D} \xi \\
	& \quad \, \times  \exp \bigg[ \int_{\Sigma}\, \big( \xi\, \tr (F^{\,\textrm{e}} ) + \ii\, \tr ( \phi^{\,\textrm{e}}\, F^{\,\textrm{e}} ) \big) \\
	& \hspace{2cm} + \frac{g^2}{2}\, \int_{\Sigma}\, \omega \, \tr ( \phi^{\,\textrm{e}}{\,} ^2 )  -  \frac{N}{ 2 g^2 }\, \int_{\Sigma}\,  B \hd\, B +  \int_{\Sigma}\, B\, ( - \xi\, N + \kappa   + \lambda\, k  ) \bigg] \ ,
\end{split}
\end{align}
where now the $F^{\,\textrm{e}} $-dependence is entirely in the second line and is linear.

At this stage, it is convenient to make a change of variable
\begin{align*}
	\phi & = \phi^{\,\textrm{e}} -\ii\, \xi \otimes \id_N \ ,
\end{align*}
giving
\begin{align*}
	- \ii\, \xi\, \tr (F^{\,\textrm{e}} ) + \tr ( \phi^{\,\textrm{e}}\, F^{\,\textrm{e}} ) &= \tr (\phi\, F^{\,\textrm{e}} ) \ , \\[4pt]
	\tr ( \phi^{\,\textrm{e}} {\,}^2 ) & =  \tr ( \phi^2 ) +2\,\ii\, \xi\,  \tr ( \phi ) - N\, \xi^2 \ ,
\end{align*}
followed by the field redefinition 
\begin{equation}
\label{changexi}
	\xi^{\prime} = \xi - \tfrac{\ii }{N}\,  \tr ( \phi ) 
\end{equation}
in order to complete the square for the Lagrange multiplier field $\xi$. Note that $\phi$ is anti-Hermitian and $\xi$ is valued in $\R$, as opposed to the other Lagrange multiplier $\lambda$ valued in $\R/\Z$; thus \eqref{changexi} is well-defined. 

With these changes, the exponent in \eqref{pathint1stO} becomes 
\begin{align*}
	 & - \frac{g^2\, N}{2}\,\int_{\Sigma}\, \omega\, \xi^{\prime}{}^2 + \ii\, \int_{\Sigma}\, \tr ( F^{\,\textrm{e}} \, \phi ) - \frac{g^2}{2}\,\int_{\Sigma}\, \omega\, \Big( \tr ( \phi^2 ) - \frac{1}{N}\, ( \ii\, \tr\, \phi )^2  \Big) \\
	 & \hspace{2cm} +  \int_{\Sigma}\, B\, \big( - \xi^{\prime}\, N - \ii\, \tr ( \phi )  + \kappa  + \lambda\, k  \big)  -  \frac{N}{ 2 g^2 }\, \int_{\Sigma}\,  B \hd\, B \ .
\end{align*}
We now integrate out the auxiliary field $\xi^{\prime}$ and produce the term 
\begin{equation*}
	\frac{N}{2 g^2}\, \int_{\Sigma}\, B \hd\, B 
\end{equation*}
that cancels the pre-existent quadratic piece in $B$.

We finally arrive at
\begin{align}
\label{pathintlast}
\begin{split}
	\mz^\kappa_{\SUN} [B] &= \int_{\mathscr{A}^{\,\textrm{e}}}\, \mathscr{D} A^{\,\textrm{e}}  \ \int_{\Omega^0(\Sigma,\mathfrak{u}(N))}\, \mathscr{D} \phi \ \int_{ \Omega ^0 (\Sigma, \R/\Z ) }\, \mathscr{D} \lambda  \  \exp \bigg(  \ii\, \int_{\Sigma}\, \tr ( F^{\,\textrm{e}} \, \phi ) \\
& \hspace{2cm} +\frac{g^2}{2}\,\int_{\Sigma}\, \omega\, \Big(  \tr ( \phi^2 ) - \frac{1}{N}\, ( \ii\, \tr\, \phi )^2  \Big)  +    \int_{\Sigma}\, B\, \big(\kappa - \ii\, \tr ( \phi )  + \lambda\, k  \big)  \bigg) \ .
	\end{split}
\end{align}

\subsubsection*{Abelianization}

The $F^{\,\textrm{e}} $-dependence in \eqref{pathintlast} is completely isolated from any other term except its linear coupling with the $\mathfrak{u}(N)$ adjoint scalar $\phi$. At this point, we run the argument of \cite{Blau:1993hj} for pure $\UN$ Yang--Mills theory with no variation, for what concerns the integration over $A^{\,\textrm{e}} $ and $\phi$.
The result is 
\begin{align}
\begin{split}
\mz^\kappa_{\SUN} [B] &= \int_{ \Omega ^0 (\Sigma, \R/\Z ) }\, \mathscr{D} \lambda \ \sum_{R^{\,\textrm{e}}  }\, (  \dim R^{\,\textrm{e}} )^{\chi } \\
& \times \exp \bigg[ - \frac{g^2}{2}\, \bigg(  C_2^{\,\textrm{e}} (R^{\,\textrm{e}} ) - \frac{Q^2}{N} + \frac{N\,(N^2-3N+2)}{12}\bigg) + \int_{\Sigma}\, B\, \big( \lambda\, k + \kappa - Q \big) \bigg] \ ,   \label{abelianizedBG}
\end{split}
\end{align}
with the sum running over isomorphism classes of irreducible representations $R^{\,\textrm{e}}$ of $\UN$. 

Observe that in \eqref{pathintlast} there is the term 
\begin{equation*}
	\int_{\Sigma}\, \omega\, \Big( \tr ( \phi^2) - \frac{1}{N}\,  ( \ii\,\tr\, \phi )^2 \Big)
\end{equation*}
in the action functional, rather than the usual $\int_{\Sigma}\, \omega\, \tr ( \phi^2 )$. This produces the term $C_2^{\,\textrm{e}} (R^{\,\textrm{e}} ) - \frac{Q^2}{N} $ in \eqref{abelianizedBG}, where the integer $Q \in \Z$ comes from $\ii\,\tr\, \phi$ after the abelianization outlined above, while $C_2^{\,\textrm{e}}$ is the $\UN$ quadratic Casimir invariant. The latter is related to the $\SUN$ Casimir invariant $C_2$ through \eqref{C2hatC2}, which is precisely the combination appearing in \eqref{abelianizedBG}. As before, we add a background counterterm involving solely the K\"ahler metric on $\Sigma$ and no dynamical field to cancel the constant decoupled factor \smash{$\frac{g^2}{24}\, N\,(N^2-3N+2)$}.

To sum up, \eqref{abelianizedBG} is naturally set to be separated into an $\SUN$ part and a $\Uu$ part through the relation \eqref{eq:R'RQ} between $\UN$ and $\SUN$ representations. We thus obtain
\begin{align}
\begin{split}
\mz_{\SUN}^{\kappa} [B] &=  \sum_{q \in \mathbb{Z}} \ \sum_{R }\,  (  \dim R )^{\chi } \ \e^{- \frac{g^2}{2}\, C_2 (R) }  \\
& \hspace{1.5cm} \times \int_{ \Omega ^0 (\Sigma, \R/\Z ) }\, \mathscr{D} \lambda  \ \exp \bigg( \int_{\Sigma}\, B\, \big( \lambda\, k + \kappa - \lvert R \rvert \big)  - 2 \pi\,  q\, \int_{\Sigma}\, \frac{N\, B}{2\pi} \bigg) \ .  \label{AbelSUN}
\end{split}
\end{align}

Recall that $k$ divides $N$ and $\frac{k\,B}{2\pi}$ will represent an integer cohomology class after integrating out the auxiliary field $\lambda$. Therefore the last term in the exponential of \eqref{AbelSUN}, which we have written separately, can be dropped because it will trivialize after integrating out $\lambda$. 
Summing over $q \in \mathbb{Z}$ first imposes the contraint 
\begin{equation*}
	\int_{\Sigma}\, \frac{N\, B}{2\pi}  \ \in \ \ii\, \mathbb{Z} \ ,
\end{equation*}
which will be overwritten by the subsequent integration over $\lambda$, imposing the more restrictive constraint \eqref{Bperiod}. The result is again \eqref{eq:ZBbg}.

\subsection{Gauging the 1-form symmetry}
\label{subsec:gauging1form}

So far we have derived the expression \eqref{eq:ZBbg} for the partition function of $\SUN$ Yang--Mills theory coupled to a background 2-form gauge field $B$ for the 1-form symmetry. To gauge this symmetry, we promote $B$ to a dynamical field and integrate over it.

Including the path integration over $B$ in \eqref{AbelSUN} and integrating over $\lambda$, we get 
\begin{align}
\begin{split}
\mz_{\SUN / \Z_k}^\kappa &= \int_{\Omega^2(\Sigma,\mathfrak{u}(1))} \, \frac{\mathscr{D}B}{2\pi} \ \mz_{\SUN}^\kappa[B]  \\[4pt]
&=   \sum_{R }\,  (  \dim R )^{\chi } \ \e^{- \frac{g^2}{2}\, C_2 (R) }  \\
& \hspace{2cm} \times \int_{ \Omega ^2 (\Sigma,\mathfrak{u}(1) ) }\, \frac{\mathscr{D} B}{2\pi} \ \sum_{\ell\in \Z }\, \delta \Big( \int_{\Sigma}\, \frac{k\,B}{2\pi} - \ii\, \ell  \Big) \ \exp \Big( \int_{\Sigma}\, B\, ( \kappa  - \lvert R \rvert) \Big) \ . \label{AbelSUN1}
\end{split}
\end{align}
We now exchange the integration over $B$ and the sum over $\ell$, a manipulation which is justified \emph{a posteriori} by the convergence of the result. After pulling out a coefficient, which gives a prefactor of $\frac{2\pi}{k}$, the delta-function is used to remove the integral over the 2-form field $B$. 

Rewriting $\ell = \beta + k\, \ell^{\prime}$ with $\beta \in \Z_k$ and $\ell^{\prime} \in \Z$, we arrive at 
\begin{align*}
\mz_{\SUN / \Z_k}^\kappa  &=  \sum_{R  }\,  (  \dim R )^{\chi } \ \e^{- \frac{g^2}{2}\, C_2 (R) } \ \frac{1}{k}\, \sum_{\beta=0} ^{k-1}\, \e^{\,2 \pi\,\ii\, \beta \,( \kappa  - \lvert R \rvert)/k } \ \sum_{\ell^{\prime}\in \Z }  \e^{\,2 \pi\,\ii\, \ell^{\prime} \,( \kappa  - \lvert R \rvert) } \\[4pt]
	& =  \sum_{R  }\, (  \dim R )^{\chi } \ \e^{- \frac{g^2}{2}\, C_2 (R) } \ \delta (  \lvert R \rvert -   \kappa \ \mathrm{mod} \ k) \ .
\end{align*}
This completes the proof of Proposition~\ref{prop:Zorbi}. Observe that the sum over $\ell^{\prime} \in \Z$ would vanish if $\kappa \notin \Z$, as expected since we have projected out the gauge non-invariant terms \eqref{thetaterm} when $\kappa \notin \Z$.

Of particular interest is the case of $N$ even, $k$ an even integer dividing $N$ and $\kappa = \frac{k}{2}$, which is the focus of Section~\ref{sec:chargeconj}.

\subsection{Comparison with combinatorial quantization}
\label{sec:WittenLoc}

In this subsection we show that our approach, based on 2-bundles over $\Sigma$, is equivalent to Witten's original derivation~\cite{Witten:1992xu} of the formula \eqref{ZorbiDelta} using the standard combinatorial approach to quantization of two-dimensional Yang--Mills theory, which involves cutting and gluing calculations along intermediate surfaces with boundary.

The idea of the proof in~\cite{Witten:1992xu} is to work directly with the gauge group $G=\SUN/\Z_k$ and consider the space of connections \smash{$\widetilde A$} on a $G$-bundle \smash{$\widetilde P\longrightarrow \Sigma$}. By deleting a point $\wp$ from $\Sigma$, the $G$-bundle \smash{$\widetilde P$} can be lifted to a principal $\SUN$-bundle $\widetilde  P^{\,\circ} $ on $\Sigma \setminus \left\{\wp \right\}$. Denoting by \smash{$\widetilde  A^{\,\circ} $} the $\SUN$-connection which is the lift of a $G$-connection \smash{$\widetilde A$} when restricted to $\Sigma \setminus \left\{\wp \right\}$, the data of the original bundle \smash{$\widetilde P \longrightarrow \Sigma$} is encoded in the pair $(\widetilde  A^{\,\circ} ,u)$, where $u\in\Z_k$ is the monodromy of $\widetilde  A^{\,\circ} $ around~$\wp$. This setup is schematically illustrated for the sphere $\Sigma=\mathbb{P}^1$ in Figure \ref{fig:monodromyWitten}.

\begin{figure}[htb]
\centering
\small
	\begin{tikzpicture}[scale=0.9]
	\node at (0,-2.6) {(a)};
  \shade[ball color = gray!40, opacity = 0.4] (0,0) circle (2cm);
  \draw (0,0) circle (2cm);
  \draw[thin] (-2,0) arc (180:360:2 and 0.6);
  \draw[dashed,thin] (2,0) arc (0:180:2 and 0.6);
  \node (p) at (-1,1) {$\bullet$};
  \node[anchor=south] (A) at (-1,2.6) {$\widetilde{A}$};
  \draw[dashed,very thick] (p) -- (A);
\end{tikzpicture}\hspace{0.25\textwidth}%
\begin{tikzpicture}[scale=0.9]
	\node at (0,-2.6) {(b)};
  \shade[ball color = gray!40, opacity = 0.4] (0,0) circle (2cm);
  \draw (0,0) circle (2cm);
  \draw[thin] (-2,0) arc (180:360:2 and 0.6);
  \draw[dashed,thin] (2,0) arc (0:180:2 and 0.6);
  \node[ellipse,dashed,thick,blue,draw] (p) at (-0.55,1) {$\times$};
  \node[anchor=west,blue] (wp) at (-.55,1) {$\wp$};
  \node[anchor=south] (A) at (-1,2.6) {$(\widetilde{A}^{\,\circ},u)$};
  \node (q) at (-1,1) {$\bullet$};
  \draw[dashed,very thick] (q) -- (A);
\end{tikzpicture}
\normalsize
\caption{(a) Schematic depiction of a $G$-connection $\widetilde{A}$ on $\mathbb{P}^1$. (b) Schematic depiction of an $\SUN$-connection $\widetilde{A}^{\,\circ}$ on $\mathbb{P}^1 \setminus \left\{ \wp \right\}$ together with its monodromy $u$.}
\label{fig:monodromyWitten}
\end{figure}

We thus learn that the space of $G$-connections is disconnected, with connected components labelled by elements $u$ of the fundamental group $\pi_1 (G) \cong\pi_0(\Z_k) = \Z_k$. Then one applies standard methods of two-dimensional Yang--Mills theory on $\Sigma\setminus\left\{\wp\right\}$ in each connected component, which amounts to multiplying the contribution of an $\SUN$-representation $R$ in the expansion \eqref{eq:MRformula} by the normalized character $\chi_R(u^{-1})/\dim R$,
and subsequently sums over $u\in\Z_k$ the contributions from all components \cite{Witten:1992xu}. The result is independent of the choice of point $\wp\in\Sigma$ due to the invariance of two-dimensional Yang--Mills theory under area-preserving diffeomorphisms.

This computation may also be understood in the following way. The obstruction to lifting a $G$-bundle \smash{$\widetilde P\longrightarrow \Sigma$} to an $\SUN$-bundle $P$ on the whole of $\Sigma$ is the lifting ${\Z_k}$-gerbe $\mathbbm{G}\longrightarrow\Sigma$ associated to the group central extension
\begin{align*}
1\longrightarrow \Z_k \longrightarrow \SUN \longrightarrow G\longrightarrow 1 \ ,
\end{align*}
which is classified by 2-cocycles $u\in H^2(\Sigma,\Z_k)\cong\Z_k$ (see e.g. \cite{Murray:2007ps}). As discussed in Subsection~\ref{sec:prelim}, it  is these $\Z_k$-gerbes, or equivalently principal $\mathrm{B}\Z_k$-bundles, which play the role of background gauge fields for the gauging of the 1-form symmetry $\Z_k^{\scriptscriptstyle (1)}$, understood as a $\text{B}\Z_k$ 2-group symmetry of the stack of $\SUN$-bundles with connection.\footnote{We are grateful to Lukas M\"uller for suggesting this description.}

The path integral coupled to a background field \smash{$\mathbbm{G}$} is computed as a path integral over the space of \smash{$\mathbbm{G}$}-twisted $\SUN$-bundles on $\Sigma$, corresponding to the 2-group extension
\begin{align*}
1\longrightarrow\SUN\longrightarrow \SUN\rtimes\mathrm{B}\Z_k\longrightarrow\mathrm{B}\Z_k\longrightarrow 1
\end{align*}
defined by the action of the 1-form symmetry. Principal \smash{$\SUN\rtimes\mathrm{B}\Z_k$}-bundles on $\Sigma$ can be equivalently described by $G$-bundles on $\Sigma$.\footnote{This is similar to the description of $\Uu$-gerbes by $\mathrm{PU}(\infty)$-bundles, where $\mathrm{PU}(\infty)=\mathrm{U}(\infty)/\Uu$ is the projective unitary group of an infinite-dimensional separable Hilbert space.} Concretely, a \smash{$\mathbbm{G}$}-twisted $\SUN$-bundle \smash{$\mathbbm{G}\times_{\mathrm{B}\Z_k}P\longrightarrow\Sigma$} can be identified with a $G$-bundle \smash{$\widetilde P\longrightarrow\Sigma$} with connection \smash{$\widetilde A$}, such that the obstruction to lifting $\widetilde P$ to an $\SUN$-bundle $P\longrightarrow\Sigma$ with connection $A$ is given by $u\in\Z_k$. 

The path integral over all \smash{$\mathbbm{G}$}-twisted $\SUN$-bundles on $\Sigma$ involves the sum over $\Z_k$-gerbes. It may be computed by summing over all possible insertions, labelled by $u\in\Z_k$, of a single topological point defect in the path integral over connections on  an  $\SUN$-bundle $P\longrightarrow\Sigma$ (as there is only one generating homology class $[\wp]$ of 0-cycles since $\Sigma$ is connected), which corresponds to the local topological operator generating the discrete 1-form symmetry $\Z_{k}^{\scriptscriptstyle (1)}$. This coincides exactly with Witten's calculation. It follows that gauging the $\mathrm{B}\Z_k$ symmetry corresponds to summing over $\Z_k$-gerbes on $\Sigma$, as well as summing over $u$-twisted $\SUN$-bundles for each $\Z_k$-gerbe with characteristic class $u$, and leads to two-dimensional Yang--Mills theory with gauge group $G=\SUN/\Z_k$.

In our formulation, we emphasize the role of gauging the 1-form symmetry by first principles path integral techniques, in the same spirit as the approach of~\cite{Witten:1992xu}. For this purpose, instead of working directly with a $G$-connection \smash{$\widetilde A$}, we have lifted the $\SUN$-bundle $P \longrightarrow \Sigma$ to a non-abelian gerbe on $\Sigma$. The pair of data $(\widetilde  A^{\,\circ} ,u)$ is replaced by the data of the 2-connection $(A,B)$ on the principal $\mathbbm{U}(N)$-bundle \eqref{eq:P2bundle}, with the constraint that the curving $\frac kN\,B$ is homomorphic to the zero curving on the $\Uu$-gerbe $ \mathbbm{P}\longrightarrow\Sigma$ under the lift of $P$ to the $\UN$-bundle \eqref{Pprimebundle}. The path integral over the subspace of such pairs $(A,B)$ in the affine bundle of connections on the non-abelian gerbe coincides with the path integral over $\SUN$-connections $\widetilde  A^{\,\circ} $ followed by the sum over monodromies~$u\in\Z_k$.

\subsection{Gauging the dual \texorpdfstring{$(-1)$-form}{(-1)-form} symmetry}
\label{sec:dualgauging}

In a $d$-dimensional quantum field theory, gauging a $p$-form symmetry produces a dual $(d-p-2)$-form symmetry. This generalizes the classical result that gauging a 0-form symmetry in two dimensions produces a dual (sometimes called `quantum') 0-form symmetry \cite{Vafa:1989ih}.

Gauging the 1-form symmetry $\sB^{\scriptscriptstyle (1)}= \Z_k ^{\scriptscriptstyle (1)}$ in two dimensions produces a $(-1)$-form symmetry $\sK^{\scriptscriptstyle (-1)}$, with 
\begin{equation*}
	\sK = \widehat{\sB} = \widehat{\Z_k} \cong \Z_k \ .
\end{equation*}	
This symmetry is generated by codimension zero topological operators, which are Wilson surfaces
\begin{equation*}
	S_{\eta} = \exp \, \eta\, \int_{\Sigma}\, B   \ , \quad \eta \in \left\{ 0,1, \dots, k-1 \right\} \ .
\end{equation*}
These operators are topological in the sense that they do not depend on the K\"ahler metric on $\Sigma$. Alternatively, one may embed the two-dimensional theory as a defect into a three-dimensional quantum field theory and use the flatness of $B$ to prove invariance of $S_{\eta}$ under small deformations of the surface by the usual arguments.

Let us assume that $m \in \mathbb{N}$ divides $k$; the limiting cases $m=1$ and $m=k$ are allowed. We now proceed to gauge the subgroup $\Z_m  \subseteq \sK $, generated by the topological symmetry operators with charge $\eta= k\, \nu /m$:
\begin{equation*}
	S_{\eta=k\,\nu/m}  = \exp \, \frac{k}{m} \, \nu \, \int_{\Sigma}\, B \ , \quad \nu \in \left\{ 0,1, \dots, m-1 \right\} \ .
\end{equation*}
One way to proceed would be to turn on a background 0-form gauge field for $\Z_m ^{\scriptscriptstyle (-1)}$ that partially undoes the gauging of $B$. Another approach for gauging such a discrete symmetry is, instead, to sum over insertions of networks of topological operators generating it.

\subsubsection*{Gauging by summing over topological surface operators}

In the present case, each topological symmetry operator fills the whole Euclidean spacetime $\Sigma$, thus we only need to sum over all possible insertions of one such operator: there is only one generating homology class $[\Sigma]$ of 2-cycles. Summing \eqref{AbelSUN} over all operator insertions gives
\begin{align*}
\sum_{\nu=0}^{m-1}\,\mz^{\kappa}_{\SUN / \Z_k} \big[ S_{\eta=k\,\nu/m} \big]  &=  \int_{ \Omega ^0 (\Sigma, \R/\Z ) }\, \mathscr{D} \lambda \ \int_{\Omega^2 (\Sigma,\mathfrak{u}(1))}\, \mathscr{D} B \ \sum_{R }\,  \left(  \dim R \right)^{\chi } \ \e^{- \frac{g^2}{2}\, C_2 (R) }  \notag \\
& \hspace{2cm} \times   \exp \bigg( \int_{\Sigma}\, B \, \big( \lambda\, k + \kappa - \lvert R \rvert \big) \bigg) \ \sum_{\nu=0}^{m-1}\, \exp \bigg( \frac{k}{m}\, \nu \, \int_{\Sigma}\, B  \bigg) \   .
\end{align*}

Integrating out $\lambda$ and then $B$ we arrive at the simple modification of \eqref{ZorbiSum} given by
\begin{align*}
\sum_{\nu=0}^{m-1}\, \mz_{\SUN / \Z_k}^{\kappa} \big[ S_{\eta=k\,\nu/m} \big]  &= \sum_{R }\,  \left(  \dim R \right)^{\chi } \ \e^{- \frac{g^2}{2}\, C_2 (R) } \ \sum_{\nu=0}^{m-1}\, \delta \Big(  \lvert R \rvert -   \kappa  - \frac{k}{m}\, \nu \ \mathrm{mod}~k \Big) \\[4pt]
	&= \sum_{R }\,  \left(  \dim R \right)^{\chi } ~ \e^{- \frac{g^2}{2}\, C_2 (R) } \ \frac{1}{k}\,\sum_{\beta=0} ^{k-1} \ \sum_{\nu=0}^{m-1}\,  \e^{ \,2 \pi \,\ii\,\frac{\beta}{k} \,( \kappa  - \lvert R \rvert ) +  2 \pi \,\ii\,\beta\, \frac{\nu}{m} } \\[4pt]
	&= \sum_{R } \,  \left(  \dim R \right)^{\chi } ~ \e^{- \frac{g^2}{2}\, C_2 (R) } \ \frac{m}{k}\,\sum_{\beta=0} ^{k-1}\, \e^{\, 2 \pi \,\ii\,\frac{\beta}{k}\, ( \kappa  - \lvert R \rvert )} \ \delta ( \beta \ \mathrm{mod}~m ) \\[4pt]
	&= \sum_{R }\,  \left(  \dim R \right)^{\chi } ~ \e^{- \frac{g^2}{2}\, C_2 (R) } \ \delta \Big(  \kappa - \lvert R \rvert  \ \mathrm{mod}~\frac{k}{m} \Big) \ .
\end{align*}

That is, we recover the theory as if we had gauged only the 1-form symmetry $\Z_{k/m} ^{\scriptscriptstyle (1)}$, as expected:
\begin{equation*}
\sum_{\nu=0}^{m-1}\, \mz_{\SUN / \Z_k}^{\kappa} \big[ S_{\eta=k\,\nu/m} \big]  = \mz_{\SUN / \Z_{k/m}}^{\kappa ~\mathrm{mod}~\frac{k}{m}}  \ .
\end{equation*}

\subsubsection*{Path integral gauging of the dual symmetry}

Let us now show how to gauge the dual symmetry $\sA^{\scriptscriptstyle (-1)}= \Z_m^{\scriptscriptstyle (-1)}$ at the level of the path integral. We do so in the limiting case $k=N$ to reduce clutter, but the argument goes through in the more general case as well. The 0-form gauge field for the $(-1)$-form symmetry is a scalar $a$. Recall from Subsection~\ref{sec:prelim} that a $(-1)$-form gauge transformation shifts $a \longmapsto a + 2 \pi\, \ii\, n$ with $n\in \mathbb{Z}$.

Looking back at the extension of $\SUN$ into $\UN$, namely neglecting the Lagrange multiplier $\lambda$ throughout the previous discussions, we would obtain a $\Uu^{\scriptscriptstyle (-1)}$ Chern--Weil symmetry with conserved 0-form current $\ast \, \tr\, F^{\,\textrm{e}} $. In the $\PSUN$ theory, this is broken down to $\Z_N^{\scriptscriptstyle (-1)}$ by the constraint that the period of $N\,B$ lies in $\ii\Z$. We can nevertheless couple the $\Uu^{\scriptscriptstyle (-1)}$ current to a source term first, and only then integrate out $\lambda$. In this way we obtain the standard minimal coupling of the 0-form gauge field $a$ for a magnetic symmetry $\sA^{\scriptscriptstyle (-1)}$ to the $\PSUN$ curvature $F^{\,\textrm{e}} $: 
\begin{equation}
\label{magcouplingaF}
	\int_{\Sigma}\, a \, \tr\, F^{\,\textrm{e}}  = N\, \int_{\Sigma}\, a \, B \ ,
\end{equation}
where we used \eqref{traceFprime}.
Inserting this term in the relevant part of \eqref{AbelSUN1}, we obtain the action functional
\begin{equation*}
	\int_{\Sigma}\,B \, \big( N\, \lambda -  N\, a + \kappa - \lvert R \rvert \big) \  .
\end{equation*}

Next we integrate over the gauge orbits $a\in\Omega^0(\Sigma,\R/\Z)$ by the $(-1)$-form symmetry. Writing $N\, a\, B = m\, a \ \frac{N}{m}\, B$, we arrive at the modification of \eqref{AbelSUN1} given by
\begin{align*}
	&  \sum_{R }\,  (  \dim R )^{\chi } \ \e^{- \frac{g^2}{2}\, C_2 (R) } \ \int_{ \Omega ^2 (\Sigma,\mathfrak{u}(1) ) }\, \frac{\mathscr{D} B}{2\pi} \ \sum_{\ell\in \Z }\, \delta \Big(\int_{\Sigma}\,  \frac{N}{m}\, \frac{B}{2\pi} - \ii \, \ell  \Big) \\
	& \hspace{7cm} \times \int_{ \Omega ^0 (\Sigma, \R/\Z ) }\, \mathscr{D} \lambda  \ \exp \Big( \int_{\Sigma}\, B\, \big( N\, \lambda + \kappa  - \lvert R \rvert\big) \Big) \ .
\end{align*}
The integration over $\lambda$ goes as before, but this time it is ineffective as a more stringent constraint has already been imposed on $B$. We conclude that the resulting theory is $\SUN/\Z_{N/m}$ Yang--Mills theory, as expected. Inverting the order by integrating out $B$ first and then $a$ would bring us back to the previous derivation.

\subsection{Line operators in two-dimensional Yang--Mills theory}
\label{sec:reading}

We state the famous result of \cite{Aharony:2013hda}, and then proceed to show how it is explicitly realized in the orbifolds of two-dimensional Yang--Mills theory considered thus far.

\begin{lem}\label{lemma:AST}
The global structure of the gauge group in pure Yang--Mills theory is detected by the spectrum of line operators in the theory.
\end{lem}

Consider pure Yang--Mills theory with gauge algebra $\mathfrak{su}(N)$, placed for simplicity on an oriented disk $\mathbb{D}$ of area $\frac{1}{2}$, so that a sphere $\mathbb{P}^1$ obtained from gluing two such disks with opposite orientations has unit area. We wish to determine the gauge group $G$. For this, we start by taking a representation $R$ of the simply-connected group with Lie algebra $\mathfrak{su}(N)$. That is, we let $R$ be an $\SUN$-representation. We write $W_R:=W_R (-\partial \mathbb{D}) $ and denote by $\langle W_R\rangle$ the path integral computed with a Wilson line in the representation $R$ inserted along the boundary $\partial \mathbb{D}$ with opposite orientation to that induced by the orientation of $\mathbb{D}$:
\begin{equation*}
	\langle W_R \rangle = \int_{\mathscr{A}} \, \mathscr{D} A \  \exp \Big(\frac{1}{2g^2 }\, \int_{\mathbb{D}}\, \tr\, F \ast F  \Big) \ \tr_R\ \mathtt{P}\, \exp \Big( \oint_{- \partial \mathbb{D}}\, A \Big) \ .
\end{equation*}

We compute 
\begin{equation}
\label{eq:WRvevdisk}
	\langle W_R \rangle = \begin{cases} \ \e^{-\frac{g^2}{4}\, C_2 (R)} \, \dim R & \text{ if $R$ is a $G$-representation} \ , \\ \ 0 & \text{ otherwise} \ . \end{cases} 
\end{equation}
Suppose we are given a list of a finite number of values of $\langle W_R \rangle$ for all representations $R$ of $\SUN$ with $\lvert R \rvert \le N$. Then we can extract the global form of the gauge group $G$ from this list. 

For example, depending on whether or not the trivial representation is in the list, we obtain
\begin{equation*}
	 \langle W_{\emptyset }\rangle  = \begin{cases} \ 1 & \Longrightarrow \ \kappa=0  \ , \\ \ 0 & \Longrightarrow \ \text{$G$ is not simply-connected and $\kappa\ne 0$} \ . \end{cases}
\end{equation*}
From the fundamental representation we infer 
\begin{equation*}
	 \langle W_{\Box }\rangle  = \begin{cases} \ N \, \e^{-\frac{g^2}{4}\, (N-1)}  & \Longrightarrow \ \kappa \in \left\{ 0,1\right\}  \ , \\ \ 0 & \Longrightarrow \ \text{$G= \SUN/\Z_k$ with (i) $k\ge 2$, $\kappa=0$ or (ii) $k>2$, $\kappa> 1$} \ . \end{cases}
\end{equation*}

Therefore, combining the information about the trivial and fundamental representations, we deduce
\begin{equation*}
	G \ = \ \begin{cases} \ \SUN & \text{ if both $\langle W_{\emptyset}\rangle \ne 0$ and $\langle W_{\Box }\rangle\ne 0$ } \ , \\  \ [\SUN/\Z_{k \ge 2}]_{\kappa =0} & \text{ if $\langle W_{\emptyset}\rangle \ne 0$ but $\langle W_{\Box }\rangle= 0$ } \ , \\ \ [\SUN/\Z_{k \ge 2}]_{\kappa =1} & \text{ if $\langle W_{\emptyset}\rangle = 0$ but $\langle W_{\Box }\rangle\ne 0$ } \ , \\ \ [\SUN/\Z_{k > 2}]_{\kappa >1} & \text{ if both $\langle W_{\emptyset}\rangle = 0$ and $\langle W_{\Box }\rangle = 0$ \ . } \end{cases}
\end{equation*}
Going further with higher-dimensional representations, one eventually finds that, among all $R$ with $\lvert R \rvert \le N$, only those with $\lvert R \rvert= \kappa \ \mathrm{mod} \ k$ yield a non-vanishing $\langle W_R\rangle  $. The pair $(k, \kappa)$ is thus read off from the non-trivial entries in the list.

This is of course a well-known result \cite{Aharony:2013hda}. The point we wish to make here is that, thanks to the formula \eqref{eq:WRvevdisk}, Lemma \ref{lemma:AST} finds a concrete and explicit realization in this model. The sphere partition function is written in the form 
\begin{equation*}
	\mz_{G} [\mathbb{P}^1] = \sum_{R} \, \langle W_R  \rangle^2 \ ,
\end{equation*}
with the sum running over all irreducible representations $R$ of $\SUN$ up to isomorphism, but $\langle W_R\rangle$ is the Wilson loop expectation value taken with respect to the group $G$.

\subsection{Charge conjugation symmetry orbifolds}
\label{sec:gaugeC}

The Lie algebra $\mathfrak{su}(N)$ possesses an outer automorphism group $\sC:=\sO=\Z_2$, with the non-trivial element denoted $\mathcal{C}$ and called charge conjugation in physics parlance. The principal extension of $\sC$ by the gauge group $\SUN$ is the disconnected group \smash{$\wSUN$}, defined by a short exact sequence of groups
\begin{equation}
\label{extSUNtilde}
	1 \longrightarrow \SUN  \longrightarrow \wSUN \longrightarrow \pi_0 \big( \wSUN \big)  \longrightarrow 1  \ ,
\end{equation}
with 
\begin{equation*}
	\pi_0 \big( \wSUN \big) \cong \sC \ . 
\end{equation*}
In physical terms, the extension \eqref{extSUNtilde} corresponds to taking the orbifold by $\sC$. Four-dimensional gauge theories with disconnected gauge group \smash{$\wSUN$} have been constructed and studied in \cite{Bourget:2018ond,Arias-Tamargo:2019jyh,Henning:2021ctv}.

An irreducible representation $\widetilde{R}$ of $\wSUN$ corresponds to a (possibly reducible) representation $R$ of $\SUN$ which satisfies $\overline{R} \cong R$, where $\overline{R}= \mathcal{C} \cdot R$ is the conjugate representation. In particular, both fundamental and anti-fundamental representations of $\SUN$ are projected out by gauging $\sC$, but the bifundamental representation $\Box\, \oplus\, \overline{\Box}$ descends to an irreducible representation of \smash{$\wSUN$}~\cite{Bourget:2018ond,Arias-Tamargo:2019jyh}.

\begin{lem}
Let $R$ be a representation of $\SUN$ that is isomorphic to its charge conjugate, $\overline{R} \cong R$. Then $R$ descends to a representation $\widetilde{R}$ of \smash{$\wSUN$}, which is irreducible if and only if $R$ is reducible.
\end{lem}

\subsubsection*{Gauging charge conjugation via topological symmetry operators}

The gauge theory resulting from gauging the outer automorphisms of a generic gauge group $G$ in two-dimensional Yang--Mills theory has been analyzed in \cite{Muller:2019tnp}. Here we focus on $G=\SUN$.

A direct way to gauge a 0-form symmetry in two dimensions is to sum over all possible insertions of networks of topological line operators that generate the symmetry \cite{Chang:2018iay}. The notion of network requires a definition of topologically equivalent classes of insertions of operators. We give a simplified definition here and refer to \cite{Muller:2019tnp} for a more careful treatment.

\begin{defin} \label{def:network}
Let $\sA$ be a finite group and consider a two-dimensional quantum field theory with 0-form global symmetry group $\sA^{\scriptscriptstyle (0)}$ on $\Sigma$. For every $\alpha \in \sA$, let $L_{\alpha}$ be a topological loop operator that generates the action of $\alpha$ and winds around $[L] \in H_1 (\Sigma, \Z)$. Fix a basis $\{ L_j \}_{j=1, \dots, 2-\chi}$ for $ H_1 (\Sigma, \Z)$. A \emph{network of topological loop operators for $\sA^{\scriptscriptstyle (0)}$}, or \emph{$\sA$-network} for short, is a collection
	\begin{equation*}
	\mathcal{L} =	\left\{ L_{j,\alpha_j} \ \big| \ \alpha_j \in \sA \ , \ j=1, \dots, 2-\chi \right\} \ .
	\end{equation*}
\end{defin}

We gauge $\sC^{\scriptscriptstyle (0)}$ by summing over the insertions of $\sC$-networks on $\Sigma$, which computes the path integral over connections on an $\SUN$-bundle $D\times_{\sC}P\longrightarrow\Sigma$ twisted by a $\sC$-bundle $D\longrightarrow\Sigma$. The aim is to illustrate an alternative approach to \cite{Muller:2019tnp}, which we still refer to for a more exhaustive analysis. The first case to consider is $\Sigma = \mathbb{P}^1$. However, the sphere does not have non-contractible cycles, therefore every network is homologically trivial. We conclude that the partition function on $\mathbb{P}^1$ is insensitive to gauging $\sC$. Then the simplest non-trivial example is the torus $\ct^2$.

\subsubsection*{The torus}

Let $\Sigma=\ct^2$. There exist four basic $\sC$-networks
\begin{equation}
\label{eq:CnetworkT2}
	\mathcal{L}_{1} := \left\{ L_{1,1} , L_{2,1}\right\} \ , \ \mathcal{L}_{\mathcal{C}}^{(1)} := \left\{ L_{1,\mathcal{C}} , L_{2,1}\right\}  \ , \ \mathcal{L}_{\mathcal{C}}^{(2)} := \left\{ L_{1,1} , L_{2,\mathcal{C}}\right\} \ , \ \mathcal{L}_{\mathcal{C}\mathcal{C}} := \left\{ L_{1,\mathcal{C}} , L_{2,\mathcal{C}} \right\}  \ ,
\end{equation}
with $(L_1, L_2)$ denoting the canonical symplectic basis of $H_1 (\ct^2, \Z)\cong\Z\oplus\Z$. All other $\sC$-networks can be decomposed into superpositions of these loop operators. 
With the notation $\ct^2 \supset \mathcal{L}$ for the torus decorated with the $\sC$-network $\mathcal{L}$, the orbifold partition function is 
\begin{equation*}
\mz_{\wSUN} [\ct^2] = \frac{1}{4}\, \Big(  \mz_{\SUN} [\ct^2 \supset \mathcal{L}_1 ]+ \sum_{j=1}^2\, \mz_{\SUN} \big[\ct^2 \supset \mathcal{L}_{\mathcal{C}}^{(j)}\big]  + \mz_{\SUN} [\ct^2 \supset \mathcal{L}_{\mathcal{C}\mathcal{C}} ] \Big) \ .
\end{equation*}
The first contribution involves the trivial network, and thus equals \smash{$ \mz_{\SUN} [\ct^2]$}. 

To compute \smash{$\mz_{\SUN} \big[\ct^2 \supset \mathcal{L}_{\mathcal{C}}^{(j)}\big] $} we cut the torus open along its $j^\text{th}$ cycle, insert the loop operator $L_{j,\mathcal{C}}$ and then glue the two sides back together. To perform the gluing we use orthonormality of the characters to compute
\begin{equation*}
	\int_{\SUN} \dd U \ \chi_R (U)^2 = \int_{\SUN} \, \dd U \ \overline{\chi_{\overline R}(U)} \, \chi_R(U) = \begin{cases} \ 1 & \text{ if } R \cong \overline{R} \ ,  \\ \ 0 & \text{ otherwise}  \ . \end{cases}
\end{equation*}
We thus get
\begin{equation}
\label{eq:ZTwithLC}
	\mz_{\SUN} \big[\ct^2 \supset \mathcal{L}_{\mathcal{C}}^{(j)}\big]  = \sum_{R}\, \e^{- \frac{g^2}{2}\, C_2 (R)} \ \delta^{\mathrm{FS}} _R \ ,
\end{equation}
where we recall that the square of the Frobenius--Schur indicator $\delta_R ^{\mathrm{FS}}$ from \eqref{eq:deltafr} vanishes unless $R \cong \overline{R}$,  in which case it gives $1$ if $R$ is irreducible. That is, these contributions  truncate the series \eqref{eq:MRformula} to real or pseudo-real irreducible representations of $\SUN$.

To better understand the contribution from the fourth $\sC$-network in \eqref{eq:CnetworkT2}, we may slightly resolve the four-way junction at the intersection of the loops $L_{1,\mathcal{C}}$ and $L_{2,\mathcal{C}}$. The result is a non-trivial loop $L_{1,\mathcal{C}} + L_{2,\mathcal{C}}$ that winds around the combination of the two generators of $H_1 (\ct^2,\Z)$. We compute its contribution below.

\subsubsection*{CP theorem}

Let $\sC = \Z_2 \cong \left\{ 1, \mathcal{C} \right\}$ be the charge conjugation symmetry group as above, and $\sP = \Z_2 \cong \left\{ 1, \mathcal{P} \right\}$ be the parity symmetry group. In two dimensions, charge conjugation $\mathcal{C}$ and parity $\mathcal{P}$ are equivalent, due to the CP theorem 
\begin{equation*}
	\mathcal{C} \circ \mathcal{P} = 1
\end{equation*}
and the $\Z_2$ relations $\mathcal{C}^2 =1=\mathcal{P}^2$. Therefore, after gauging $\sC$ we expect to obtain the same partition function as that gotten by projecting the original theory onto its $\sP$-invariant sector. We shall now demonstrate this explicitly.

Consider again $\Sigma=\ct^2$ and insert any of the $\sC$-networks from \eqref{eq:CnetworkT2}. The trivial $\sC$-network gives back $\mz_{\SUN}[\ct^2]$. When the network \smash{$\mathcal{L}_{\mathcal{C}}^{(1)}$} is inserted, we can transport the gauge connection $A$ around $L_2$ to obtain \smash{$A^{\dagger} =-A \in \mathfrak{su}(N)$}, and likewise for \smash{$\mathcal{L}_{\mathcal{C}}^{(2)}$}. This is the same result we obtain when considering two-dimensional Yang--Mills theory on the Klein bottle $\mathbb{K}$. Therefore introducing a topological line operator $L_{\mathcal{C}}$ wrapping a non-trivial 1-cycle $L$ has the same net effect as cutting $\ct^2$ open along $L$ and closing it with a pair of cross-caps, as illustrated in Figure \ref{fig:crosscap}. 

\begin{figure}[htb]
\centering
\begin{equation*}
\begin{tikzpicture}[tqft/flow=east,baseline=-0.5ex]
			\begin{scope}[tqft/boundary upper style={draw,thick},tqft/boundary lower style={draw,thick}]
			\node[tqft/cylinder,draw] (c) at (0,0) {};
			\end{scope}
			\node[ellipse,draw,thick,colC,align=center] (L) at (0,0) {\hspace{1pt}\\ \vspace{-6.5pt}\hspace{1pt}};
			\node[anchor=north] (C1) at (L.south east) {$L_{\mathcal{C}}$};
	\end{tikzpicture} \ = \ \begin{tikzpicture}[tqft/flow=east,baseline=-0.5ex]
			\begin{scope}[tqft/boundary upper style={draw,thick},tqft/boundary lower style={draw,thick}]
			\node[tqft/cylinder,draw] (c) at (-1.5,0) {};
			\node[ellipse,draw,thick,colC,align=center] (L) at (c.outgoing boundary 1) {\hspace{1pt}\\ \vspace{-6.5pt}\hspace{1pt}};
			\draw[thick,colC] (L.north west) -- (L.south east);
			\draw[thick,colC] (L.north east) -- (L.south west);
			\node[tqft/cylinder,draw] (d) at (1.5,0) {};
			\node[ellipse,draw,thick,colC,align=center] (M) at (d.incoming boundary 1) {\hspace{1pt}\\ \vspace{-6.5pt}\hspace{1pt}};
			\draw[thick,colC] (M.north west) -- (M.south east);
			\draw[thick,colC] (M.north east) -- (M.south west);
			
			\node[anchor=north] (cc) at (0,-0.5) {\footnotesize cross-caps};
			\draw[thin] (-0.15,-0.5) -- (-0.45,-0.2);
			\draw[thin] (0.15,-0.5) -- (0.45,-0.2);
			\end{scope}
	\end{tikzpicture}
\end{equation*}
\vspace{-5mm}
\caption{Inserting a topological line operator $L_{\mathcal{C}}$ wrapping a 1-cycle $L \subset \ct^2$ is equivalent to cutting $\ct^2$ open along $L$ and then closing it with a pair of cross-caps. The black boundaries in the picture are glued together.}
\label{fig:crosscap}
\end{figure}

By a simple generalization of the formula for $\mz_{\SUN}[\mathbb{RP}^2]$ given in \cite{Wang:2020jgh} to non-orientable surfaces with more than one cross-cap, we indeed find that $\mz_{\SUN} [\mathbb{K}]$ equals \eqref{eq:ZTwithLC}, providing a consistency check for this derivation. Finally, the contribution from the fourth network in \eqref{eq:CnetworkT2} equals that of the trivial network, because transporting $A$ along the sum of the two non-trivial cycles with holonomy $-1$ around each yields a trivial overall holonomy. 

By summing over the insertions of $\sC$-networks we thus obtain the equality 
\begin{equation*}
	\mz_{\wSUN} [\ct^2] = \tfrac{1}{2}\, \mz_{\SUN} [\ct^2] +  \tfrac{1}{2}\, \mz_{\SUN} [\mathbb{K}] \ .
\end{equation*}
The right-hand side is the sum of contributions from the orientable and non-orientable closed surfaces of Euler characteristic $\chi=0$, whence
\begin{equation*}
	 \mz_{\wSUN}[\ct^2] = \mz_{\SUN \rtimes \sP} [\ct^2] \ .
\end{equation*}

\section{Spontaneous breaking of charge conjugation symmetry}
\label{sec:chargeconj}

In this section we discuss the spontaneous breaking of the discrete 0-form symmetry ${\sC}^{\scriptscriptstyle (0)} = \Z_2^{\scriptscriptstyle (0)}$ in the orbifolds of $\SUN$ Yang--Mills theory by discrete 1-form symmetries, generalizing earlier considerations of~\cite{Aminov:2019hwg}.

\subsection{Spontaneous symmetry breaking in U(1) Yang--Mills theory}

As a warm-up and for completeness, we begin by stating the result of spontaneous breaking of charge conjugation symmetry in $\Uu$ Yang--Mills theory. This has been known previously, see in particular \cite[Section~2.1]{Komargodski:2017dmc}.

\begin{prop}
In pure $\Uu$ Yang--Mills theory with $\theta=\pi$, charge conjugation symmetry is spontaneously broken.
\end{prop}

One easy way to prove this is using the partition function 
\begin{align*}
	\mz^{\theta}_{\Uu} & = \sum_{q \in \Z }\, \e^{- \frac{g^2}{2}\, q^2 - \ii\, \theta\, q} = \sqrt{\frac{g^2}{2\pi}} \  \sum_{n \in \Z}\, \e^{- \frac{ 2\pi^2}{g^2}\, \left( n - \frac{\theta}{2\pi}\right)^2} \ , 
\end{align*}
where the second equality follows from Poisson resummation and defines the instanton expansion (after including moduli of flat connections when $\chi\leq0$). At $\theta=0$, charge conjugation $q\longmapsto-q$ sends $n\longmapsto-n$ and there is a unique symmetric vacuum state $n=0$. At $\theta=\pi$, there are two degenerate vacua $n=0$ and $n=1$. Acting with charge conjugation $q\longmapsto-q$, which now sends $n\longmapsto-n+1$, the two vacuum states are exchanged and the symmetry is spontaneously broken.

\subsubsection*{Wilson loops in U(1) Yang--Mills theory}

Let 
\begin{equation*}
	W_r (L)= \exp\, r\, \oint_L \, A
\end{equation*}
be a Wilson line in $\Uu$ Yang--Mills theory, labelled by the irreducible $\Uu$ representation $r$, classified by the integer which we denote with the same symbol $r \in \Z$, and supported on a closed oriented curve $L\subset\Sigma$. Acting with a generic element of the $\Uu^{\scriptscriptstyle (1)}$ symmetry, parametrized by $\vartheta\in[0,2\pi)$, the Wilson line acquires a phase 
\begin{equation*}
	W_r(L) \ \longmapsto \ \e^{\, \ii\, r\, \vartheta } \ W_r(L) \ .
\end{equation*}

On the other hand, Wilson lines are also charged under charge conjugation $\sC^{\scriptscriptstyle (0)}$:
\begin{equation*}
	\mathcal{C} : \ W_r(L) \ \longmapsto \ W_{-r}(L) \ .
\end{equation*}
The pairs of charge conjugate Wilson lines $W_{\pm r}(L)$ have the same 1-form symmetry charge $(-1)^r$ under the subgroup $\Z_2^{\scriptscriptstyle (1)} \subset \Uu^{\scriptscriptstyle (1)}$ at $\vartheta=\pi$.

\subsection{Spontaneous symmetry breaking in \texorpdfstring{$\SUN/\Z_k$}{SU(N)/Zk} Yang--Mills theory}
\label{subsec:CbreakingSUNZk}

It was shown in \cite{Aminov:2019hwg} that the expression \eqref{ZorbiDelta} at $k=2$ and $\kappa=1$ implies two degenerate vacua that are exchanged under charge conjugation. Therefore charge conjugation symmetry is spontaneously broken in the $\SUN/\Z_2$ theory. It is straightforward to extend the argument to the $\SUN/\Z_k$ theory when $k$ is any power of $2$ and $\kappa = \frac{k}{2}$.

\begin{prop}
Let $k \ge 2$ be a power of $2$ and $N$ a multiple of $k$. In pure $\SUN/\Z_k$ Yang--Mills theory with $\theta_{\kappa}$-angle given by $\kappa=\frac{k}{2}$, charge conjugation symmetry is spontaneously broken.
\end{prop}

\subsubsection*{Degenerate vacua}

The action of charge conjugation on the representation ring $\mathfrak{R}(\SUN)$ has been reviewed in Subsection~\ref{sec:repreview}. In particular, the operation $\mathcal{C}$ maps the single-column tableau of $\frac{k}{2}$ boxes into the single-column tableau of $N-\frac{k}{2}$ boxes, both of which have $\frac{k}{2}$ mod$~k$ boxes. For example, if $N=8$ and $k=4$, the charge conjugation map is depicted by
\begin{equation*}
	\begin{ytableau} \ \\ \ \\ *(colC) \ \\ *(colC) \ \\ *(colC) \ \\ *(colC) \ \\ *(colC) \ \\ *(colC) \ \end{ytableau}
\end{equation*}

By direct computation of the Casimir term, the latter two representations are degenerate vacua if $\kappa=\frac{k}{2}$, thus charge conjugation is spontaneously broken in all such $\SUN/\Z_k$ theories.

\subsubsection*{Wilson loops in ${\boldsymbol \SUN}$ Yang--Mills theory}

Let us take a step back and consider $\SUN$ Yang--Mills theory, prior to orbifolding. Similarly to the abelian case, there exists a subgroup $\Z_2^{\scriptscriptstyle (1)} \subset \Z_N^{\scriptscriptstyle (1)}$ of the 1-form symmetry under which pairs of Wilson lines $W_R(L)$ and $W_{\overline R}(L)$ that are exchanged under $\mathcal{C}$ have equal charge $(-1)^{|R|}$. This is a direct consequence of the relation
\begin{equation*}
	\lvert \overline{R} \rvert = R_1\,N - \lvert R \rvert \ .
\end{equation*}

Taking the orbifold by this $\Z_2^{\scriptscriptstyle (1)}$ with discrete $\theta$-angle $\theta_\kappa$, the Wilson lines with 1-form charge $(-1)^\kappa$ descend to Wilson lines of the $\SUN/\Z_2$ theory, while the others are projected out. The interesting case is $\kappa=1$, in which the trivial representation of $\SUN$ drops out. Then, as we have seen, there are two degenerate vacua exchanged under charge conjugation.

\subsubsection*{Spontaneous symmetry breaking for orthogonal gauge algebras}

One may attempt a similar analysis for other gauge groups $G$. We consider two-dimensional Yang--Mills theory with simple gauge group that integrates the gauge algebra $\mathfrak{g} = \mathfrak{so}(N)$ for $N\in 2\N$. The discrete $\theta$-angles are described in Appendix \ref{sec:othergauge}. 

We perform an explicit analysis of the orbifolds and their vacua in Appendix \ref{app:breaksoN}, akin to the $\mathfrak{su}(N)$ case. The upshot is that charge conjugation symmetry is spontaneously broken in $\mathrm{SO}(6)/\Z_2$ Yang--Mills theory, while the theory possesses a unique $\mathcal{C}$-invariant vacuum for every $N >8$. 

\section{Non-invertible symmetry orbifolds}
\label{sec:NIhigher}

The 1-form symmetry originating from the centre $\Z_N \subset \SUN$ is only the invertible part of a much larger set of non-invertible 1-form symmetries of two-dimensional Yang--Mills theory \cite{Nguyen:2021naa}. 

\begin{defin}
A symmetry is \emph{non-invertible} if it is generated by a set of topological defect operators that do not form a group under fusion. 
\end{defin}

In two spacetime dimensions, finite non-invertible 0-form symmetries are, by definition, generated by topological line operators that form a (unitary) fusion category \cite{Chang:2018iay}. For example, Definition~\ref{def:network} immediately extends to non-invertible symmetries by relaxing the group condition: The finite group $\sA$ is replaced with its representation category $\mathfrak{Rep} (\sA)$, and the symmetry parameter $\alpha$ is taken to be an object of $\mathfrak{Rep} (\sA)$. These generalized symmetries have been extensively studied, see e.g.~\cite{Chang:2018iay,Fuchs:2007tx,Thorngren:2019iar,Thorngren:2021yso,Komargodski:2020mxz,Kikuchi:2021qxz,Kikuchi:2022gfi,Huang:2021zvu,Inamura:2022lun,Chang:2022hud,Lin:2022dhv,Nagoya:2023zky,Bhardwaj:2023idu,Damia:2024xju} for an incomplete list of works in continuum quantum field theory. A rigorous approach to gauging non-invertible symmetries was recently given in \cite{Decoppet:2022dnz,Carqueville:2023qrk,Choi:2023vgk,Perez-Lona:2023djo,Diatlyk:2023fwf}. 

On the contrary, non-invertible 1-form symmetries are more elusive. 
In this section we undertake a systematic study of the gauging of the non-invertible symmetry of two-dimensional Yang--Mills theory and the ensuing effect on Wilson loops. Specifically, we
\begin{itemize}
\item introduce a generalized $\theta$-term in the theory after gauging the non-invertible symmetry; and
\item show how the spontaneous breaking of charge conjugation symmetry descends to the theory after gauging the non-invertible symmetry.
\end{itemize}
As we discuss below, these apply equally well after gauging only a subset of the non-invertible 1-form symmetry.

\subsection{The non-invertible 1-form symmetry}

It was shown in \cite{Nguyen:2021naa} that pure $\SUN$ Yang--Mills theory in two dimensions admits a vast class of zero-dimensional topological symmetry operators, defined as follows. Let $\SUN^{\#}$ denote the set of conjugacy classes in $\SUN$. Take a point $\wp \in \Sigma$ and a conjugacy class $[U_0] \in \SUN^{\#}$, for an arbitrary element $U_0 \in \SUN$. Delete $\wp$ and require that the gauge connection $A$ has holonomy in $[U_0]$ around the puncture. The path integral over such gauge fields gives rise to a zero-dimensional gauge-invariant disorder operator $V_{\wp} [U_0]$, analogous to a Gukov--Witten operator. It is topological thanks to the invariance of two-dimensional Yang--Mills theory under area-preserving diffeomorphisms, and therefore we henceforth omit the subscript $\wp$ indicating its location. It is gauge-invariant by construction, as only the conjugacy class of the holonomy is fixed rather than the holonomy itself.

Let us briefly review how these local topological operators act on the Wilson lines. Consider for simplicity a disk $\mathbb{D}$ of area $\frac{1}{2}$ with a Wilson loop in the representation $R$ running along the boundary with opposite orientation, as in Subsection \ref{sec:reading}. We recall that the path integral with the Wilson loop inserted is denoted $\langle W_R\rangle$ and evaluates to \eqref{eq:WRvevdisk}. If we insert a disorder operator $V [U_0]$ at an arbitrary point in the interior of the disk, a standard computation shows that
\begin{equation}
\label{eq:noninv1form}
	\langle V [U_0] \, W_R \rangle = \chi_R (U_0) \ \e^{- \frac{g^2}{4}\, C_2 (R)} \ ,
\end{equation}
which in general differs from the correlator without the insertion of the topological symmetry operators. The two invariant cases are $U_0= \id_N$ (trivial topological operator) and $R= \emptyset $ (trivial Wilson line). 

The topological symmetry operators thus constructed are generically non-invertible \cite{Nguyen:2021naa}. As a concrete example, if $N \in 4\N$ then one of the operators is associated to the conjugacy class $[U_0]$ with representative 
\begin{equation*}
	U_0 = \text{diag} (\underbrace{ 1, \dots, 1}_{N/2} , \underbrace{ -1, \dots, -1}_{N/2}) \ \in \ \SUN \ .
\end{equation*}
For this choice, the correlator \eqref{eq:noninv1form} vanishes for a large set of representations, including the fundamental representation. Therefore the topological symmetry operator annihilates the corresponding Wilson loops, and hence does not admit an inverse. 

Demanding invertibility and imposing the special unitary constraint on $U_0$, we are left with operators associated to the conjugacy classes of elements $U_0 \in \sZ(\SUN)= \Z_N$, thus recovering the usual 1-form centre symmetry $\sZ^{\scriptscriptstyle (1)}$ as the invertible subset of this larger set of symmetries. The standard group law agrees with the multiplication induced by the fusion product of topological symmetry operators.

\subsection{Gauging the non-invertible 1-form symmetry}
\label{sec:orbiNI}

We have seen in \eqref{eq:noninv1form} that inserting a topological pointlike operator $V [U_0]$ in the correlation functions of Wilson loops projects out all the representations $R$ for which $\chi_R (U_0) =0$. Gauging the full non-invertible 1-form symmetry is enforced by integrating over insertions $\left\{ V [U_0] \ | \ [U_0] \in \SUN^{\#} \right\}$ in the partition function on $\Sigma$. The integral is a Molien--Weyl projector which has the net effect of projecting out all representations except the trivial one. 

While the outcome is in agreement with general arguments, it is not possible to derive it in the path integral formalism. The obstruction is the absence of a definition of the 2-form gauge field $B$ when the symmetry is non-invertible.

\subsubsection*{Gauging subsets of the non-invertible 1-form symmetry}

Exactly as it is possible to gauge any subgroup $\sB \subseteq \sZ$, it should be possible to gauge certain collections within the non-invertible 1-form symmetry; the analogous statement for gauging non-invertible 0-form symmetries was recently derived in \cite{Perez-Lona:2023djo}. Indeed, it is possible to gauge the symmetry classified by conjugacy classes $H^{\#} \subseteq \SUN^{\#}$ for every normal subgroup $H \subseteq \SUN$. This is done explicitly by integrating \eqref{eq:noninv1form} over $[U_0] \in H^{\#}$. In this case the integral is again a Molien--Weyl projector, but we ought to take into account the branching rule under $\SUN\longrightarrow H$ to decompose the $\SUN$-representation $R$ into $H$-representations. Every $R$ whose branching rule yields the trivial representation of $H$ (among others) will survive the gauging of the subset of non-invertible symmetries classified by $H^{\#}$.

\subsubsection*{Generalized $\boldsymbol\theta$-term for the non-invertible 1-form symmetry}

There are various ways to gauge the invertible 1-form symmetry, landing on different theories labelled by the discrete parameter $\kappa$. We thus propose a generalization of this concept to non-invertible symmetries.

When gauging the standard 1-form symmetry $\sZ^{\scriptscriptstyle (1)}$ in Section \ref{sec:orbi}, we proceeded by introducing a discrete parameter $\kappa$ and weighing the action functional accordingly. We sum over insertions of local topological operators labelled by $\e^{\,2 \pi\,\ii\, \beta /N} \in \sZ$, for $\beta \in \left\{0 ,1, \dots, N-1 \right\}$, with weight \cite{Sharpe:2014tca}
\begin{equation*}
	\chi_{\kappa} ^{\sZ} (\e^{\,2 \pi\,\ii\, \beta/N} ) = \e^{\,2 \pi\,\ii\, \kappa\, \beta/N}
\end{equation*}
given by the pairing of $\e^{\,2 \pi\,\ii\, \beta /N} \in \sZ$ with a character $\chi_{\kappa} ^{\sZ}$ of $\sZ$ labelled by $\kappa \in \left\{ 0, 1, \dots, N-1\right\}$. That is, a $\theta$-term corresponds to turning on a pairing between the local topological operators that generate the symmetry $\sZ$ and the character lattice of $\sZ$.

In this way, we are led to propose an extension of the notion of $\theta$-term to non-invertible symmetries. For the local operators $V [U_0]$ classified by $\SUN^{\#}$, we ought to turn on a pairing between $\SUN^{\#}$ and the character lattice $\mathfrak{C}_{\Z}(\SUN)$ of $\SUN$. The character lattice is the image of the ring homomorphism $\mathfrak{R} (\SUN) \longrightarrow \mathfrak{C} (\SUN)$ with \smash{$R \longmapsto \chi_{R}$}, under which the character is fixed by $R \in \mathfrak{R} (\SUN)$. We conclude 

\begin{prop}
For gauging the non-invertible 1-form symmetry, the generalized discrete parameter $\kappa$ is an isomorphism class of irreducible representations of $\SUN$. 
\end{prop}

\subsubsection*{The pairing}

The local topological operators $V[U_0]$ are classified by $\SUN^{\#}$. The Wilson loops $W_R$, charged under the non-invertible symmetry, are classified by the classes $R \in \mathfrak{R}(\SUN)$ in the Grothendieck ring $\mathfrak{R}(\SUN)$ of the category of $\SUN$-representations. Taking correlation functions $\langle V[U_0]\, W_R \rangle$ defines a pairing $\langle \, \cdot \,,\, \cdot \, \rangle$ between $\SUN^{\#}$ and $\mathfrak{R}(\SUN)$:
\begin{equation*}
	\langle U_0 , R \rangle := \langle V[U_0]\, W_R \rangle \ .
\end{equation*}

This is valid for every Riemann surface $\Sigma$, even though we only explicitate the expressions for the disk for clarity. In this case, the pairing is given by \eqref{eq:noninv1form} and it differs from the canonical pairing by the exponential of the Casimir invariant. We may also introduce a pairing $\langle \, \cdot \, \vert \, \cdot \, \rangle$ between generators of $\mathfrak{R}(\SUN)$ according to
\begin{equation*}
	\langle R_1 \vert R_2 \rangle := \int_{\SUN} \, \dd U_0 \ \langle R_1 , U_0 \rangle \, \langle U_0 , R_2 \rangle = \int_{\SUN} \dd U_0 \ \langle V[U_0]\, W_{R_1} \rangle \, \langle V[U_0]\, W_{R_2} \rangle \ ,
\end{equation*}
where we used bi-invariance of the Haar measure to extend integration over conjugacy classes $\SUN^{\#}$ to $\SUN$.
When $\Sigma$ is topologically the disk $\mathbb{D}$, it is given by 
\begin{equation}
\label{pairingR1R2}
	\langle R_1 \vert R_2 \rangle = \e^{- \frac{g^2}{2}\, C_2 (R_2)} \ \delta_{\overline{R_1}\,,\,R_2} \ ,
\end{equation}
where we used orthonormality of the basis of characters. Again, it differs from the canonical pairing through the dependence on the Casimir invariant.

\subsubsection*{Gauging with generalized $\boldsymbol\theta$-term}

We have seen that gauging the non-invertible 1-form symmetry with generalized discrete $\theta$-angle labelled by $\kappa$ dictates to integrate over insertions of operators $V[U_0]$ weighted by the character $\chi_{\kappa} (U_0)$. Taking again the disk topology for concreteness, this yields 
\begin{align*}
\int_{\SUN}\,\dd U_0 \ \chi_\kappa(U_0) \, \langle U_0,R\rangle &=	\int_{\SUN}\,\dd U_0 \ \chi_{\kappa} (U_0) \, \chi_R (U_0) \ \e^{- \frac{g^2}{4}\, C_2 (R)} \\[4pt]
&= \e^{- \frac{g^2}{4}\, C_2 (R)} \ \delta_{\overline{\kappa}\,,\,R} = \e^{\,\frac{g^2}{4}\, C_2 (\kappa)} \ \langle \kappa \vert R \rangle \ ,
\end{align*}
where for the first equality we used \eqref{eq:noninv1form}, for the second equality we used orthonormality of the characters, and for the last equality we used the pairing \eqref{pairingR1R2}. We thus arrive at a projection onto $R \cong \overline{\kappa}$. Had we defined the Wilson loop with orientation $\partial \mathbb{D}$, instead of $-\partial \mathbb{D}$, we would get the projection onto $R \cong \kappa$. From the considerations so far we arrive at

\begin{prop}\label{prop:gaugeNIkappa}
Gauging the non-invertible 1-form symmetry with generalized $\theta$-angle labelled by $\kappa$ corresponds to acting with the projector $\langle \kappa \vert$ on the Grothendieck ring $\mathfrak{R}(\SUN)$.
\end{prop}

As a consequence of the non-invertible 1-form symmetry, two-dimensional Yang--Mills theory `decomposes' into infinitely many invertible field theories, labelled by the irreducible representations. Gauging with generalized $\theta$-angle selects one such invertible theory. From this viewpoint, Proposition \ref{prop:gaugeNIkappa} extends the decomposition results in \cite{Sharpe:2019ddn} from the invertible to the full 1-form symmetry.

\subsubsection*{Another perspective on the generalized ${\boldsymbol \theta}$-term}

We have introduced a representation theoretic definition of $\theta$-terms for the non-invertible symmetry. We now rephrase these findings in the formalism of Kapustin--Seiberg \cite{Kapustin:2014gua}, see also \cite{Bhardwaj:2022kot,Bhardwaj:2022maz}.

Let us start again with $\SUN$ Yang--Mills theory on $\Sigma$ and its 1-form symmetry $\sB^{\scriptscriptstyle (1)} \subseteq \Z_N$. Following \cite{Kapustin:2014gua,Bhardwaj:2022kot}, another presentation of the discrete $\theta$-angle in the orbifold corresponds to stacking the $\SUN$ theory (before gauging) with a two-dimensional topological field theory, whose partition function is equal to
\begin{equation*}
	\exp \left( \kappa\, \int_{\Sigma}\, B \right) \ .
\end{equation*}
Different choices of parameter $\kappa$ correspond to different couplings for the background gauge field $B$. Gauging the diagonal 1-form symmetry of the juxtaposed system produces the orbifold $\SUN/\Z_k$ Yang--Mills theory with discrete $\theta$-angle $\theta_{\kappa}$.

The difficulty in extending this approach is the lack of notion of background gauge fields for non-invertible symmetries in general. Our previous analysis yields a solution in the particular case of the non-invertible 1-form symmetry of two-dimensional Yang--Mills theory. Conjugacy classes $[U_0] \in \SUN^{\#}$ play the role of background gauge fields for this non-invertible symmetry. The possible ways of defining a partition function for these background fields are in one-to-one correspondence with characters of irreducible representations of $\SUN$.
 
Consider the set $\mathfrak{T}$ consisting of two-dimensional topological field theories possessing this non-invertible 1-form symmetry and differing by the choice of coupling. $\mathfrak{T}$ is endowed with the fusion product $\otimes$ (fusion of two-dimensional spacetimes of theories) and the direct sum $\oplus$ (disjoint union of theories). From the above discussion, we infer that $(\mathfrak{T}, \otimes, \oplus)$ acquires the structure of a ring, identified with the character lattice $\mathfrak{C}_{\Z}(\SUN)$, and hence with the Grothendieck ring~$\mathfrak{R} (\SUN)$.

\subsubsection*{Wilson loops in non-invertible symmetry orbifolds}

The study of Wilson loops in orbifolds by the non-invertible 1-form symmetry generalizes the program initiated in \cite{Aharony:2013hda}. By collecting the expectation values of all the Wilson loops, it is in principle possible to reconstruct which part of the non-invertible symmetry has been gauged, and with which generalized $\theta$-angle. In particular, if $\langle W_{\emptyset}\rangle \ne 0 $ we already know that the gauging has been performed at $\kappa= \emptyset$ (trivial representation). 

Suppose, for example, that we start computing $\langle W_R \rangle$ by successively increasing $\lvert R \rvert$ and find
\begin{equation*}
	\begin{cases} \ \langle W_R \rangle \ne 0 & \text{ if } R \in \left\{ \emptyset \,,\, \Box \,,\, \Box\!\Box  \,,\, \Box\!\Box\!\Box \,,\, \dots \right\} \ , \\[4pt] \ \langle W_R \rangle = 0 & \text{ otherwise} \ .  \end{cases}
\end{equation*}
This is compatible with the branching rules for $\SUN \longrightarrow \mathrm{SU}(N-1)$. Note that the surviving Young diagrams are also compatible with an $\mathrm{SU}(2)$ or $\mathrm{SO}(3)=\mathrm{PSU}(2)$ gauge theory, among others. Nevertheless, the values of the non-vanishing $\langle W_R \rangle$ will make clear that the theory is a non-invertible gauging of an $\SUN$ theory, because $\dim R$ and $C_2 (R)$ are evaluated for the $\SUN$-representations associated to the single-row Young diagrams. For example, when $N>2$
\begin{equation*}
	\langle W_{\Box\!\Box\cdots\Box} \rangle \ \ne \ \langle W_{\Box\!\Box\cdots\Box} \rangle \big\vert_{G=\mathrm{SU}(2)} \ .
\end{equation*}

In conclusion, from the list of the expectation values of Wilson loops, one can reconstruct the global form of the gauge group even after the projection due to gauging a non-invertible symmetry.

\subsection{Spontaneous symmetry breaking in the non-invertible symmetry orbifold}
\label{sec:SSBNI}

We now assume $N \in 2\N$. The spontaneous breaking of charge conjugation symmetry generalizes to orbifolds of $\SUN$ Yang--Mills theory by a subset of the non-invertible 1-form symmetry. For the invertible part, the subgroup $\Z_2 \subset \Z_N$ for $N \in 2\N$ consists of the elements \smash{$\{ \e^{\, 2 \pi\,\ii\, \beta/N}\}_{\beta =0, N/2} \cong \left\{ 1, -1 \right\}$}, which are mapped into themselves by $\mathcal{C}$. Gauging this $\Z_2 ^{\scriptscriptstyle (1)}$ leaves behind elements that are not fixed under the action of $\mathcal{C}$. This argument applies to the gauging of every $\Z_k ^{\scriptscriptstyle (1)}$ with $k$ a power of 2 that divides $N$, where some of the elements that survive the gauging are not fixed under $\mathcal{C}$ and therefore are able to distinguish a representation from its conjugate.

The generalization of this idea to the non-invertible 1-form symmetry is to gauge the $\mathcal{C}$-invariant collection of all the symmetries. Since the non-invertible topological operators are labelled by conjugacy classes in $\SUN^{\#}$, the $\mathcal{C}$-invariant subset consists of those operators labelled by conjugacy classes in $\SON^{\#}$. The centre $\Z_2 = \sZ(\SON)$ is precisely the invertible 1-form symmetry group $\Z_2 ^{\scriptscriptstyle (1)}$ that we have gauged to find spontaneous breaking of charge conjugation symmetry in Subsection~\ref{subsec:CbreakingSUNZk}.

A generalized $\theta$-term that extends $\kappa = 1 \!\!\mod 2$ to the non-invertible symmetry is achieved by choosing $\kappa$ to be the vector representation of $\SON$. By their branching rules, both $\Box$ and $\overline{\Box}$ survive the projection, thus we conclude that charge conjugation symmetry is spontaneously broken in the orbifold.

\begin{prop}
Consider two-dimensional Yang--Mills theory with gauge group $\SUN$, $N\in 2\N$. In the orbifold theory obtained by gauging the non-invertible 1-form symmetry $\SON^{\#}$ with a generalized $\theta$-angle given by the character of the vector representation of $\SON$, charge conjugation symmetry is spontaneously broken.
\end{prop}

\section{Anomalies}
\label{sec:anomalies}

In this section we begin by giving an exhaustive analysis of mixed anomalies in $\Uu$ and $\UN$ Yang--Mills theories. Then we discuss mixed anomalies present in the orbifolds of $\SUN$ Yang--Mills theory involving the $(-1)$-form symmetry. The main result is Proposition~\ref{AnomCL}, which we prove directly in the two-dimensional gauge theory. 

\subsection{Charge conjugation and lower form gauge transformations}

Let $\mathsf{\Gamma}^{\scriptscriptstyle (-1)} $ be a $(-1)$-form symmetry group, which we will assume to be either $\Uu$ or a cyclic subgroup $\Z_k\subset\Uu$. A $\mathsf{\Gamma}^{\scriptscriptstyle (-1)} $ transformation in two dimensions is generated by a Wilson surface labelled by $\eta \in \mathsf{\Gamma}$. The insertion of these operators in the partition function and correlation functions shifts the $\theta$-parameter as $ \frac{\theta}{2\pi} \longmapsto \frac{\theta}{2\pi} + \eta$. We have also observed in Subsection \ref{sec:dualgauging} (and will find again below) that, after turning on a 0-form gauge field, it shifts $\frac{\theta}{2\pi}$ by the corresponding background field.

In the language of the $(-1)$-form symmetry, the periodicity of $\theta$ becomes invariance under a $(-1)$-form gauge transformation, when a background for the $(-1)$-form symmetry is activated. Conversely, when the theory fails to be invariant under a shift of $\theta$ by its periodicity after some background gauge field is turned on, we interpret it as a breaking of the $(-1)$-form symmetry. These ideas will be explicitly realized momentarily.

Having set the stage to treat $(-1)$-form symmetries, we can now state the obvious

\begin{prop}\label{thm1}
Consider two-dimensional pure Yang--Mills theory with multiply-connected gauge group $G$, i.e. $\pi_1 (G) \ne 1 $, which admits outer automorphisms. Let $\mathcal{C}$ be the non-trivial element in $\sC\subseteq\mathsf{Out}(G)$, and $\mathcal{C}^{\prime}$ the operator $\mathcal{C}$ followed by a $(-1)$-form gauge transformation. The theory is invariant under $\mathcal{C}$ at $\theta =0$ and under $\mathcal{C}^{\prime}$ at $\theta = \pi$.
\end{prop}

The only novelty of this statement is to pinpoint the role of the $(-1)$-form symmetry, which however has important consequences. Indeed, the mixed anomalies involving charge conjugation in the literature typically rely on the non-invariance under shifts of $\theta$. We claim that the anomalous nature of $\mathcal{C}^{\prime}$ originates in an anomalous $(-1)$-form symmetry.

\subsection{Anomalies in U(\texorpdfstring{$N$}{N}) Yang--Mills theory}
\label{sec:U1UN}

\subsubsection*{Charge conjugation anomaly in U(1) Yang--Mills theory}

We consider two-dimensional $\Uu$ Yang--Mills theory and state here the result of \cite[Section~2]{Komargodski:2017dmc}. 

\begin{prop}\label{pr:U1CZNanom}
	Two-dimensional pure $\Uu$ Yang--Mills theory at $\theta = \pi$ suffers from a mixed anomaly between $\mathcal{C}^{\prime}$ and the 1-form symmetry.
\end{prop}

The proof is straightforward. The $\theta$-term 
\begin{equation*}
	\frac{1}{2}\, \int_{\Sigma}\, F 
\end{equation*}
is not invariant under background $\Uu^{\scriptscriptstyle (1)}$-gauge transformations, as these shift the action functional by $\pi\,\ii\,n$ for $n\in \Z$. This is fixed by the usual replacement $F \longmapsto F-B$, but now the counterterm is not invariant under $\mathcal{C}^{\prime}$. In contrast to the flipping of sign of $F$, the non-invariance of $B$ cannot be fixed by a shift $\theta \longmapsto \theta + 2 \pi $, whereby
\begin{equation*}
	\frac{\theta}{2\pi}\,\int_{\Sigma}\, B  \ \longmapsto \  \frac{\theta}{2\pi}\, \int_{\Sigma}\, B + \int_{\Sigma}\, B \ , 
\end{equation*}
because the last summand is not integrally quantized. In conclusion, one cannot preserve both charge conjugation and $1$-form symmetry.

\subsubsection*{Anomaly in the space of couplings in $\boldsymbol\Uu$ Yang--Mills theory}

It was shown in \cite[Section~4]{Cordova:2019jnf} that pure $\Uu$ Yang--Mills theory suffers from an anomaly in the space of coupling constants. This means the impossibility of turning on a background for $\Uu^{\scriptscriptstyle (1)}$ while at the same time promoting $\theta$ to a position-dependent coupling, i.e. an axion field, and preserving background gauge invariance.

\subsubsection*{Mixed lower-higher form  anomaly in $\boldsymbol\Uu$ Yang--Mills theory}

We now revisit the two anomalies just presented. The difference in our approach with respect to~\cite{Cordova:2019jnf} is that we keep $\theta$ fixed and turn on a background gauge field $a$ for $\Uu^{\scriptscriptstyle (-1)}$.

\begin{prop}\label{pr:U1PManom}
	Two-dimensional pure $\Uu$ Yang--Mills theory suffers from a mixed anomaly between the $\Uu^{\scriptscriptstyle (1)}$ 1-form symmetry and the $\Uu^{\scriptscriptstyle (-1)}$ $(-1)$-form symmetry.
\end{prop}

Consider the $\theta$-term with a 2-form background $\Uu^{\scriptscriptstyle (1)}$ gauge field $B$ turned on. Introducing the background $\Uu^{\scriptscriptstyle (-1)}$-gauge field $a$, the non-invariant part of the action functional is 
\begin{equation*}
	\int_{\Sigma}\, (F - B)\, \left( \frac{\theta}{2\pi} + \ii\, a \right) \ . 
\end{equation*}
We now leave $\theta$ unchanged but act with a gauge transformation of $a$. There is a shift by $\int_\Sigma\, B$ that is not properly quantized, thus spoiling the invariance.

We conclude that, even at $\theta= \pi$, the mixed anomaly involving $\mathcal{C}^{\prime}$ is a consequence of the impossibility of preserving gauge invariance under both $\Uu^{\scriptscriptstyle (1)}$ and $\Uu^{\scriptscriptstyle (-1)}$. This is a corollary of Proposition \ref{pr:U1PManom} combined with Proposition \ref{thm1}. We stress that turning on the $\Uu^{\scriptscriptstyle (-1)}$ gauge field $a$ breaks the symmetry under naive charge conjugation $\mathcal{C}$ explicitly, but in exactly the same way as any $\theta \ne 0$ does. The crux of the matter is whether or not $\mathcal{C}$ maps the theory to a $\Uu^{\scriptscriptstyle (-1)}$ gauge equivalent theory.

Let us emphasize that this result is analogous to the mixed electric-magnetic anomaly of Maxwell theory in four dimensions. The electric 1-form symmetry and the magnetic $(d-3)$-form symmetry generally both participate in the anomaly in $d$ dimensions. For $d=4$ both are 1-form symmetries, while for $d=2$ the magnetic symmetry is a $(-1)$-form symmetry.

\subsubsection*{${\boldsymbol \UN}$ Yang--Mills theory}

It is straightforward to extend the argument of $\Uu$ Yang--Mills theory to higher rank gauge groups $\UN$ for $N>1$. Turning on the background gauge field $B$ for the 1-form symmetry $\Uu^{\scriptscriptstyle (1)}$, the action functional becomes 
\begin{equation*}
	\frac{1}{2g^2}\, \int_{\Sigma}\, \tr \big[( F - B \otimes \id_N ) \ast ( F - B \otimes \id_N )\big]  + \frac{\theta}{2\pi}\, \int_{\Sigma}\, \tr  \left( F - B \otimes \id_N \right) \ ,
\end{equation*}
which fails to be invariant under shifts $\frac{\theta}{2\pi} \longmapsto \frac{\theta}{2\pi} + n$ for $n\in\Z$. However, in contrast to the $N=1$ case, if
\begin{equation*}
	N\, \int_\Sigma\, B \ \in \ 2 \pi\, \ii\, \Z \ , 
\end{equation*}
the $2\pi$-periodicity of $\theta$ is preserved. In other words, in the $\UN$ theory for $N>1$, a subgroup $\Z_N^{\scriptscriptstyle (1)} \subset \Uu^{\scriptscriptstyle (1)}$ is anomaly-free. This non-anomalous $\Z_N^{\scriptscriptstyle (1)}$ is identified with the 1-form symmetry of $\SUN \subset \UN$. 

\subsection{Mixed lower-higher form anomaly}

We will now show

\begin{prop}\label{AnomCL}
Let $k \ge 2$ and $N$ a multiple of $k$, with $k^2 \notin N \Z$. Pure $\SUN/\Z_k$ Yang--Mills theory on $\Sigma$ suffers from a mixed anomaly between the $(-1)$-form symmetry $\Z_k^{\scriptscriptstyle (-1)}$ and the 1-form symmetry $\Z_{N/k} ^{\scriptscriptstyle (1)}$.
\end{prop}

In the $\SUN$ and $\PSUN$ theories, there cannot be such an anomaly, because the $(-1)$-form symmetry in the former and the 1-form symmetry in the latter are trivial, hence trivially anomaly-free.

Let us quickly recall the setup of Section \ref{sec:orbi}. The $\SUN/\Z_k$ curvature $F^{\,\textrm{e}}$ decomposes into the $\SUN$ curvature $F$ and the 2-form gauge field $B$.
The theory has the electric 1-form symmetry \smash{$\check{\sB}^{\scriptscriptstyle (1)} = \Z_{N/k} ^{\scriptscriptstyle (1)}$}, which we denote with a check to distinguish it from $\sB^{\scriptscriptstyle (1)} = \Z_k ^{\scriptscriptstyle (1)}$ of Section \ref{sec:orbi}, and the magnetic $(-1)$-form symmetry $\sK^{\scriptscriptstyle (-1)}= \Z_k ^{\scriptscriptstyle (-1)}$. The background gauge fields are respectively the 2-form $\check{B}$ and the scalar $K$.

The coupling of the electric 2-form field $\check{B}$ to $F^{\,\textrm{e}}$ shifts 
\begin{equation*}
	F^{\,\textrm{e}}  \longmapsto F^{\,\textrm{e}} - \check{B} \otimes \id_N \ .
\end{equation*}
On the other hand, $K$ enters through the magnetic coupling 
\begin{equation}
\label{KFterm}
 \frac{k}{N}\,\int_{\Sigma}\, K \, \tr\, F^{\,\textrm{e}} \ .
\end{equation}
The coefficient here, which generalizes the standard coupling \eqref{magcouplingaF} to the case $k < N$, follows from \eqref{FprimeCW}, or equivalently from the fact that, by embedding $\sK^{\scriptscriptstyle (-1)}$ into the $\Uu^{\scriptscriptstyle (-1)}$ Chern--Weil symmetry, the conserved current is $\ast\, \frac{k}{N}\,\tr\, F^{\,\textrm{e}} $. In any case \eqref{KFterm} is invariant under background $\sK^{\scriptscriptstyle (-1)}$-gauge transformations, by recalling that $\int_{\Sigma}\, \tr\, F^{\,\textrm{e}}  \in 2 \pi\, \frac{N}{k}\, \Z$.

When both background gauge fields are turned on, the pertinent part of the action functional reads 
\begin{equation*}
\frac{k}{N}\,\int_{\Sigma}\, K\, \tr ( F^{\,\textrm{e}} - \check{B} \otimes \id_N ) = - k\, \int_{\Sigma}\, K\, \check{B} + 2 \pi\, \ii\, n \ ,
\end{equation*}
where $n\in\Z$.
We arrive at the natural modification of the action functional in \eqref{AbelSUN1} given by
\begin{equation}
\label{mixanomalous}
	\int_{\Sigma}\, \check{B}\,  \left( - k\, K + \kappa  \right) \ , 
\end{equation}
where we do not write explicitly all the gauge-invariant terms that define the theory, and only focus on the anomalous part. This term is invariant under a background $\check{\sB}^{\scriptscriptstyle (1)}$-gauge transformation. 

However, it is not invariant under a background $\sK^{\scriptscriptstyle (-1)}$-gauge transformation unless $\int_\Sigma\, k\, \check{B} \in 2 \pi\, \ii\, \Z$. This is not generically the case, and there are three possibilities:
\begin{enumerate}[1)]
\item If $k= \frac{N}{k}$, then the coefficient in \eqref{mixanomalous} is just enough to preserve invariance under both $\sB^{\scriptscriptstyle (1)}$ and $\sK^{\scriptscriptstyle (-1)}$ background gauge transformations. More generally, if $k$ is an integer multiple of $\frac{N}{k}$, i.e.~$k^2 \in N\, \mathbb{Z}$, then \eqref{mixanomalous} is also invariant under $\sK^{\scriptscriptstyle (-1)}$ background gauge transformations.
\item If $k^2\notin N\Z$, then \eqref{mixanomalous} is improperly quantized and is not invariant under general $\sK^{\scriptscriptstyle (-1)}$ background gauge transformations.
\item If $\mathrm{gcd}(k, \frac Nk) =m> 1$, then a subgroup $\Z_m ^{\scriptscriptstyle (1)} \subset \check{\sB}^{\scriptscriptstyle (1)}$ is anomaly-free.
\end{enumerate}
The lack of background gauge invariance if both background fields are turned on indicates the presence of a mixed anomaly.

\subsection{Mixed 2-group anomaly}
\label{sec:CandZ1}

As we have seen, charge conjugation $\sC^{\scriptscriptstyle (0)}$ acts on the 1-form symmetry $\Z_{N}^{\scriptscriptstyle (1)}$. 
We can thus consider the non-trivial symmetry extension
\begin{align*}
1\longrightarrow\Z_N\longrightarrow\Z_N \rtimes \Z_2\longrightarrow\sC\longrightarrow 1
\end{align*}
in the category of groups. This is the
semi-direct product group $\Z_N \rtimes \Z_2$ in the automorphism 2-group $\mathsf{Aut}(\SUN)\,/\!\!/\,\SUN$ of the stack of $\SUN$-bundles with connection, where the left factor is the 1-form symmetry and the right factor is charge conjugation. 

As sets
\begin{align*}
	\Z_N \cong \big\{ \e^{\, 2 \pi\,\ii\, \beta /N} \big\}_{\beta =0,1, \dots, N-1} \qquad , \qquad
	\sC \cong \{ 1, \mathcal{C} \} \ ,
\end{align*}
and $\Z_N \rtimes \Z_2$ contains the same elements as the set $\Z_N \times \sC$. The group law in $\Z_N \rtimes \Z_2$ is given by
\begin{align*}
	(\e^{\, 2 \pi \, \ii \,  \beta_1 /N} , 1) \cdot (\e^{\, 2 \pi \, \ii \,  \beta_2 /N} , 1) & = (\e^{\, 2 \pi \, \ii \,  (\beta_1+\beta_2) /N} , 1) \ , \\[4pt]
	(\e^{\, 2 \pi \, \ii \,  \beta_1 /N} , 1) \cdot (\e^{\, 2 \pi \, \ii \,  \beta_2 /N} , \mathcal{C}) & = (\e^{\, 2 \pi \, \ii \,  (\beta_1+\beta_2) /N} , \mathcal{C}) \ , \\[4pt]
	(\e^{\, 2 \pi \, \ii \,  \beta_1 /N} , \mathcal{C}) \cdot (\e^{\, 2 \pi \, \ii \,  \beta_2 /N} , 1) & = (\e^{\, 2 \pi \, \ii \,  (\beta_1-\beta_2) /N} , \mathcal{C}) \ , \\[4pt]
	(\e^{\, 2 \pi \, \ii \,  \beta_1 /N} , \mathcal{C}) \cdot (\e^{\, 2 \pi \, \ii \,  \beta_2 /N} , \mathcal{C}) & = (\e^{\, 2 \pi \, \ii \,  (\beta_1-\beta_2) /N} , 1) \ ,
\end{align*} 
for all $\beta_1, \beta_2 \in \left\{  0, 1, \dots, N-1 \right\}$.
 
The symmetry defects for $\sC^{\scriptscriptstyle (0)}$ are topological line operators of type $\eta\in\{0,1\}$, while the symmetry defects for $\Z_{N}^{\scriptscriptstyle (1)}$ are local topological operators of type $\beta\in\{0,1,\dots,N-1\}$. The action of the 0-form symmetry on local operators restricts to give an action on the generators of the 1-form symmetry. By the general argument in \cite{Tachikawa:2017gyf}, gauging either $\sC^{\scriptscriptstyle (0)}$ or $\Z_{N}^{\scriptscriptstyle (1)}$ will produce a mixed anomaly between the remaining symmetry and the Pontryagin dual symmetry that emerges after the gauging. We can make this rather explicit in the case at hand when $N \in 2 \N$.

Starting from pure $\SUN$ Yang--Mills theory on $\Sigma$, gauging the 1-form symmetry $\sZ^{\scriptscriptstyle (1)}=\Z_N^{\scriptscriptstyle (1)}$ produces the dual $(-1)$-form symmetry \smash{$\widehat\sZ{}^{\scriptscriptstyle (-1)}=\Z_N^{\scriptscriptstyle (-1)}$}, where \smash{$\widehat\sZ\cong\Z_N$} is the Pontryagin dual of $\sZ=\Z_N$. The orbifold theory is pure $\PSUN$ Yang--Mills theory on $\Sigma$ and the symmetry group is now the direct product $\Z_N\times\Z_2$ with the remaining 0-form symmetry $\sC^{\scriptscriptstyle (0)}$. Gauging \smash{$\Z_{N/k}^{\scriptscriptstyle (-1)} \subseteq \widehat\sZ{}^{\scriptscriptstyle (-1)}$} can be achieved through the magnetic coupling of a background 0-form gauge field $a$ to the $\PSUN$ curvature $ F^{\,\textrm{e}} $:
\begin{align*}
\frac{N/k}{N}\int_\Sigma\,a \ \tr\, F^{\,\textrm{e}} =\frac{N}{k}\,\int_\Sigma\,a\,B \ .
\end{align*}
Charge conjugation acts on this term as $\mathcal{C}:B\longmapsto -B$, as a result of the original non-trivial extension. This is not invariant and, by a computation completely analogous to the one below \eqref{mixanomalous}, the two configurations before and after applying $\mathcal{C}$ are not equivalent unless $k^2 \in N \Z$. Hence there is no way in general to gauge the $(-1)$-form symmetry \smash{$\Z_{N/k}^{\scriptscriptstyle (-1)}$} while preserving the 0-form symmetry $\sC^{\scriptscriptstyle (0)}$.

On the other hand, gauging the 0-form symmetry $\sC^{\scriptscriptstyle (0)}$ in $\SUN$ Yang--Mills theory yields the dual 0-form symmetry $\widehat\sC{}^{\scriptscriptstyle (0)}=\Z_2^{\scriptscriptstyle (0)}$. The orbifold theory is now \smash{$\wSUN$} Yang--Mills theory with symmetry group the direct product $\Z_N\times\Z_2$ with the remaining 1-form symmetry $\Z_N^{\scriptscriptstyle (1)}$. As illustrated in Figure~\ref{fig:2dSSL}, when a local operator of type $\beta$ is dragged across a line of type $\eta$, it acquires a phase factor $\lambda_\beta(\eta)$, where \smash{$\lambda\in Z^1(\mathrm{B}\Z_N,\widehat{\sC}\,)$} is a 1-cocycle corresponding to the non-trivial extension class. This 1-cocycle therefore becomes a mixed anomaly between \smash{$\widehat{\sC}^{\scriptscriptstyle (0)}$} and $\Z_N^{\scriptscriptstyle (1)}$, once \smash{$\sC^{\scriptscriptstyle (0)}$} is gauged~\cite{Benini:2018reh}.

\begin{figure}[htb]
\centering
\begin{tikzpicture}
\draw[gray,thick,fill,opacity=0.5] (-4,0) -- (-1,0) -- (-1,1.5) -- (-4,1.5) -- (-4,0);
\draw[Plum,thick] (-2.5,0) -- (-2.5,1.5);
\node[black] at (-3,0.75) {$\bullet$};

\node[black,anchor=east] at (-3,0.75) {$\wp_\beta$};
\node[Plum,anchor=south] at (-2.5,1.5) {$L_{\eta}$};

\node at (0.25,0.75) {$ \ = \ \e^{\,2\pi\,\ii\,\lambda_\beta(\eta)}$};
\draw[gray,thick,fill,opacity=0.5] (4.5,0) -- (1.5,0) -- (1.5,1.5) -- (4.5,1.5) -- (4.5,0);
\draw[Plum,thick] (3,0) -- (3,1.5);
\node[black] at (3.5,0.75) {$\bullet$};
\node[black,anchor=west] at (3.5,0.75) {$\wp_\beta$};
\node[Plum,anchor=south] at (3,1.5) {$L_{\eta}$};
\end{tikzpicture}
\caption{The mixed anomaly between $\widehat{\sC}^{\scriptscriptstyle (0)}$, where $\sC^{\scriptscriptstyle (0)}$ is generated by the topological line defects $L_\eta$, and $\Z_N^{\scriptscriptstyle (1)}$ generated by the topological point defects $\wp_\beta$.}
\label{fig:2dSSL}
\end{figure}

\section{Applications}
\label{sec:applications}

We conclude in this final section by briefly describing some simple implications of our results for other field theories. For further phenomenological implications, especially in higher dimensions, see~\cite{Aloni:2024jpb}.

\subsection{The massive Schwinger model}
\label{sec:Schwinger}

Let us discuss our results for the abelian case in relation with the massive Schwinger model. This is $\Uu$ gauge theory with a massive Dirac fermion in two dimensions. Let $m$ denote the bare mass of the fermion field and $g$ the gauge coupling as before. At $\theta=\pi$, the phase diagram as a function of the dimensionless parameter $\hat{m}:=\frac{m}{g}$ presents a critical point $\hat{m}_{\text{cr}}$ \cite{Coleman:1976uz,Dempsey:2023gib}. For $\hat{m}<\hat{m}_{\text{cr}}$ the theory has a unique gapped vacuum, while in the large mass phase $\hat{m}>\hat{m}_{\text{cr}}$ the vacuum is two-fold degenerate and charge conjugation symmetry is spontaneously broken \cite{Coleman:1976uz,Dempsey:2023gib}.

From our findings it is clear that, for very large mass, charge conjugation symmetry must be spontaneously broken at $\theta=\pi$. Indeed we expect that, when $\frac{m}{g} \gg 1$, one can integrate out the massive fermion and obtain pure $\Uu$ Yang--Mills theory in a first approximation. Using the bosonization of the massive Schwinger model \cite{Coleman:1976uz}, one can show that the leading irrelevant operator preserves both $\sC^{\scriptscriptstyle (0)}$ and $\Uu^{\scriptscriptstyle (1)}$. Since $\sC$ is spontaneously broken at $\theta=\pi$ in pure Yang--Mills theory, so it must also be in the massive Schwinger model when the mass is very large. In fact, after bosonization our argument becomes identical to the one used in \cite[Section~2.2]{Komargodski:2017dmc} for the two-dimensional abelian Higgs model.

\subsection{Two-dimensional adjoint quantum chromodynamics}
\label{sec:AdjQCD2}

The study of quantum chromodynamics in two dimensions (QCD$_2$) with adjoint quarks has a long history, see for example \cite{Gross:1995bp} for an influential early work and \cite{Cherman:2019hbq,Komargodski:2020mxz} for recent developments. The action is that of pure $\SUN$ Yang--Mills theory coupled to a Majorana fermion $\psi$ in the adjoint representation of $\SUN$. A major outcome of \cite{Komargodski:2020mxz} is that this theory possesses a number of non-invertible 0-form symmetries, which are explicitly broken by a mass term for $\psi$.

In this subsection we discuss implications of our analysis to orbifolds of this theory. In two Euclidean spacetime dimensions the gamma-matrices are
\begin{equation}
\label{eq:gammamatrix}
	\gamma^1 = \left( \begin{matrix} 0 & 1 \\ 1 & 0 \end{matrix}\right) \quad , \quad \gamma^2 = \left( \begin{matrix} 0 & \ii \\ -\ii & 0 \end{matrix}\right) \quad , \quad \gamma^5 = \left( \begin{matrix} -1 & 0 \\ 0 & 1 \end{matrix}\right) \ ,
\end{equation}
where, in the notation of \cite{Komargodski:2020mxz}, $\gamma^5=\ii\, \gamma^1\, \gamma^2$ implements a chirality transformation.

\subsubsection*{Adjoint QCD$_{\boldsymbol 2}$ with large quark mass}

By giving a large mass $m$ to the adjoint fermion, such that $\frac{m^2}{g^2} \gg 1 $, the theory develops a large mass gap and the low energy dynamics is captured by the appearance of massive particles on top of the vacua provided by pure Yang--Mills theory. This analysis was already performed in \cite{Komargodski:2020mxz} when the gauge group is $\SUN$ or $\PSUN$. Here we consider instead the case $\SUN/\Z_k$ for $k$ a power of $2$ that divides $N$. Since the quark field is in the adjoint representation, the 1-form symmetry is preserved. We can thus gauge the subgroup $\Z_k$ with discrete $\theta_\kappa$-angle given by $\kappa=\frac{k}{2}$. Due to the large mass, the vacua are those already found in earlier sections. In particular, $\sC$ is spontaneously broken.

Our earlier analysis also leads to a prediction for the particle spectrum in $\SUN/\Z_k$ adjoint QCD${}_2$. Let us exemplify this for $k=2$ and $\kappa=1$, with the vacua corresponding to the representations $\Box$ and $\overline{\Box}$. It is possible to act with a single adjoint fermion field on each vacuum. This is because the tensor product of the fundamental and adjoint representations expands into irreducible representations according to
\begin{equation*}
\ytableausetup{centertableaux,smalltableaux}
\Box \ \otimes \ \begin{ytableau} \ & \  \\ \  \\ \  \\ \ \\ \ \end{ytableau} \ \cong \ \Box \ \oplus \ \begin{ytableau} \ & \  \\ \ & \ \\ \  \\ \ \\ \ \end{ytableau} \ \oplus \ \begin{ytableau} \ & \ & \  \\ \ \\ \  \\ \ \\ \ \end{ytableau}
\end{equation*}
and the fundamental representation $\Box$ appears on the right-hand side. The same is true for the anti-fundamental representation $\overline{\Box}$, for which the conjugate representations appear on the right-hand side. 

The particle spectrum is thus created on top of each degenerate vacuum by acting with an arbitrary number of creation operators for the adjoint quark $\psi$, and then forming gauge singlets. The two towers of states are exchanged by charge conjugation.

\subsubsection*{Massless adjoint QCD$_{\boldsymbol 2}$}

Let us now step back to the simply-connected gauge group $\SUN$. In the opposite limit $m=0$, in which the quark field $\psi$ is massless, the theory acquires a $\Z_2^{\scriptscriptstyle (0)}$ chiral symmetry generated by the matrix $\gamma^5$ in \eqref{eq:gammamatrix}. There is a mixed anomaly between the chiral $\Z_2^{\scriptscriptstyle (0)}$ and the 1-form symmetry $\Z_N^{\scriptscriptstyle (1)}$. We shall now reinterpret this anomaly in the light of our earlier analysis.

By gauging the full $\Z_N^{\scriptscriptstyle (1)}$ we land on $\PSUN$ gauge theory with discrete parameter $\kappa$. It was shown in \cite[Section~3]{Komargodski:2020mxz} that the action of the chiral symmetry shifts $\kappa \longmapsto \kappa + \frac{N}{2}$ if $N$ is even. We reinterpret this fact by saying that, in order to remain within the same theory, the action of chiral symmetry must be accompanied by a $(-1)$-form background gauge transformation. This is because the theories with parameters $\kappa$ and $\kappa + \frac{N}{2}$ are not equivalent. Therefore $\PSUN$ adjoint QCD$_2$ possesses a higher structure mixing $\Z_2^{\scriptscriptstyle (0)}$ and the $(-1)$-form symmetry. Gauging the $(-1)$-form part, by the argument in Subsection \ref{sec:CandZ1} applied to the chiral symmetry, we predict a mixed anomaly between $\Z_2^{\scriptscriptstyle (0)}$ and the dual symmetry $\Z_N^{\scriptscriptstyle (1)}$, recovering the known results in our language.

\subsection{\texorpdfstring{$\mathrm{T}\,\overline{\mathrm{T}}$}{TT}-deformation}

The composite operator $\mathrm{T}\,\overline{\mathrm{T}}$ induces an irrelevant deformation of any two-dimensional quantum field theory with a conserved stress-energy tensor $\mathrm{T}$. It is thus not expected to modify the low energy behaviour of the theory. As a consequence, every symmetry of the original quantum field theory, be it invertible or non-invertible, is either explicitly broken by the term $\mathrm{T}\,\overline{\mathrm{T}}$ in the action functional, or otherwise it should be preserved at every point of the irrelevant flow triggered by the $\mathrm{T}\,\overline{\mathrm{T}}$-deformation. 
If the global (possibly non-invertible) symmetry suffers from an anomaly, one invokes the 't~Hooft anomaly matching condition to argue that the full symmetry structure is preserved. Indeed, the anomaly is captured by topological terms, which do not couple to the metric and thus are insensitive to the deformation by the operator $\text{T}\,\overline{\text{T}}$. We now proceed to show explicitly how the latter statement works within the context of our earlier results.

The irrelevant perturbation of two-dimensional Yang--Mills theory by the operator $\text{T}\,\overline{\mathrm{T}}$ was considered in \cite{Conti:2018jho,Santilli:2018xux,Ireland:2019vvj,Santilli:2020qvd,Gorsky:2020qge,Griguolo:2022hek,Griguolo:2022vdx}. Let $G$ be a gauge group with Lie algebra $\mathfrak{g}=\mathfrak{su}(N)$. The $\mathrm{T}\,\overline{\mathrm{T}}$-deformed partition function on a closed, connected and oriented Riemann surface $\Sigma$ is 
\begin{equation}
\label{eq:ZYMTT}
	\mz_G ^{\mathrm{T}\,\overline{\mathrm{T}}} [\Sigma] = \sum_{R\, \in\, \mathfrak{R}(\SUN)_{\tau}} \,  (\dim R)^{\chi} \ \e^{-  \frac{g^2}{2}\, C_2 ^{\mathrm{T}\,\overline{\mathrm{T}}} (R; \tau)} \ ,
\end{equation}
where $\tau \ge 0$ is the deformation parameter. In \eqref{eq:ZYMTT} we have defined 
\begin{equation}
\label{eq:TTdefC2}
	C_2 ^{\mathrm{T}\,\overline{\mathrm{T}}} (R; \tau) := \frac{ C_2 (R) }{1 - \frac{\tau}{N^3}\, C_2 (R)} \ ,
\end{equation}
while the instanton analysis of~\cite{Griguolo:2022hek} shows that the sum over irreducible $\SUN$-representations $R$ is restricted to the subset
\begin{equation*}
	\mathfrak{R}( \SUN )_{\tau} := \big\{ R\in\mathfrak{R}(\SUN) \ \big\vert \ \tau\, C_2 (R) < N^3  \big\} \ ,
\end{equation*}
which guarantees convergence of the series \eqref{eq:ZYMTT}.

Following step by step the derivation of \cite{Santilli:2020qvd} we find that the $\mathrm{T}\,\overline{\mathrm{T}}$-deformation is compatible with the orbifold. The $\mathrm{T}\,\overline{\mathrm{T}}$-deformed partition function of $\SUN/\Z_k$ Yang--Mills theory on $\Sigma$ with discrete $\theta$-angle $\kappa$ is simply
\begin{equation*}
	\mz_{\SUN/\Z_k} ^{\mathrm{T}\,\overline{\mathrm{T}}, \kappa } [\Sigma]  = \sum_{R\,\in\,\mathfrak{R}( \SUN )_{\tau}}\, (\dim R)^{\chi} \ \e^{-  \frac{g^2}{2}\, C_2 ^{\mathrm{T}\,\overline{\mathrm{T}}} (R; \tau)} ~\delta \left( \lvert R \rvert - \kappa\!\!\!\! \mod k \right) \ .
\end{equation*}
From this expression and the fact that the deformed Casimir \eqref{eq:TTdefC2} yields the same vacua as the undeformed Casimir invariant, we may state

\begin{cor}\label{cor:TTSSB}
Let $k \ge 2$ be a power of $2$ and $N \in k \N$. In $G=\SUN/\Z_k$ (resp. $G=\Uu$) $\mathrm{T}\,\overline{\mathrm{T}}$-deformed Yang--Mills theory at $\kappa=\frac{k}{2}$, hence $\theta_{\kappa}= \pi$ (resp. $\theta=\pi$), charge conjugation symmetry is spontaneously broken.
\end{cor}

This result was expected, because $\mathrm{T}\,\overline{\mathrm{T}}$ is an irrelevant operator and thus should not change the low energy behaviour of a system, while the spontaneous breaking of $\sC$ is a statement about the vacua of the theory. An alternative point of view is to say that gauging the 1-form symmetry $\Z_k^{\scriptscriptstyle (1)}$ corresponds to inserting networks of topological point defects, which are insensitive to the deformation of the metric brought about by the $\mathrm{T}\,\overline{\mathrm{T}}$-deformation.

\subsubsection*{Non-invertible 1-form symmetry}

One main aspect of the theory with partition function \eqref{eq:ZYMTT} is that the $\mathrm{T}\,\overline{\mathrm{T}}$-deformation preserves the exact Casimir scaling of the Wilson loops in Yang--Mills theory. As a consequence, the non-invertible 1-form symmetry discovered in \cite{Nguyen:2021naa} persists in the $\mathrm{T}\,\overline{\mathrm{T}}$-deformed theory. 

\subsubsection*{Spontaneous symmetry breaking in the \texorpdfstring{$\boldsymbol{\mathrm{T}\,\overline{\mathrm{T}}}$}{TT}-deformed massive Schwinger model}

Combining Corollary \ref{cor:TTSSB} with, for instance, the observations in Subsection \ref{sec:Schwinger}, we predict that charge conjugation symmetry is spontaneously broken, in the large mass phase, in the $\mathrm{T}\,\overline{\mathrm{T}}$-deform{-}ation of  the massive Schwinger model.

\subsection{Lower form symmetry in higher dimensions}
\label{sec:lowhighd}

A version of Proposition \ref{thm1} applies to four and higher even dimensions. The conclusions of this subsection overlap with the recent results of \cite{Aloni:2024jpb}.

\subsubsection*{The Gaiotto--Kapustin--Komargodski--Seiberg anomaly}

In \cite{Gaiotto:2017yup}, Gaiotto--Kapustin--Komargodski--Seiberg considered four-dimensional pure $\SUN$ Yang--Mills theory and observed a mixed anomaly between CP symmetry and the 1-form symmetry $\Z_N ^{\scriptscriptstyle (1)}$. The anomaly is signalled by the breakdown of the periodicity $\theta \sim \theta + 2 \pi$ when a background 2-form gauge field for $\Z_N ^{\scriptscriptstyle (1)}$ is turned on.

Here we consider the four-dimensional theory on a closed four-manifold $M_4$ with metric whose Hodge star operator is denoted $\ast$. We use the $\theta$-term 
\begin{equation*}
	- \ii\, \theta\, \int_{M_4}\, c_2 (\mathrm{ad}\,P) 
\end{equation*}
where $c_2 (\mathrm{ad}\,P)$ is the second Chern class of the adjoint bundle associated to a principal $\SUN$-bundle $P \longrightarrow M_4$. Since $c_2 (\mathrm{ad}\, P)$ is closed, we can consider the Chern--Weil $(-1)$-form symmetry $\Uu^{\scriptscriptstyle (-1)}$ with conserved 0-form current $\ast\, c_2 (\mathrm{ad}\, P)$. Turning on a background $0$-form gauge field $a$ as a source for this current, the $\theta$-term becomes 
\begin{equation}
\label{GKKSterm}
	-\ii\,\int_{M_4}\, c_2 (\mathrm{ad}\,P) \, \left( \theta  + \ii\, a \right) \ ,
\end{equation}
where now the periodicity $\theta \sim \theta + 2 \pi$ is reinterpreted as the invariance under $(-1)$-form gauge transformations. 

After coupling this system to a background 2-form gauge field $B$, \eqref{GKKSterm} ceases to be invariant under shifts $\theta \longmapsto \theta + 2 \pi$. In the language of $(-1)$-form symmetries, $B$ breaks invariance under background $(-1)$-form gauge transformations. The Gaiotto--Kapustin--Komargodski--Seiberg anomaly is thus a mixed anomaly between the 1-form and $(-1)$-form symmetries.

\subsubsection*{Lower form symmetry in even dimensions}

The previous discussion generalizes in the following way. For any closed $2n$-dimensional manifold $M_{2n}$, with metric and corresponding Hodge star operator $\ast$, we consider a principal $G$-bundle $P \longrightarrow M_{2n}$ whose adjoint bundle has non-trivial $n^{\text{th}}$ Chern class: $c_n (\mathrm{ad}\,P) \ne 0$. Then $G$ Yang--Mills theory on $M_{2n}$ admits the $\theta$-term 
\begin{equation*}
	\ii^{1-n} \, \theta \, \int_{M_{2n}}\, c_n (\mathrm{ad}\,P) \ ,
\end{equation*}
or its appropriate modification as in \eqref{kappaB} for a discrete parameter $\kappa$. The $2\pi$-periodicity of $\theta$, or alternatively the periodic identification $\kappa \sim \kappa + k$, can be interpreted as the gauge invariance under $(-1)$-form Chern--Weil symmetries associated to the conserved 0-form current $\ast\, c_n (\mathrm{ad}\,P)$.

Turning on a background 2-form gauge field $B$ for the 1-form symmetry $\sZ^{\scriptscriptstyle (1)}:=\sZ (G)$ modifies $c_n (\mathrm{ad}\,P)$ by replacing $F \longmapsto F-B \otimes \id$ in its representation as a Chern--Weil form. Generically, the mixed terms in the expansion involving
\begin{equation*}
	 \frac{\ii\,\theta}{(2 \pi)^n} \,  \int_{M_{2n}}\, B^j \wedge \tr\, F^{l} 
\end{equation*}
may spoil the periodicity of the $\theta$-parameter. By coupling the conserved current $\ast\, c_n (\mathrm{ad}\,P)$ to a background 0-form gauge field $a$, we recast the breakdown of periodicity as a mixed anomaly between the 1-form symmetry and the $(-1)$-form symmetry.

\begin{appendix}

\section{Notation and conventions}
\label{app:notation}

\noindent $\Sigma : \ $ A closed, connected and oriented Riemann surface on which the field theory lives.

\noindent $\chi : \ $ The Euler characteristic of $\Sigma$.

\noindent $\omega : \ $ The K\"ahler form on $\Sigma$, normalized to unit area $\int_{\Sigma}\, \omega = 1$.

\noindent $S : \ $ A surface operator.

\noindent $L : \ $ A line or loop operator.

\noindent $G : \ $ A Lie group, taken to be the gauge group.

\noindent $\mathfrak{g} : \ $ The Lie algebra of $G$, called the gauge algebra.

\noindent $\sZ (G) : \ $ The centre of $G$.

\noindent $\mathfrak{R}(G) : \ $ The Grothendieck ring of the representation category of $G$.

\noindent $\chi_R : \ $ The character of a representation $R\in\mathfrak{R}(G)$.

\noindent $\sA, \sB,\sK : \ $ Finite abelian groups, taken to be global symmetries.

\noindent $\sC : \ $ The $\Z_2$ subgroup of outer automorphisms in $\mathsf{Out}(G)$ (when non-trivial), alias charge conjugation.

\noindent $\wSUN : \ $ The principal extension of the Lie group $\SUN$ by $\sC$.

\noindent \smash{$\mathbbm{X}_1 \doublerightarrow{\ \ }{ \ } \mathbbm{X}_0: \ $} A groupoid with maps to the source and target objects in $\mathbbm{X}_0$ of a morphism in~$\mathbbm{X}_1$.

\noindent $\mathrm{B}H : \ $ The delooping groupoid \smash{$ H \doublerightarrow{\ \ }{ \ } 1$} of a group $H$.

\noindent $\mathbbm{U}(N) : \ $ The Lie 2-group central extension of the Lie group $\SUN$ by $\text{B}\Uu$.

\noindent $A : \ $ A connection on a $G$-bundle, alias a gauge field.

\noindent $F : \ $ The curvature of $A$.

\noindent $B : \ $ A curving on a gerbe, alias a 2-form gauge field. 

\noindent $\id_N : \ $ The $N{\times} N$ identity matrix. 

\noindent $\mathscr{A} : \ $ The affine space of $G$-connections on $\Sigma$.

\noindent $\mz_G : \ $ The partition function of the $G$ gauge theory on $\Sigma$. 

\bigskip

We work in the conventions in which the Chern--Weil form representing the first Chern class of a bundle is given by the curvature of a connection divided by $2 \pi$, that is $c_1 = \tr\big(\frac{F}{2\pi}\big)$, so that $\int_{\Sigma}\,\tr\big(\frac{F}{2\pi}\big) \in \mathbb{Z}$. Conventions for gerbes are adapted with same factors of $\frac{1}{2\pi}$. Our normalization and conventions for the fields follow \cite{Blau:1993hj}, and in particular the gauge field $A$ is anti-Hermitian. The imaginary unit $\ii = \sqrt{-1}$ appearing in the action functionals is adapted to this normalization.

To avoid confusion, we specify a $p$-form symmetry group with a superscript~${}^{\scriptscriptstyle (p)}$. When a group is indicated without subscript, we refer to its group and topological properties, regardless of its action in the field theory.
The notation $\Z_k$ indicates the cyclic subgroup of $\Uu$ consisting of the $k^{\text{th}}$ roots of unity, and we denote a $p$-form cyclic symmetry as $\Z_k ^{\scriptscriptstyle (p)} \subset \Uu^{\scriptscriptstyle (p)}$. To distinguish an element $x \in \mathfrak{u}(1)$ tangent to $\Uu$ at a $k^{\text{th}}$ root of unity from the root itself, we write $x \in \left\{0,1, \dots , k-1 \right\}$.

The symbol $\cong$ is used to denote isomorphisms of objects in a particular category (of groups, rings, vector spaces, etc.). For example, we write $R_1 \cong R_2$ if and only if the two representations $R_1$ and $R_2$ are isomorphic.

\section{Orthogonal gauge algebras}
\label{app:Ortho}

\subsection{\texorpdfstring{$\theta$-terms}{theta-terms} and anomalies}
\label{sec:othergauge}

We consider the family of theories with gauge algebra $\mathfrak{so}(N)$ where $N$ is even, and analyse the possible $\theta$-angles and anomalies. The ensuing analysis is inspired by~\cite{Hsin:2020nts}, and adapts the ideas therein to two dimensions.

Consider the Lie group $\SpiN$. Its centre is given by
\begin{equation*}
	\sZ (\SpiN) = \begin{cases} \ \Z_2 \times \Z_2 \ , & \quad N=0 \!\!\!\mod 4 \ , \\  \ \Z_2 \ ,  & \quad N=1 \!\!\!\mod 4 \ , \\ \ \Z_4 \ , & \quad N= 2 \!\!\!\mod 4 \ , \\ \ \Z_2 \ ,  & \quad N=3 \!\!\!\mod 4 \ . \end{cases}
\end{equation*}
Closely related is the group $\SON$, of which $\SpiN$ is an extension by the group $\sA = \Z_2$:
\begin{equation*}
	1 \longrightarrow \sA \longrightarrow \SpiN \longrightarrow \SON \longrightarrow 1 \ .
\end{equation*}
Principal $\SON$-bundles $P\longrightarrow\Sigma$ are characterized by their second Stiefel--Whitney class $w_2(P) \in H^2 (\Sigma, \sA)$, obstructing the lift of $P$ to a $\SpiN$-bundle.

On the other hand, the group $\ON$ is an extension of $\SON$ by a different  group $\sC= \Z_2$:
\begin{equation*}
	1 \longrightarrow \sC \longrightarrow \ON \longrightarrow \SON \longrightarrow 1  \ .
\end{equation*}
This $\Z_2$ group is identified with charge conjugation. The obstruction to reducing an $\ON$-bundle $P\longrightarrow\Sigma$ to an $\SON$-bundle is its first Stiefel--Whitney class $w_1(P) \in H^1 (\Sigma, \sC)$. One can also combine the two $\Z_2$ extensions of $\SON$ to form the group $\PiN$.

Finally, for even $N$, one can further take the quotient $\mathrm{P}\SON = \SON/\Z_2$, which corresponds to gauging the intrinsic 1-form symmetry.

For $N=2 \!\!\mod 4$, we can start from $\SpiN$ Yang--Mills theory, based on a principal $\SpiN$-bundle $P\longrightarrow\Sigma$, and gauge the 1-form symmetry $\sZ^{\scriptscriptstyle (1)} = \Z_4 ^{\scriptscriptstyle (1)}$ in two steps. There is a non-trivial group extension 
\begin{equation}
\label{Z2Z4Z2}
	1 \longrightarrow \sA \longrightarrow \Z_4 \longrightarrow \sB \longrightarrow 1 
\end{equation}
with $\sA\cong \sB \cong \Z_2$ (we denote them with different symbols to avoid confusion). By gauging $\sA$ we obtain $\SON$ Yang--Mills theory. 

In more detail, we turn on the background 
\begin{equation*}
	\beta := w_2 ^{\sB}(P) + 2\, w_2 ^{\sA}(P) \ \in  \ H^2 (\Sigma , \Z_4 ) \ ,
\end{equation*}
where $w_2 ^{\sA}(P)$ corresponds to the Stiefel--Whitney class we want to sum over to pass from $\SpiN$ to $\SON$ gauge theory, and $w_2 ^{\sB}(P)$ is the characteristic class of the bundle $P$ for the remaining $\sB^{\scriptscriptstyle (1)} = \Z_2 ^{\scriptscriptstyle (1)}$ 1-form symmetry. In the notation of the main text, we embed $\sA$ and $\sB$ into $\Uu$ to obtain the identifications 
\begin{equation*}
	w_2 ^{\sA}(P) = \frac{2 \, a}{2\pi} \qquad , \qquad w_2 ^{\sB}(P) = \frac{2 \, B}{2\pi}
\end{equation*}
for 2-form gauge fields $a$ and $B$, respectively. A $\Z_4$-gauge transformation shifts $\beta$ by an element of $H^2 (\Sigma, 4\Z)$, corresponding to the standard gauge transformation of the fields $a$ and $B$.

Coupling the $\SpiN$ gauge theory to such a background allows for a discrete $\theta$-angle classified by $\Z_4$, which we write as 
\begin{equation*}
	2 \pi\, \frac{\kappa}{4}\, \int_{\Sigma}\,  \beta = \pi\, \kappa\, \int_{\Sigma}\, \left( \frac{1}{2}\, w_2 ^{\sB}(P) + w_2 ^{\sA}(P)  \right)  \ .
\end{equation*}
Here $\kappa $ has periodicity $\kappa \sim \kappa +4$. To gauge $\sA^{\scriptscriptstyle (1)}$ we sum over topological classes of bundles labelled by $w_2 ^{\sA}(P) \in H^2 (\Sigma , \sA)$.

Having gauged $\sA^{\scriptscriptstyle (1)}$, the $\SON$ gauge theory has a Pontryagin dual $\sK ^{\scriptscriptstyle (-1)} = \widehat{\sA}{}^{\scriptscriptstyle (-1)} $ $(-1)$-form symmetry. We want to couple the theory to a 0-form gauge field $K$ for this $\sK ^{\scriptscriptstyle (-1)} = \Z_2^{\scriptscriptstyle (-1)} $ $(-1)$-form symmetry. By reasoning along the lines of \cite[Section~4]{Hsin:2020nts}, but now with the magnetic symmetry being a $(-1)$-form symmetry, we find the coupling 
\begin{equation*}
	\pi\, \int_{\Sigma}\, w_2^{\sA}(P)\,  \left( \kappa -2\, \ii\, K  \right) + \frac{\pi}{2}\, \int_{\Sigma}\,  w_2 ^{\sB}(P)\, \left( \kappa -2\, \ii\, K  \right) \ ,
\end{equation*}
with the understanding that $w_2^{\sA}(P)$ is summed over while $w_2 ^{\sB}(P)$ and $K$ are background fields which are kept fixed. The first term is gauge-invariant, and it is also invariant under $(-1)$-form background gauge transformations. The second term, however, is not invariant under $\sK ^{\scriptscriptstyle (-1)} $ background gauge transformations. This is a manifestation of the mixed anomaly of \cite{Tachikawa:2017gyf}, except that here it involves the lower form symmetry.

\subsection{Orbifolds and spontaneous symmetry breaking}
\label{app:breaksoN}

We study orbifolds of two-dimensional Yang--Mills theory with gauge group $G$ whose gauge algebra is $\mathfrak{g} = \mathfrak{so}(N)$, and analyze the fate of charge conjugation symmetry.

Let us focus on $G=\SON$ with $N\in 2 \N$; throughout this subappendix only we write $N=2r$. Its tensor representations are in correspondence with Young diagrams of row lengths at most $r$, where the last row length comes with a sign. Spinor representations can also be put in correspondence with Young diagrams whose row lengths are shifted by a half-integer. Of particular interest to us are the following representations:
\begin{itemize}
\item The vector representation $\Box$, with Casimir $C_2 (\Box) =N-1 $. 
\item The spinor representation $\mathsf{s}$ corresponding to $R_i = \frac{1}{2}$ for all $i=1, \dots, r$, with Casimir $C_2 (\mathsf{s}) =\frac{N}{8}\,(N-1) $.
\item The cospinor representation $\mathsf{c}$ corresponding to $R_i = \frac{1}{2}$ for all $i=1, \dots, r-1$ and $R_r=-\frac{1}{2}$, with Casimir $C_2 (\mathsf{c}) =\frac{N}{8}\,(N-1) $.
\end{itemize}
The spinor and cospinor representations have equal quadratic Casimir invariant as well as equal dimension, and are exchanged under charge conjugation.

We now gauge the 1-form symmetry $\sB^{\scriptscriptstyle (1)} = \Z_2 ^{\scriptscriptstyle (1)}$ and turn on a discrete $\theta$-angle with $\kappa =1$ and periodicity $\kappa \sim \kappa +2$. When $N=2 \!\!\mod 4$, and hence $r$ is odd, the three representations listed above survive the gauging. When $N=6$, the spinor and cospinor representations have lowest quadratic Casimir invariant, thus we obtain a degenerate vacuum and charge conjugation symmetry is spontaneously broken in $\mathrm{PSO}(6)$ Yang--Mills theory. However, for $N>8$ the vector representation has lower quadratic Casimir invariant.

We note that for $N=8$, the vector, spinor and cospinor representations of $G=\mathrm{SO}(8)$ all have equal quadratic Casimir invariant and equal dimension, as a consequence of the enhancement of the outer automorphism group of the Lie algebra $\mathfrak{so}(8)$ from the cyclic group $\Z_2$ of order two to the symmetric group $S_3$ of degree three, i.e. $\mathrm{SO}(8)$-triality. It would be interesting to engineer a non-abelian orbifold of $\mathrm{SO}(8)$ which removes the trivial representation, so that the vacuum is three-fold degenerate and the triality $S_3$ is spontaneously broken.

\end{appendix}

\bibliography{YMsymmetry}

\end{document}